\documentclass[11pt]{article}
\usepackage{mathrsfs}
\usepackage{amsmath}
\usepackage{amsfonts}
\usepackage{amssymb, amsmath, cite}
\usepackage{color}
\usepackage{graphicx}
\usepackage{float}

\newtheorem{theorem}{Theorem}

\newcommand\be{\begin{equation}}
\newcommand\ee{\end{equation}}
\newcommand\ber{\begin{eqnarray}}
\newcommand\eer{\end{eqnarray}}
\newcommand\berr{\begin{eqnarray*}}
\newcommand\eerr{\end{eqnarray*}}\newcommand\eq{\eqref}
\newcommand\bea{\begin{eqnarray}}
\newcommand\eea{\end{eqnarray}}

\newcommand\bfR{\mathbb{R}}\newcommand\dd{\mbox{d}}
\newcommand\e{\mathrm{e}}\newcommand\x{{\bf x}}
\newcommand\pa{\partial}
\newcommand{\vp}{\varphi}
\newcommand{\nn}{\nonumber}

\newcommand{\Lm}{{\cal L}_{\mbox{\tiny M}}}
\newcommand{\lb}{\label}
\newcommand{\vep}{\varepsilon}

\title{Electromagnetic Asymmetry, Relegation of Curvature Singularities\\ of Charged Black Holes, and Cosmological Equations of\\ State in View of the Born--Infeld Theory}

\author{
Yisong Yang\footnote{Email address: yisongyang@nyu.edu}\\Courant Institute of Mathematical Sciences\\ New York University\\New York, New York 10012, USA
}

\date{}

\begin{document}
\maketitle
\begin{abstract}
It is shown that, in a generic sense, and 
unlike what for electric point charges, the Born--Infeld nonlinear electrodynamics excludes monopoles as finite-energy magnetic
point charges, thus spelling out an electromagnetic asymmetry. Moreover, it is demonstrated, in a systematic way, that the
 curvature singularities of finite-energy charged black holes in the context of the Born--Infeld theory 
 may effectively be relegated
or in some cases removed under a critical mass-energy condition,
which has been employed successfully in earlier concrete studies. Furthermore, it is illustrated through numerous
examples 
considered here that, when 
adapted to describe scalar-wave matters known as k-essences, the Born--Infeld formalism provides a fertile ground for cosmological
applications,
including achieving accelerated dark-energy expansions and acquiring adequate field-theoretical realizations of the equations of state of various cosmic fluid models.
\medskip

\medskip

\medskip


\end{abstract}

\section{Introduction}

The original motivation of the work of Born and Infeld \cite{BI} in extending the Maxwell theory of electromagnetism
to that of a nonlinear setting is to overcome the puzzle of energy divergence of an electric point charge which serves
to model the electron classically as a point particle. Their idea came from considering the action principle of a particle
in free relativistic motion that naturally imposes an upper bound to the velocity of the particle, such that, when an analogous 
formalism is introduced in the context of electromagnetism, one is led to the Born--Infeld theory, which succeeds in resulting in a
bounded electric field so that an electric point charge miraculously carries a finite energy as desired. In contemporary theoretical
physics, the Born--Infeld theory also plays an active and important role. For example, it is shown 
 to arise in string and brane theories \cite{CM,FT,Gibbons,Ts1,Ts2}, and its idea is applied to modify the Einstein--Hilbert action in an
attempt \cite{DG} to regularize gravity theory. See \cite{JHOR} for a recent comprehensive survey on subsequent development
and progress of the subject, including an extensive collection of literature.
In this work, we aim to demonstrate that the Born--Infeld theory may be explored to uncover an asymmetry between
 electricity and magnetism,  effectively produce charged black hole solutions free of curvature singularities when coupled with the Einstein equations, and provide a field-theoretical interpretation in view of scalar-wave matters for an arbitrarily given 
equation of state of a cosmological fluid, among other subjects. Specifically, it will be shown that imposing an upper bound to
the electric field is not necessary to acquire a finite-energy electric point charge, although generically, a theory of nonlinear electrodynamics renders energy divergence for a magnetic point charge. In other words, there is an asymmetry between
electric and magnetic point charges, in general, in nonlinear electrodynamics.  Moreover, it will be seen how
 finite-energy electric and magnetic
point charges may be used to relegate or remove curvature singularities  in a systematic formalism. Furthermore, it will be illustrated through examples that the Born--Infeld scalar-wave
matters provide a wide range of cosmological applications when interpreted as k-essences. Below we elaborate on these subjects separately in some detail.

First, recall that the Maxwell theory enjoys a perfect symmetry between its electric and magnetic sectors
through the so-called electromagnetic duality. Such a symmetry is carried over to the situations of electric and magnetic point
charges so that both suffer the same energy divergence.
Such a symmetry is preserved by the Born--Infeld theory \cite{BI} in which electric and magnetic point charges both enjoy
the same energy convergence or finiteness. In other words, in both Maxwell and  Born--Infeld theories,
electric and magnetic point charges are given equal footing energetically, and the theories offer no 
mechanism to rule out the occurrence of a magnetic point charge, i.e., a monopole \cite{Curie,Dirac}. Although the notion
of monopoles is fascinating and important in field theory physics \cite{GO,Pre,Ra,Wein}, such magnetically charged point particles have never been
observed in nature or laboratory in isolation, except for some of their condensed-matter-system simulations \cite{Gib,Raj,Yak}.
The idea of the Born--Infeld theory opens the opportunity of modifying the Maxwell theory to break the electromagnetic
symmetry so that a finite-energy electric point charge is maintained but a finite-energy magnetic point charge is excluded.
In fact, such models are known to exist in literature. For example, we will see that in the $\arcsin$-model \cite{K3,K4} and a subclass of the
fractional-powered model \cite{K5},  monopoles are not allowed, although electric point charges are allowed, in view of the
finite-energy condition. Instead of considering individual models, here, we shall show that such a breakdown of electric and
magnetic point charge symmetry, referred to as electromagnetic asymmetry, may be regarded as a {\em generic} property of nonlinear electrodynamics. More precisely, it will be established that, for any nonlinear electrodynamics governed by a polynomial 
function, the theory always accommodates finite-energy electric point charges but excludes magnetic ones, although unlike
what is seen in the classical Born--Infeld model, no
upper bound for electric field may be imposed in the current context. The word `generic' is used to refer to the fact that the set of 
polynomials is dense in the function space of nonlinear Lagrangian functions in view of the Stone--Weierstrass theorem \cite{Stone,Yosida} such that
any model of nonlinear electrodynamics may be approximated in a suitable sense by a sequence of models governed by polynomials.
As a consequence, we may conclude that monopoles are generically ruled out with regard to the finite-energy condition. These results are summarized in Theorem \ref{th1}.

Next, we explore the implications of finite-energy electric and magnetic point charges to the construction of charged black holes
with regard to the associated curvature singularities. It is well known that, in terms of the spherical coordinates, the 
Kretschmann curvature (say), $K$, of the Schwarzschild solution, near the center of the black hole, is of the type $K\sim r^{-6}$. 
For the Reissner--Nordstr\"{o}m charged black hole solution, on the other hand, the curvature singularity is elevated to the type
$K\sim r^{-8}$. In view of the Born--Infeld theory formalism, it is readily seen that such an elevation is due to the divergence of
energy of an electric or magnetic point charge in the linear Maxwell electrodynamics. Part of the goal of the present work is to show
that the finite-energy point charge structure in the Born--Infeld type nonlinear electrodynamics in general gives rise to 
non-elevated curvature singularities, $K\sim r^{-6}$. In other words, in view of curvature singularities, finite mass and finite
electromagnetic energy present themselves at an equal footing. Furthermore, we show that, when a critical mass-energy condition is fulfilled,
the Schwarzschild type curvature singularity may be relegated or even eliminated in a systematic way.
In this regard, the much studied Bardeen black hole \cite{AG1,AG2,BV,Paula} and the Hayward black hole \cite{F,Hay,Kumar} belong to this category of the
regular black hole
solutions of the Einstein equations coupled with the Born--Infeld type nonlinear electrodynamics for which the critical mass-energy
condition is embedded into the specialized forms of the Lagrangian or Hamiltonian densities.
In the context of our general and unified treatment, we shall see that the Bardeen and Hayward black holes are among some broad
families of regular charged black hole solutions. Thus such a treatment leads us to unveil new families of regular charged black hole
solutions beyond the Bardeen and Hayward solutions. On the other hand, however, a common feature shared by the Bardeen
and Hayward black holes is that the form of their Lagrangian action density can only accommodate one sector of electromagnetism,
that is, either electric field or magnetic field is allowed to be present to model a point charge, but not both, due to the sign
restriction under the radical root operation, in a sharp contrast to the classical Born--Infeld theory \cite{BI} which accommodates both
electricity and magnetism at an equal footing. Hence it will be of interest to find by taking advantage of our general formalism a model of nonlinear electrodynamics that accommodates both electric and magnetic point charges and at the same time gives rise to regular charged black hole solutions
as in the Bardeen and Hayward models. Indeed, such a model may be obtained by taking the large $n$ limit in a 
naturally formulated binomial model, which is a special situation of the polynomial model. Surprisingly, the binomial model does not
allow a finite-energy magnetic point charge but its large $n$ limit, which assumes the form of an exponential model considered earlier
by Hendi in
\cite{H1,H2} in other contexts, accommodates both electric and magnetic point charges, and is free of sign restriction. 
Within such a
framework, it becomes a ready task to uncover other models which may offer regular charged black hole solutions. For example,
another somewhat more complicated exponential model considered by Kruglov in \cite{K} is also shown to give rise to regular magnetically charged
black hole solutions and at the same time accommodates a finite-energy electric point charge. Interestingly, Ma \cite{Ma} obtained
 regular magnetically charged black hole solutions arising from a rational-function model of Kruglov \cite{K7,K8,K9} in which both
electric and magnetic point charges of finite energies are allowed. For this model, we shall see that electrically charged black holes
with a relegated curvature singularity, with $K\sim r^{-4}$, in the same setting are also present. This rich collection of results is put together as Theorem \ref{th2}.

Recall that, in order to obtain a theoretical interpretation of the observed accelerated expansion of the universe attributed to
the existence of dark energy, a real scalar-wave field, often referred to as `quintessence', is introduced to 
explain the rise of  a hidden propelling force \cite{CDS,Car,Dv,RP,Tsu},  or  the `fifth force'.
The quintessence is assumed to be governed canonically as in the usual Klein--Gordon model such that the kinetic energy density term is minimal and the potential energy density may be adjusted to give rise to the desired dynamic evolution of the
universe \cite{Kep,Linde1,Linde2,Tos}.
When the kinetic density term is taken to be an adjustable nonlinear function of the canonical kinetic energy density,
the scalar-wave matter is referred
to as the k-essence \cite{CGQ,DT,JMW,PL}. This nonlinear kinetic dynamics is of the form of the
 Born--Infeld theory and may be used to provide a field-theoretical interpretation of an
arbitrarily given equation of state relating the pressure and density of a cosmological fluid. Over such a meeting ground,
we examine the cosmic fluid contents or interpretations of several generalized Born--Infeld type models considered earlier in the paper. In particular, we
shall obtain a full description of the dynamics of the cosmological evolution driven by the exponential model,
as a kind of k-essence, used to generate
regular charged black hole solutions. In this situation, the equation of state is found to be determined in terms of the Lambert $W$
function which gives rise to a big-bang cosmology characterized by infinite curvature, density, and pressure at the start of time,
and vanishing density and pressure at  time infinity, such that the ratio of the pressure and density monotonically interpolates the
dust and stiff-matter fluid models. Besides, the adiabatic squared speed of sound of the fluid of the exponential model is seen to stay
within a physically correct range. These results are summarized as Theorem \ref{th3}.

The content of the rest of the paper is organized as follows. In Section 2, we introduce the formalism of the generalized Born--Infeld
electrodynamics and fix the notation. In Section 3, we present a discussion of the generalized Born--Infeld theory governed by
an arbitrary polynomial function. We show that in all nonlinear situations an electric point charge will carry a finite energy but the 
energy of a magnetic point charge always diverges. Thus, energetically and generically, we see that the Born--Infeld type nonlinear electrodynamics provides a
mechanism to exclude monopoles. In Section 4, a systematic study is carried out for the quadratic model which is a minimal
polynomial model. For an electric point charge, we compute the induced electric field, energy, and free charge distribution, and
estimate the implied electron radius along the work of Born and Infeld \cite{BI}. For a magnetic point charge, we show that, in
addition to the energy, its free magnetic charge determined by the induced magnetic intensity field is also divergent. In other
words, the quadratic model rejects a monopole by the divergence of both its  energy and free magnetic charge, which serves
as a further evidence of the electromagnetic asymmetry in nonlinear electrodynamics. In Section 5, we consider a few generalized
Born--Infeld models which have been used in literature to produce charged black hole solutions and accelerated expanding
solutions in cosmic evolution problems. Here we examine the issues of energy convergence and divergence of electric and magnetic point charges
which are useful later when we address the question of charged black hole solutions with relegated curvature singularities at
various levels. In Section 6, we study a family of the Born--Infeld type fractional-powered models labeled by an exponent
$0<p<1$ such that the classical radical root model of Born and Infeld corresponds to $p=\frac12$. It will be shown that $p$
gives rise to a division for the range of $p$ so that electric point charges always carry finite energies but magnetic point charges
can carry finite energies only when $p$ satisfies $0<p<\frac34$. In Section 7, we consider dyonic point charges.  In the models 
considered, the energy always diverges. Nevertheless, we will aim to describe how the energy divergence happens with regard to
passing from a dyonic point charge to a magnetic one or an electric one. In particular, we will observe that electromagnetic asymmetry often occurs at a dyonic level near the center of the charge but symmetry is resumed asymptotically away from the
center of the charge. Furthermore, we will also see that, energetically, a dyonic point charge diverges faster than a sole charge
near the center of the charge. In Section 8, we present a general discussion of electrically charged black hole solutions and illustrate
how the finiteness of electric energy enables us to place mass and electric energy at an equal footing with regard to the
associated curvature singularity, which relegates the singularity of the classical Reissner--Nordstr\"{o}m black hole carrying
a divergent energy. In Section 9, we exhibit a series of examples of electrically charged black holes which all display the 
anticipated relegated curvature singularities due to the finiteness of electric energies. As a by-product, we recognize a critical
mass-energy condition under which the curvature singularities at the center of the black holes are further relegated 
to various extents beyond that
of the Schwarzschild black hole. In Section 10, we present a general discussion of dyonically charged black hole solutions and formulate an expression that gives rise to the black hole metric factor in terms of the prescribed electric and magnetic charges. This
formalism is useful for our later developments in search for regular charged black hole solutions in a systematic way. We 
emphasize by
 presenting a few generalized Reissner--Nordstr\"{o}m charged black hole solutions arising from the Born--Infeld type
electrodynamics
 that, unlike that in the classical Born--Infeld model, electromagnetic asymmetry occurs in general even asymptotically near spatial
infinity. In Section 11, we study how curvature singularities of magnetically charged black hole solutions may be effectively
relegated. Due to the way how the magnetic sector presents itself in an electromagnetic Lagrangian action density, magnetic charges are readily prescribed to render the associated Hamiltonian energy density in a directly computable form which
determines
the black hole metric factor explicitly. A series of examples will be worked out to demonstrate how curvature singularities are
relegated in general and under a special critical mass-energy condition similar to that in the electric charge situation.
In particular, we will show that, under the critical mass-energy condition, magnetically charged black hole solutions
arising from the exponential model \cite{K} and the $\arctan$-model \cite{K2}, introduced by Kruglov, are singularity free.
In a sharp contrast, we also show by the polynomial model that curvature singularity may be elevated or aggravated in an
arbitrary manner. In Section 12, we consider the Bardeen type black holes. The Bardeen black hole \cite{Bardeen} is known to 
provide an example of
a charged black hole solution free of curvature singularity.  Later, Ay\'{o}n--Beato and Garcia \cite{AG1,AG2} formulated a
Born--Infeld theory interpretation of such a solution. Such a solution,
in fact, is an example of charged black holes
subject to the critical mass-energy condition, in the context of generalized Born--Infeld theory. Indeed, in this section, we will
extend the studies in \cite{AG1,AG2} to the case of a general family of models so that the Bardeen black hole is an example
among an integrable class of models made significant by the Chebyshev theorem \cite{CGY,CGY2,CGLY,MZ,T}.
In Section 13, we consider the Hayward black hole \cite{BP,F,Hay,Kumar} and its extensions. As that for the Bardeen black hole, 
we show that there is a full family of regular black hole solutions 
arising from a generalized Born--Infeld theory of a similar structure as that for the Bardeen black hole. Moreover,
unlike the Bardeen black hole family discussed in the previous section, the Hayward family of 
solutions may all be constructed through an explicit integration, without the Chebyshev type obstruction seen in Section 12.
In Section 14, we consider an exponential Born--Infeld type model introduced by Hendi \cite{H1,H2} which allows both electric and magnetic point charges of
finite energies. We shall obtain electrically charged black hole solutions with a relegated curvature singularity 
and magnetically charged black hole solutions free of singularity, both under the critical mass-energy condition as described in the earlier sections. Furthermore, we show that the electrically charged black hole solutions satisfy the strong
energy condition, which explains why
the singularity is relegated but not removed, and the magnetically charged ones violate the strong energy condition, which 
explains why the singularity disappears, in agreement of the Hawking--Penrose singularity theorems \cite{HE,MTW,Wald}.
In Section 15, we apply the idea of generalized Born--Infeld theory to the study of cosmic evolution of
an isotropic and homogeneous universe. More specifically,  we adapt some models of
nonlinear electrodynamics to scalar-wave matters along
the concept of
 k-essence cosmology and work out the associated equations of state for the underlying cosmological fluids described. In
particular, we present a detailed study of the exponential model, considered in the previous section to produce regular
charged black hole solutions, as a prototype example, and show that it offers a full range of desired dynamical behavior including the big-bang scenario
and dark-energy evolution, expressed exactly in terms of the Lambert $W$ function. Interestingly, this model is shown to
undergo a transition from dust to stiff matters in a monotonic manner. As a by-product, we shall see
by calculating the associated adiabatic squared speed of sound, for example, that the k-essence realizations of the Born--Infeld
models may serve as a convenient checkpoint regarding their physical content or meaningfulness. In Section 16, we briefly summarize
the results of this work and make some concluding comments.

\section{General formalism and notation}

We work on the Minkowski spacetime with the temporal and spatial coordinates $x^0=t, (x^i)={\bf x}$, and metric $(\eta_{\mu\nu})=\mbox{diag}
(1,-1,-1,-1)$, so that the speed of light is unity.
Use the notation
\be\lb{2.1}
\Lm=-\frac14 F_{\mu\nu}F^{\mu\nu}
\ee
for the usual Maxwell electromagnetic action density induced from a gauge field $A_\mu$ with $F_{\mu\nu}=\pa_\mu A_\nu-
\pa_\nu A_\mu$ being the electromagnetic fields such that ${\bf E}=(E^i)=(F^{i0})$ and ${\bf B}=(B^i)=\left(-\frac12\vep^{ijk}F_{jk}\right)$ are the electric and magnetic fields, respectively, or in a matrix form,
\be\lb{2.2}
(F^{\mu\nu})=\left(\begin{array}{cccc}0&-E^1&-E^2&-E^3\\ E^1&0&-B^3&B^2\\E^2&B^3&0&-B^1\\E^3&-B^2&B^1&0\end{array}
\right).
\ee
We now take the point of view that the action density \eq{2.1} is the leading-order approximation of the a general action
density
\be\lb{2.3}
{\cal L}=f(\Lm),
\ee
where $f(s)$ is an analytic function of the real variable $s\in\bfR$ satisfying the normalization condition
\be\lb{2.4}
f(0)=0,\quad f'(0)=1.
\ee

Consider an applied source current $(j^\mu)$. Then the full action density reads
\be\lb{2.5}
{\cal L} -A_\mu j^\mu.
\ee
Thus, variation of the gauge field leads to the Euler--Lagrange equations associated with \eq{2.5}:
\be\lb{2.6}
\pa_\mu P^{\mu\nu}=j^\nu,
\ee
where
\be\lb{2.7}
(P^{\mu\nu})=(f'(\Lm)F^{\mu\nu})=\left(\begin{array}{cccc}0&-D^1&-D^2&-D^3\\ D^1&0&-H^3&H^2\\D^2&H^3&0&-H^1\\D^3&-H^2&H^1&0\end{array}
\right),
\ee
with ${\bf D}=(D^i)$ and ${\bf H}=(H^i)$ being the electric displacement and magnetic intensity fields, respectively, given by
\be
D^i=P^{i0},\quad H^i=-\frac12\vep^{ijk}P_{jk},
\ee
so that \eq{2.6} assumes the Maxwell equation form
\be\lb{2.9}
\nabla\cdot{\bf D}=\rho,\quad \frac{\pa{\bf D}}{\pa t}-\nabla\times{\bf H}=-{\bf j},
\ee
where $\rho=j^0$ and ${\bf j}=(j^i)$ are the charge and current densities, respectively. On the other hand, the Bianchi identity
$\pa_\mu \tilde{F}^{\mu\nu}=0$, where
\be
\tilde{F}^{\mu\nu}=\frac12\vep^{\mu\nu\alpha\beta}F_{\alpha\beta},\quad (\tilde{F}^{\mu\nu})=
\left(\begin{array}{cccc}0&-B^1&-B^2&-B^3\\ B^1&0&E^3&-E^2\\B^2&-E^3&0&E^1\\B^3&E^2&-E^1&0\end{array}
\right),
\ee
is the dual of $F_{\mu\nu}$, gives rise to the other two Maxwell equations,
\be\lb{2.11}
\nabla\cdot{\bf B}=0,\quad \frac{\pa{\bf B}}{\pa t}+\nabla\times {\bf E}={\bf 0}.
\ee

Furthermore, the energy-momentum tensor of the generalized theory may be calculated to be
\be
T_{\mu\nu}=-f'(\Lm)F_{\mu\alpha}\eta^{\alpha\beta}F_{\nu\beta}-\eta_{\mu\nu}f(\Lm),
\ee
such that the Hamiltonian energy density reads
\be\lb{2.13}
{\cal H}=T_{00}=f'(\Lm){\bf E}^2 -f(\Lm),\quad \Lm=\frac12\left({\bf E}^2-{\bf B}^2\right).
\ee

In the Born--Infeld theory \cite{BI}, we have
\be\lb{2.14}
f(s)=b^2\left(1-\sqrt{1-\frac2{b^2} s}\right),
\ee
where $b>0$ is the Born parameter. Thus \eq{2.13} gives us
\be
{\cal H}=b^2\left(\frac1{\sqrt{1-\frac1{b^2}({\bf E}^2-{\bf B}^2)}}-1\right)+\frac{{\bf B}^2}{\sqrt{1-\frac1{b^2}({\bf E}^2-{\bf B}^2)}},
\ee
which is familiar and again positive definite. In general,  \eq{2.13} is not guaranteed to be positive definite.

In view of \eq{2.2} and \eq{2.7}, we obtain the equations relating the pairs ${\bf D}, {\bf H}$ and ${\bf E}, {\bf B}$:
\be\lb{2.16}
{\bf D}=\vep({\bf E},{\bf B}){\bf E}, \quad {\bf B}=\mu({\bf E},{\bf B}){\bf H},
\ee
where $\vep({\bf E},{\bf B}), \mu({\bf E},{\bf B})$ are the field-dependent dielectrics and permeability coefficients given by
\be\lb{2.17}
\vep({\bf E},{\bf B})=f'(\Lm), \quad \mu({\bf E},{\bf B})=\frac1{f'(\Lm)}.
\ee

In short, the governing equations describing the nonlinear electromagnetism defined by the 
generalized Lagrangian action density are comprised of \eq{2.9}, \eq{2.11}, and \eq{2.16}. The first equation in \eq{2.11} indicates
that $\bf B$ is solenoidal such that there is a vector field $\bf A$ serving as a potential for $\bf B$. That is, ${\bf B}=\nabla\times{\bf A}$. Inserting this into the second equation in \eq{2.11}, we get
\be
\nabla\times\left({\bf E}+\frac{\pa{\bf A}}{\pa t}\right)={\bf 0}.
\ee
Hence there is a real scalar field $\phi$ such that 
\be\lb{2.18}
{\bf E}+\frac{\pa {\bf A}}{\pa t}=-\nabla\phi.
\ee

In view of \eq{2.18}, we see that the remaining governing equations, given in \eq{2.9}, become
\bea
\nabla\cdot\left(f'\left(\frac12\left[\left|\nabla\phi+\frac{\pa{\bf A}}{\pa t}\right|^2-|\nabla\times{\bf A}|^2\right]\right)\left(
\nabla\phi+\frac{\pa{\bf A}}{\pa t}\right)\right)&=&-\rho,\lb{2.20}\\
\frac{\pa}{\pa t}\left(f'\left(\frac12\left[\left|\nabla\phi+\frac{\pa{\bf A}}{\pa t}\right|^2-|\nabla\times{\bf A}|^2\right]\right)\left(
\nabla\phi+\frac{\pa{\bf A}}{\pa t}\right)\right)\quad\quad &&\nn\\
+\nabla\times \left(f'\left(\frac12\left[\left|\nabla\phi+\frac{\pa{\bf A}}{\pa t}\right|^2-|\nabla\times{\bf A}|^2\right]\right)\nabla\times{\bf A}\right)&=&{\bf j},\lb{2.21}
\eea
which is a closed coupled system of two non-homogeneous equations with unknowns $\phi$ and $\bf A$.

\section{Energy convergence and divergence  of  electric and magnetic point charges in polynomial models}\label{s3}
\setcounter{equation}{0}
\setcounter{theorem}{0}

First, for simplicity and clarity, we begin with the case that $f$ in \eq{2.3}  is a polynomial of the form
\be\lb{3.1}
f(s)=s+\sum_{k=2}^n a_k s^k, \quad a_2,\dots,a_n\in\bfR,\quad a_n\neq0.
\ee
We regard \eq{3.1} as an important general model because by the celebrated Stone--Weierstrass density theorem \cite{Stone,Yosida} any function in the space
$C[a,b]$ (the space of continuous functions over the interval $[a,b]$ equipped with the standard norm) may be approximated by polynomials.

Consider the electrostatic field generated from a point charge placed at the origin such that in \eq{2.9} we have
\be
\rho=4\pi q\delta(\x), \quad q>0,\quad {\bf j}={\bf0}.
\ee
Thus we have ${\bf H}={\bf 0}$ and
\be\lb{3.3}
{\bf D}=\frac{q\x}{r^3},\quad \x\neq{\bf0},\quad r=|\x|.
\ee
On the other hand, from \eq{2.16}, \eq{2.17}, and \eq{3.1}, we have 
\bea\lb{3.4}
{\bf D}^2&=&\left(f'(s)\right)^2 {\bf E}^2,\quad s=\frac{{\bf E}^2}2\nn\\
&=&\left(1+\sum_{k=2}^n\frac{ka_k}{2^{k-1}}{\bf E}^{2(k-1)}\right)^2 {\bf E}^2,\quad a_n\neq0.
\eea
In view of \eq{3.3} and \eq{3.4}, we have
\be\lb{3.5}
{\bf E}^2=\mbox{O}\left(r^{-\frac4{2n-1}}\right),\quad r\to{0}.
\ee
Thus, by virtue of \eq{3.5}, we see that the Hamiltonian density \eq{2.13} enjoys the property
\be\lb{3.6}
{\cal H}=\mbox{O}\left(r^{-\frac{4n}{2n-1}}\right),\quad r\to0.
\ee
Moreover, using \eq{3.3} in \eq{3.4} again, we have
\be\lb{3.7}
{\bf E}^2=\mbox{O}\left(r^{-4}\right),\quad r\to\infty.
\ee
Therefore \eq{2.13} leads to
\be\lb{3.8}
{\cal H}=\mbox{O}\left(r^{-4}\right),\quad r\to\infty.
\ee
Combining \eq{3.6} and \eq{3.8}, we arrive at the finite energy conclusion $\int_{\bfR^3} {\cal H}\,\dd\x<\infty$ 
for an electric point charge as anticipated as far as $n\geq2$.

Next, in the general situation, we assume that the function $f$ is smooth and we expand it following \eq{3.1} into
\be\lb{3.9}
f(s)=s+\sum_{k=2}^{n-1} a_k s^k+a_n(s) s^n, \quad a_2,\dots,a_{n-1}\in\bfR,
\ee
where $a_n(s)$ is real-valued function. Similar to \eq{3.4}, we now have
\be\lb{3.10}
{\bf D}^2=\left(1+\sum_{k=2}^{n-1}\frac{ka_k}{2^{k-1}}{\bf E}^{2(k-1)}+\left[n a_n\left(\frac{{\bf E}^2}2\right)
+a_n'\left(\frac{{\bf E}^2}2\right)\frac{{\bf E}^2}2\right]\left[\frac{{\bf E}^2}2\right]^{n-1}\right)^2 {\bf E}^2.
\ee
Thus, it is clear that the assumption
\be
\liminf_{s\to\infty}|na_n(s)+a_n'(s)s|\geq C_0
\ee
for some positive constant $C_0$ leads to \eq{3.5} as before. Consequently, for $s=\frac{{\bf E}^2}2$ near spatial infinity, the lower
order terms in $\cal H$ are still of no concern. More precisely, assume
\be\lb{3.12}
(2n-1)a_n(s)+2 a_n'(s)s=\mbox{O}(s^\sigma),\quad \mbox{as }s\to\infty.
\ee
Then we have, in view of \eq{2.13}, \eq{3.5}, and \eq{3.12}, we arrive at the asymptotic behavior
\be
{\cal H}\sim s^{\sigma +n}=\mbox{O}(r^{-\frac{4(n+\sigma)}{2n-1}}),\quad r\to{0}.
\ee
Therefore, the convergence of the energy at $\x={\bf 0}$ imposes the condition
\be
\sigma<\frac n2-\frac34,\quad n\geq2.
\ee

Hence we conclude  that a finite-energy electric point charge naturally spells out the nonlinear `correction' terms to the action density generating
function $f(s)$ as expressed in \eq{3.1}.

We now consider a magnetic monopole. In this situation the electric sector is trivial but the magnetic field $\bf B$ is generated from
a point magnetic charge $g>0$ resting at the origin so that
\be
\nabla\cdot {\bf B}=4\pi g \delta(\x),\quad g>0,
\ee
which renders the solution
\be\label{4.34}
{\bf B}=\frac{g\x}{r^3},\quad\x\neq{\bf0}.
\ee
Hence, inserting ${\bf E}={\bf0}$ and \eq{4.34} into \eq{2.13} and observing \eq{3.1} or more generally \eq{3.9}, we see that
${\cal H}\sim r^{-4}$ regardless of the details of the model. In other words, for a magnetic point charge, we always get
 a divergent energy around the origin.

Thus, we see that, energetically, the simple but general model \eq{3.1} breaks the electric and magnetic symmetry and accepts an electric point charge but
rejects a magnetic point charge. Such a phenomenon is not seen in the classical Born--Infeld theory \cite{BI}.

We may summarize our results in this section as follows.

\begin{theorem}\lb{th1}
A nonlinear theory of electromagnetism of an arbitrary polynomial type \eq{3.1} always accommodates a finite-energy electric point charge but rejects a finite-energy magnetic point charge. Therefore,
in view of the Stone--Weierstrass density theorem that any continuous function over a closed interval may be approximated by a sequence of polynomials, we conclude that, generically,
a nonlinear theory of electromagnetism allows a finite-energy electric point charge but does not allow a magnetic monopole, thereby establishing an electromagnetic asymmetry.
\end{theorem}

In the next section, we specialize on the quadratic case, $n=2$, of the model \eq{3.1} in detail.

\section{Quadratic nonlinearity situation}
\setcounter{equation}{0}

It will be instructive and of interest to work out explicitly the simplest nonlinear situation when \eq{3.1} assumes the quadratic form
\be\lb{4.1}
f(s)=s+a s^2, \quad a>0,
\ee
which has also been studied earlier in \cite{C2015,De,GS,K2007,K2017} and is of independent interest.
To this end, we insert \eq{4.1} into \eq{3.4} or
\be
{\bf D}^2=(1+a{\bf E}^2)^2 {\bf E}^2
\ee
to obtain
\be\lb{4.3}
{\bf E}^2=\frac1a\left(\frac{h({\bf D}^2)}6+\frac2{3h({\bf D}^2)}-\frac23\right),\quad h({\bf D}^2)
=\left(8+108 a{\bf D}^2+12\sqrt{81 a^2 {\bf D}^4+12 a{\bf D}^2}\right)^{\frac13}.
\ee
Using \eq{3.3}, we see that \eq{4.3} renders
\be\lb{4.4}
{\bf E}^2=\mbox{O}(r^{-\frac43}),\quad r\to0,
\ee
which in view of \eq{2.13} and \eq{4.1} gives us
\be\lb{4.5}
{\cal H}=\mbox{O}(r^{-\frac83}),\quad r\to0.
\ee
The asymptotics \eq{4.4} and \eq{4.5} are special cases of \eq{3.5} and \eq{3.6}, respectively. The asymptotics of $\bf E$ and 
$\cal H$ at spatial infinity are clearly given by \eq{3.7} and \eq{3.8}.

From \eq{2.13} and \eq{4.1}, we have
\be\lb{4.6}
a{\cal H}=\frac12 a{\bf E}^2 +\frac34 (a{\bf E}^2)^2,
\ee
where
\be\lb{4.7}
a{\bf E}^2=\frac{p(r)}6+\frac2{3p(r)}-\frac23,\quad p(r)=h({\bf D}^2)=\left(8+\frac{108aq^2}{r^4}+12\sqrt{\frac{81 a^2 q^4}{r^8}+ \frac{12 aq^2}{r^4}}\right)^{\frac13},
\ee
by using \eq{3.3}. These expressions appear complicated. Fortunately, the dependence of the function $p(r)$ on the parameters
$a$ and $q$ may be scaled away by setting 
\be\lb{4.8}
r=(2a)^{\frac14}q^{\frac12}\rho,\quad \rho>0,
\ee
where the parameter $a$ is comparable to the quantity $\frac1{2b^2}$ in the Born--Infeld theory \cite{BI}, 
as shown in \eq{2.14}, such that the energy of the point charge may be computed to yield
\be
E=4\pi \int_0^\infty {\cal H}\,r^2\dd r=8\pi \, (2a)^{-\frac14}q^{\frac32}\int_0^\infty  a{\cal H}\,\rho^2\dd\rho
\approx 4\pi  (2a)^{-\frac14}q^{\frac32}(2.939282832),
\ee
 in view of \eq{4.6}--\eq{4.8}, by utilizing a MAPLE 10 integration package, agreeing with the result obtained in \cite{C2015,K2017}.

In spherical coordinates $(r,\theta,\varphi)$, $\bf E$ is radially given by ${\bf E}=(E^r,0,0)$ with
\be\lb{4.10}
E^r=-\frac{\dd \phi}{\dd r}=-\phi'(r).
\ee
Thus, inserting \eq{4.7}, we obtain
\bea
\phi(r)&=&\frac{q}{r_0}\,h\left(\frac r{r_0}\right),\quad r_0=(2a)^{\frac14}q^{\frac12},\lb{4.11}\\
h(\rho)&=&\sqrt{2}\int^\infty_{\rho}\left(\frac{p_0(\eta)}6+\frac2{3p_0(\eta)}-\frac23\right)^{\frac12}\dd\eta,\quad \rho=\frac r{r_0},\lb{4.12}\\
p_0(\rho)&=&\left(8+\frac{54}{\rho^4}+6\sqrt{\frac{81}{\rho^8}+ \frac{24 }{\rho^4}}\right)^{\frac13},\lb{4.13}
\eea
which resembles a parallel result derived in \cite{BI} since the integrand in \eq{4.12} has the property
\be
\sqrt{2}\left(\frac{p_0(\eta)}6+\frac2{3p_0(\eta)}-\frac23\right)^{\frac12}=\frac1{\sqrt{1+\eta^4}}+\frac3{8\eta^{10}}+\mbox{O}\left({\eta^{-14}}\right)\quad\mbox{as }\eta\to\infty,
\ee
such that
\be
\phi(r)=\frac qr-\frac{q r_0^4}{10r^5}+\mbox{O}\left(r^{-9}\right)\quad\mbox{as }r\to\infty,
\ee
which says the theory coincides with that given by the Coulomb law asymptotically. Moreover, in terms of the quantity $r_0$ in
\eq{4.11}, we have
\be\lb{4.16}
\frac E{4\pi}=(2.9393)\,\frac{q^2}{r_0}.
\ee
In \cite{BI}, a similarly obtained corresponding numerical factor, replacing $2.9393$ above, is $1.2361$.

On the other hand, however, the expressions \eq{4.11} and \eq{4.12} lead to
\be
\phi(0)\approx(4.408757352)\frac{q}{r_0}.
\ee
This result deviates from that of the Coulomb potential, of course, which diverges like $\frac1r$ as $r\to0$, and that of the Born--Infeld potential, which assumes the 
value \cite{BI}:
\be
\frac q{r_0}\int_0^\infty \frac{\dd \eta}{\sqrt{1+\eta^4}}=\frac14 B\left(\frac14,\frac14\right)\frac q{r_0}\approx (1.854074677)\frac q{r_0}.
\ee

In view of \eq{4.10}--\eq{4.13}, we obtain after a lengthy computation the asymptotics of the free charge density 
\bea
\rho_{\mbox{\tiny free}}=\frac1{4\pi }\nabla\cdot{\bf E}=\frac1{4\pi r^2}\frac{\dd}{\dd r}(r^2 E^r)&=&\frac{q r_0^4}{2\pi r^7}+\mbox{O}\left(r^{-11}\right)\quad \mbox{as }r\to\infty,\lb{4.19}\\
&=&\frac{2^{\frac13}q}{3\pi r_0^{\frac43}r^{\frac53}}+\mbox{O}(r^{-\frac13})\quad\mbox{as }r\to0.\lb{4.20}
\eea
The result \eq{4.19} coincides with that in \cite{BI} but the result \eq{4.20} deviates from that in \cite{BI} which finds
$\rho_{\mbox{\tiny free}}\sim r^{-1}$ as $r\to0$.

To compute the total free charge, we need to evaluate
\be
q_{\mbox{\tiny free}}=\int_{\bfR^3}\rho_{\mbox{\tiny free}}\,\dd \x,
\ee
in which the integrand $\rho_{\mbox{\tiny free}}$ appears complicated. Fortunately, the asymptotic behavior of $E^r$ is readily
determined through \eq{4.11}--\eq{4.13} to yield
\be\lb{4.22}
E^r=\frac{2^{\frac13} q}{r^{\frac43}_0 r^{\frac23}}+\mbox{O}\left(r^{\frac23}\right)\quad\mbox{as }r\to0; \quad E^r=\frac q{r^2}+\mbox{O}\left(r^{-6}\right)\quad \mbox{as }r\to\infty.
\ee
In view of the left-hand side of \eq{4.19} and the estimates in \eq{4.22}, we get
\be\lb{4.23}
q_{\mbox{\tiny free}}=(r^2 E^r)^{r=\infty}_{r=0}=q.
\ee
 Thus, as in the Born--Infeld theory \cite{BI}, the total free charge $q_{\mbox{\tiny free}}$
generated from the electric field induced from a point charge coincides with the
prescribed value of the point charge $q$, although the former is given by a continuously distributed charge density but the latter
by a measure concentrated at a single point. This fact is an interesting feature of the theory.

\begin{figure}[h]
\begin{center}
\includegraphics[height=6cm,width=8cm]{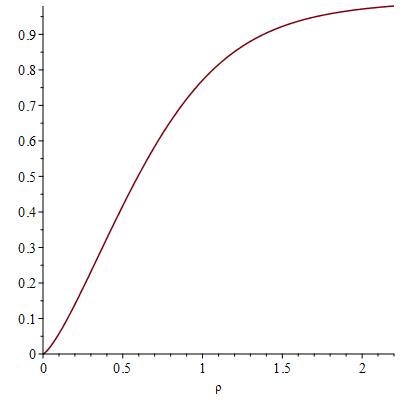}
\caption{A plot of the free electric charge $q_{\mbox{\tiny free}}(r)$ in the ball of radius $r$ around the point charge given as in \eqref{x4.26} relative to its limit value $q$ against  
the ratio $\rho=\frac r{r_0}$. It is seen that about 80\% of the free charge is contained in the ball of radius $r_0$ and the charge
quickly approaches its full-space limiting value when the ball of radius is beyond the threshold $r_0$. }
\label{F1}
\end{center}
\end{figure}

Although the free electric charge density $\rho_{\mbox{\tiny free}}$ appears too complicated to present here, we may
solve \eq{2.16} with \eq{3.3} and \eq{4.1} to get
\bea
E^r&=&\frac{\sqrt{2}q}{r_0^2}\left(\frac{p(r)^{\frac13}}6-\frac2{p(r)^{\frac13}}\right),\\
p(r)&=&\frac{108}{\sqrt{2}}\left(\frac{r_0}r\right)^2+12\sqrt{\frac{81}2\left(\frac{r_0}r\right)^4+12}.\lb{x4.25}
\eea
Hence the free charge contained in $\{|\x|\leq r\}$ is
\be\lb{x4.26}
q_{\mbox{\tiny free}}(r)=\int_{|\x|\leq r}\rho_{\mbox{\tiny free}}\,\dd \x=r^2 E^r=\sqrt{2}q\left(\frac r{r_0}\right)^2
\left(\frac{p(r)^{\frac13}}6-\frac2{p(r)^{\frac13}}\right),
\ee
where $p(r)$ is given in \eq{x4.25}. As a consequence of \eq{x4.26}, we have
\be
q_{\mbox{\tiny free}}(r)=2^{\frac13}q\left(\frac r{r_0}\right)^{\frac43}+\mbox{O}(r^{\frac83})\,\mbox{ as }r\to0;\quad
q_{\mbox{\tiny free}}(r)=q-\frac q2\left(\frac{r_0}r\right)^4+\mbox{O}\left(r^{-8}\right)\mbox{ as }r\to
\infty,
\ee
which are consistent with \eq{4.22} and \eq{4.23}. At $r=r_0$ and $r=2r_0$, the free charges given by \eq{x4.26} are
\be
q_{\mbox{\tiny free}}(r_0)=(0.7709)\,q,\quad q_{\mbox{\tiny free}}(2r_0)=(0.9714)\,q,
\ee
respectively, within 4 decimal places. In Figure \ref{F1}, we plot the quantity $\frac {q_{\mbox{\tiny free}}(r)}q$ against $\rho=\frac r{r_0}$, which shows
that the free charge contained in $\{|\x|\leq r\}$ converges to its limit value $q$ rapidly as $r\to\infty$.

As noted earlier, the numerical factor on the right-hand side of  \eq{4.16} and that in the same calculation of the point-charge energy lead to the ratio
\be
\tau_0=\frac{2.9391}{1.2361}=2.37772.
\ee
Hence, inserting the electron energy and charge values as in \cite{BI}, we obtain
\be
r_0=\tau_0(2.28\times 10^{-13}\mbox{cm})=5.4212\times 10^{-13}\mbox{cm},
\ee
as an estimate for the electron radius. Consequently, we may use \eq{4.11} to find
\be
a=\frac { r^4_0}{2e^2}=\frac{r_0^4}{2\times (4.8032\times 10^{-10})^2}=1.8719\times 10^{-31},
\ee
in the electrostatic system of units, or esu, which is a tiny quantity. This result suggests how the Maxwell electromagnetism may be
viewed as the leading-order approximation of the nonlinear theory considered here as in the Born--Infeld theory. As a comparison,
we have
\be
\frac{e}{r_0^2}=\frac1{\sqrt{2a}}=1.634345\times 10^{15},
\ee
which is of a similar magnitude of the parameter $b$ in \cite{BI}, sometimes also referred to as the Born parameter.

On the other hand, in view of \eq{4.34} with \eq{4.1} and \eq{2.16}, we see that the magnetic intensity field $\bf H$ of a magnetic
point charge is
\be
{\bf H}=\left(1-\frac{ag^2}{r^4}\right)\frac{g\x}{r^3},\quad \x\neq{\bf0}; \quad H^r=\frac\x r\cdot{\bf H}=\left(1-\frac{ag^2}{r^4}\right)\frac{g}{r^2},
\ee
which leads to a divergent total free magnetic charge as well, in addition to its divergent energy, since
\be
g_{\mbox{\tiny free}}=\frac1{4\pi}\int_{\bfR^3}\nabla\cdot{\bf H}\,\dd x=\int_0^\infty\frac{\dd}{\dd r}\left(r^2 H^r\right)\,\dd r
=g\left(1-\frac{ag^2}{r^4}\right)_{r=0}^{r=\infty}=\infty.
\ee
Consequently, we see that, both energetically and in terms of free magnetic charge, parallel to the notion of free electric charge, the theory \eq{4.1}, or more generally, \eq{3.1}, rejects  a magnetic monopole.
This simple feature is not present in the Born--Infeld theory in which a point magnetic charge is of finite energy as a point electric charge.
\medskip

It will be interesting to consider static solutions of the equations in a general setting. In this case, \eq{2.11} indicates that 
$\bf E$ is conservative and $\bf B$ is solenoidal. Hence there are 
time-independent real scalar field $\phi$ and vector field $\bf A$ such that 
\be
{\bf E}=-\nabla\phi,\quad {\bf B}=\nabla\times{\bf A}.
\ee
Thus, using these in \eq{2.16} and observing \eq{4.1}, we see that \eq{2.9} assumes the form
\bea
\nabla\cdot\left([1+a(|\nabla \phi|^2-|\nabla\times {\bf A}|^2)]\nabla\phi\right)&=&-\rho,\\
\nabla\times\left([1+a(|\nabla \phi|^2-|\nabla\times {\bf A}|^2)]\nabla\times{\bf A}\right)&=&{\bf j},
\eea
which are highly nonlinearly coupled. Of course, these equations are a limiting case of \eq{2.20}--\eq{2.21}. In the special situation either ${\bf j}={\bf 0}$ or $\rho=0$, it is consistent to set
${\bf A}={\bf0}$ or $\phi=0$, respectively, such that the system reduces into
\be
\nabla\cdot\left([1+a|\nabla\phi|^2]\nabla\phi\right)=-\rho,
\ee
which is a generalized Poisson equation, or its vector field version,
\be
\nabla\times\left([1-a|\nabla\times{\bf A}|^2]\nabla\times{\bf A}\right)={\bf j},
\ee
respectively, both of independent analytic interest.

\section{Some other models}
\setcounter{equation}{0}

In this section, we consider some well-developed model examples as illustrations and 
 preparation of the subsequent investigations on issues such as 
electromagnetic asymmetry and black hole singularities. To put these
examples in our current perspectives, we also reformulate and review some details of the results
as we move forward.

We first consider the Lagrangian density \eq{2.3} given analytically by
\be\lb{6.1}
f(s)=s\e^{\beta s},\quad\beta>0,
\ee
as in \cite{K}. The function \eq{6.1} does not meet our condition assumed in Section \ref{s3}.

To proceed, we assume an electrostatic situation and use \eq{2.16} to get
\be
{\bf D}=\left(1+\frac12\beta{\bf E}^2\right)\e^{\frac12\beta{\bf E}^2}{\bf E}.
\ee
Inserting the point charge configuration of $\bf D$ given in \eq{3.3}, we see that the nontrivial radial component $E^r$ of $\bf E$
satisfies
\be\lb{6.3}
\frac{q^2}{r^4}=\left(1+\frac12\beta (E^r)^2\right)^2 \e^{\beta (E^r)^2}(E^r)^2.
\ee
Hence we have
\be\lb{6.4}
E^r=\frac{q}{r^2}\quad\mbox{as }r\to\infty. 
\ee
On the other hand, using \eq{6.1} in \eq{2.13}, we see that the energy density is
\be\lb{6.5}
{\cal H}=\frac12 {\bf E}^2(1+\beta{\bf E}^2)\e^{\frac12\beta {\bf E}^2}.
\ee
Hence the energy near infinity clearly converges. To analyze what happens near $r=0$, we note from \eq{6.3} that $E^r$
blows up as $r\to0$. Hence we conclude that
\be\lb{6.6}
\e^{\beta (E^r)^2}\leq \frac1{r^4},\quad 0<r\leq r_1,
\ee
for some small $r_1>0$. Solving \eq{6.6}, we get
\be\lb{6.7}
(E^r)^2\leq -\frac4\beta\ln r,\quad 0<r\leq r_1.
\ee
Applying  \eq{6.6}--\eq{6.7} in \eq{6.5}, we obtain
\be\lb{6.8}
{\cal H}\leq\frac2\beta\ln\frac1r\left(1+4\ln\frac1r\right)\frac1{r^2}, \quad 0<r\leq r_1,
\ee
which yields the convergence of the energy near $r=0$ as well. As a consequence of the asymptotics stated in \eq{6.4} and
\eq{6.7}, we immediately deduce the result that the free electric charge is again $q$.

We next consider a magnetic point charge given by \eq{4.34}. Inserting \eq{2.16} with \eq{6.1}, we have
\be
H^r=\left(1-\frac{\beta g^2}{2r^4}\right)\e^{-\frac{\beta g^2}{2r^4}}\frac g{r^2}.
\ee
Hence
\be
g_{\mbox{\tiny free}}\equiv \frac1{4\pi}\int_{\bfR^3}\nabla\cdot{\bf H}\,\dd x=\int_0^\infty \frac{\dd}{\dd r}(r^2 H^r)\,\dd r=g,
\ee
indicating that the total free magnetic charge over $\bfR^3$ coincides with that prescribed as a point magnetic charge.
Furthermore, from \eq{2.13} and \eq{4.34}, we get the energy density
\be
{\cal H}=\frac{{\bf B}^2}2\e^{-\frac12\beta{\bf B}^2}=\frac{g^2}{2r^4}\e^{-\frac{\beta g^2}{2r^4}},
\ee
which gives a finite total energy over the space:
\be\lb{x5.12}
E=4\pi\int_0^\infty {\cal H}r^2\,\dd r=2\pi g^2\int_0^\infty\frac1{r^2}\e^{-\frac{\beta g^2}{2r^4}}\,\dd r
=\frac{\pi^2 g^{\frac32}}{(2\beta)^{\frac14}\Gamma\left(\frac34\right)},
\ee
as obtained earlier in \cite{K2017a}.

Another example is when the Lagrangian density \eq{2.3} is defined by \cite{K2}
\be\lb{6.12}
f(s)=\frac1\beta\arctan (\beta s),\quad \beta>0.
\ee
Thus, in the electrostatic situation, \eq{2.16} gives us the relation
\be
{\bf D}=\frac{{\bf E}}{1+\frac14\beta^2{\bf E}^4},
\ee
from which we may derive the bound
\be
|{\bf D}|\leq\frac34\left(\frac43\right)^{\frac14}\frac1{\sqrt{\beta}}.
\ee
In particular, a point electric charge in the form \eq{3.3}  cannot exist. On the other hand,
in the magnetostatic situation, \eq{2.16}
 renders
\be\lb{6.15}
{\bf H}=\frac{\bf B}{1+\frac14\beta^2 {\bf B}^4}.
\ee
Thus, if $\bf B$ describes a 
point magnetic charge given by \eq{4.34}, then \eq{6.15} yields the nontrivial radial component of $\bf H$:
\be
H^r=\frac{\left(\frac r{r_0}\right)^6}{\sqrt{\beta}\left(\left(\frac r{r_0}\right)^8+\frac14\right)},\quad r_0=\beta^{\frac14}g^{\frac12},
\ee
such that
\be
H^r=\frac g{r^2},\quad r\to\infty;\quad H^r=\frac{4r^6}{\beta^2 g^3},\quad r\to0.
\ee
Therefore it is clear that the total free magnetic charge $g_{\mbox{\tiny free}}=\frac1{4\pi}\int_{\bfR^3}\nabla\cdot{\bf H}\,\dd x$
coincides with $g$ as before. Explicitly, the free magnetic charge density reads
\be
\rho_{\mbox{\tiny free}}=\frac1{4\pi}\nabla\cdot{\bf H}=\frac1{4\pi r^2}\frac{\dd}{\dd r}(r^2 H^r)=\frac g{2\pi r_0^3}\frac{\left(\frac r{r_0}\right)^5}{
\left(\left(\frac r{r_0}\right)^8+\frac14\right)^2},
\ee
resulting in $\rho_{\mbox{\tiny free}}\sim r^5$ for $r\ll r_0$, $\rho_{\mbox{\tiny free}}\sim {r^{-11}}$ for $r\gg r_0$, and
the free magnetic charge contained in $\{|\x|\leq r\}$:
\be
g_{\mbox{\tiny free}}(r)=\int_{|\x|\leq r}\rho_{\mbox{\tiny free}}\,\dd\x=\frac{g}{1+\frac14 \left(\frac {r_0}r\right)^8}.
\ee
 So $g_{\mbox{\tiny free}}(r)$ is highly localized with $g_{\mbox{\tiny free}}(r_0)=\frac45 g$
and $g_{\mbox{\tiny free}}(2r_0)=\frac{2^{10}}{2^{10}+1}g$. Besides, using \eq{6.12} in \eq{2.13}, we see that the energy density is
\be
{\cal H}=\frac1\beta \arctan\left(\frac12\beta {\bf B}^2\right),
\ee
giving rise to a total finite energy with $\bf B$ stated in \eq{4.34}. In other words, the model \eq{6.12} does not permit
a point electric charge but a monopole.

A model that is in close resemblance to the classical Born--Infeld theory \cite{BI} and involves an $\arcsin$-type nonlinearity \cite{K3,K4}
is given by
\be\lb{6.18}
f(s)=\frac1\beta\arcsin(\beta s),\quad\beta>0,
\ee
where a truncation to the range of $s$, $-1<\beta s<1$,  is observed.

In the electrostatic situation, we are led from \eq{2.16} to the equation
\be
{\bf D}=\frac{\bf E}{\sqrt{1-\frac14\beta^2{\bf E}^4}},
\ee
which may be resolved to render
\be\lb{6.20}
{\bf E}^2=\frac2{\beta^2}\left(\sqrt{\beta^2+\frac1{{\bf D}^4}}-\frac1{{\bf D}^2}\right),
\ee
which ensures the condition $\frac12\beta {\bf E}^2<1$ automatically.
For the point electric charge field given in \eq{3.3}, we get from \eq{6.20} the results \cite{K2016}
\be\lb{6.21}
(E^r)^2=\frac2\beta-\frac{2r^4}{\beta^2 q^2}+\mbox{O}(r^8),\quad r\to0;\quad E^r=\frac q{r^2}+\mbox{O}\left(r^{-6}\right),\quad r\to\infty.
\ee
Thus, as in the Born--Infeld theory, the electric field is finite at the point charge. Using \eq{6.18} in \eq{2.13}, we see that the energy density reads
\be\lb{6.22}
{\cal H}=\frac{{\bf E}^2}{\sqrt{1-\frac14\beta^2 {\bf E}^4}}-\frac1\beta\arcsin\left(\frac12\beta{\bf E}^2\right).
\ee
From \eq{6.21} and \eq{6.22}, we have
\be\lb{6.23}
{\cal H}=\mbox{O}\left(\frac1{r^2}\right),\quad r\to0;\quad {\cal H}=\frac{q^2}{2r^4},\quad r\to\infty,
\ee
which leads to  finiteness of energy. Moreover, the estimates \eq{6.21} also enables us to show that the free total electric
charge $q_{\mbox{\tiny free}}=\frac1{4\pi}\int_{\bfR^3}\nabla\cdot{\bf E}\,\dd x$ is the same as
the prescribed charge $q$. 

However, unlike that in the classical Born--Infeld theory, a magnetic point charge is {\em not} permitted in the $\arcsin$-model since the restriction $\frac12\beta{\bf B}^2<1$
cannot be observed with \eq{4.34}. Later, we shall show that a dyonic point charge is not permitted either energetically in the
$\arcsin$-model.

In view of \eq{4.10}, \eq{3.3}, and \eq{6.20}, we see that the electric potential field $\phi$ is determined by
\bea
\phi(r)&=&\frac q{r_0}h\left(\frac r{r_0}\right),\quad r_0=\beta^{\frac14} q^{\frac12},\\
h(\rho)&=&\sqrt{2}\int^\infty_\rho\frac1{(\sqrt{1+\eta^8}+\eta^4)^{\frac12}}\,\dd\eta,\quad r=\beta^{\frac14}q^{\frac12}\rho=r_0\rho.
\eea
Besides, as in \cite{BI}, it is of interest to estimate the parameter $\beta$ in the $\arcsin$-model. For this purpose, we insert \eq{3.3}
into \eq{6.20} and use \eq{6.22} to get
\bea\lb{6.26}
E&=&\int_{\bfR^3}{\cal H}\,\dd x=4\pi\int_0^\infty {\cal H}\,r^2\,\dd r\nn\\
&=&4\pi \beta^{-\frac14}q^{\frac32}\int_0^\infty \left(\frac{2p_0(\rho)}{\sqrt{1-p_0^2(\rho)}}-\arcsin p_0(\rho)\right)\,\rho^2\,\dd\rho,\quad
p_0(\rho)=\frac1{\sqrt{1+\rho^8}+\rho^4}.
\eea
Thus, using MAPLE 10, the integral in \eq{6.26} can be computed to yield
\be\lb{6.27}
\frac E{4\pi}=(1.52244813)\beta^{-\frac14} q^{\frac32}=(1.52244813)\,\frac{q^2}{r_0}.
\ee
See also \cite{K2016}.
Hence, by \eq{6.27} and applying the electron data as in\cite{BI}, we arrive at the estimates
\bea
r_0&=&\frac{1.52244813}{1.2361}\times 2.28\times 10^{-13}=2.808172264\times 10^{-13}\mbox{ cm},  \\
\beta&=&\frac{r_0^4}{e^2}=2.695464653\times 10^{-32}\mbox{ esu},
\eea
for the radius of the electron and the parameter $\beta$, respectively. Note that, despite of the fact that the $\arcsin$-model does not allow a magnetic point charge, the estimate for the radius of the electron is rather close to that obtained by Born and Infeld in
\cite{BI}.

It will be interesting to obtain the free charge density explicitly as a comparison. For this purpose, we insert \eq{3.3} into
\eq{6.20} to get
\be\lb{xx5.34}
E^r=\frac{\sqrt{2} q}{\left(\sqrt{r_0^8+r^8}+r^4\right)^{\frac12}}.
\ee
Thus,  we find
\be\lb{xx31}
\rho_{\mbox{\tiny free}}=\frac1{4\pi r^2}\frac{\dd}{\dd r}\left(r^2 E^r\right)=\frac{\sqrt{2}q}{2\pi r_0^3
\left(\frac r{r_0}\right)\sqrt{\left(\frac r{r_0}\right)^8+1}\left(\sqrt{\left(\frac r{r_0}\right)^8+1}+\left(\frac r{r_0}\right)^4\right)^{\frac32}},
\ee
with the asymptotic properties $\rho_{\mbox{\tiny free}}\sim \frac1r$ for $r\ll r_0$ and $\rho_{\mbox{\tiny free}}\sim r^{-11}$ for $r\gg r_0$. This result shows that the free charge density decay much faster than that in the Born--Infeld
model for which we have $\rho_{\mbox{\tiny free}}\sim r^{-7}$ for $r\gg r_0$. 
Integrating  \eq{xx31}, we obtain by \eq{xx5.34}
\be
q_{\mbox{\tiny free}}(r)=\int_{|\x|\leq r}\rho_{\mbox{\tiny free}}\,\dd\x=\frac{\sqrt{2}q}{\left(1+\sqrt{1+\left(\frac{r_0}r\right)^8}\right)^{\frac12}}.
\ee
 For example, we have 
\be
q_{\mbox{\tiny free}}(r_0)=\frac{\sqrt{2}q}{(1+\sqrt{2})^{\frac12}}\approx 0.9012 \,q,\quad q_{\mbox{\tiny free}}(2r_0)=\frac{\sqrt{2}q}{\left(1+\frac{\sqrt{257}}{16}\right)^{\frac12}}\approx 0.9995\, q,
\ee
indicating a much more concentrated free charge distribution.

A well-known Born--Infeld type model with a logarithmic interaction density \cite{AM,Fe,Gaete,K6,Soleng} is of the form
\be\lb{6.30}
f(s)=-b^2\ln\left(1-\frac s{b^2}\right),\quad b>0.
\ee
Thus, in the electrostatic situation, the  relation \eq{2.16} leads to
\be
{\bf D}=\frac{\bf E}{1-\frac1{2b^2}{\bf E}^2},
\ee
which may be resolved to give us the nontrivial radial component of $\bf E$:
\be\lb{6.32}
E^r=b\left(\sqrt{2+\frac{b^2 r^4}{q^2}}-\frac{br^2}q\right),
\ee
as obtained earlier in \cite{K6}, whose asymptotics are
\be\lb{6.34}
E^r=\sqrt{2}b +\mbox{O}(r^2),\quad r\to0;\quad E^r=\frac{q}{r^2}+\mbox{O}\left(r^{-6}\right),\quad r\to\infty.
\ee
Using \eq{6.30} in \eq{2.13}, we also have
\be\lb{6.35}
{\cal H}=\frac{{\bf E}^2}{1-\frac1{2b^2}{\bf E}^2}+b^2\ln\left(1-\frac{{\bf E}^2}{2b^2}\right).
\ee
Thus, in view of \eq{6.34} and \eq{6.35}, the finiteness of the energy follows. Besides, \eq{6.34} implies that the free total electric
charge generated from $\bf E$ is $q_{\mbox{\tiny free}}=\frac1{4\pi}\int_{\bfR^3}\nabla\cdot{\bf E}\,\dd x=q$.

Moreover, in
the magnetostatic situation, the equation \eq{2.16} and \eq{6.30} yield
\be\lb{6.36}
{\bf H}=\frac{\bf B}{1+\frac1{2b^2}{\bf B}^2},
\ee
which gives rise to the nontrivial radial component of ${\bf H}$ to be
\be
H^r=\frac{gr^2}{r^4+\frac{g^2}{2b^2}},
\ee
when $\bf B$ in \eq{6.36} is the magnetic field of a point magnetic charge  given as in \eq{4.34}. The expression immediately leads to the result that the free total magnetic charge  $g_{\mbox{\tiny free}}=\frac1{4\pi}\int_{\bfR^3}\nabla\cdot{\bf H}\,\dd x$
coincides with the prescribed point magnetic charge in \eq{4.34}.  From \eq{2.13}, we get the 
magnetostatic energy density
\be
{\cal H}=b^2\ln\left(1+\frac{{\bf B}^2}{2b^2}\right),
\ee
whose integral over $\bfR^3$ is of course finite when $\bf B$ is defined by \eq{4.34}. Thus a point magnetic charge is allowed in
the model.

Using \eq{6.32}, we can calculate the electric potential of a point charge to be
\bea
\phi(r)&=&\frac{q}{r_0}h\left(\frac r{r_0}\right),\quad r_0=\sqrt{\frac qb},\\
h(\rho)&=&2\int_\rho^\infty\frac{\dd\eta}{\sqrt{2+\eta^4}+\eta^2},\quad r=r_0\rho.
\eea
Thus, inserting \eq{6.32} into \eq{6.35}, we see that the total energy of a point electric charge is given by
\bea
\frac E{4\pi}&=&\int_0^\infty {\cal H}\,r^2\,\dd r= r_0^3\int_0^\infty {\cal H}\,\rho^2\,\dd\rho\nn\\
&=&\frac{q^2}{r_0}\int_0^\infty\left(\frac{2p_0(\rho)}{1-p_0(\rho)}+\ln(1-p_0(\rho))\right)\rho^2\,\dd\rho,\lb{x5.47}\\
p_0(\rho)&=&\frac2{(\sqrt{2+\rho^4}+\rho^2)^2},\lb{x5.48}
\eea
which give us the numerical value
\be
\frac{E}{4\pi}=(1.384853)\,\frac{q^2}{r_0},
\ee
which is again rather close to that in the classical Born--Infeld result.
Therefore, applying the data in \cite{BI} for the electron, we obtain the following estimates for the radius of the electron and
the parameter $b$, respectively,
\bea
r_0&=&\frac{1.384853}{1.2361}\times2.28\times 10^{-13}=2.55437654\times 10^{-13}\mbox{ cm},\\
b&=&\frac{e}{r_0^2}=7.36140712\times 10^{15}\mbox{ esu}.
\eea
Both quantities are in close proximity of those obtained in the Born--Infeld calculation \cite{BI}. In other words, the model \eq{6.30} may be regarded as within a small variation of the classical theory.

From \eq{6.32}, we obtain the free electric charge density
\bea\lb{6.45}
\rho_{\mbox{\tiny free}}&=&\frac{1}{4\pi r^2}\frac{\dd}{\dd r}(r^2 E^r)\nn\\
&=&\frac{2q}{\pi r_0^3\left(\frac r{r_0}\right)\sqrt{2+\left(\frac r{r_0}\right)^4}\left(\sqrt{2+\left(\frac r{r_0}\right)^4}+
\left(\frac r{r_0}\right)^2\right)^2},
\eea
such that $\rho_{\mbox{\tiny free}}\sim \frac1r$ for $r\ll r_0$ and $\rho_{\mbox{\tiny free}}\sim r^{-7}$ for $r\gg r_0$,
which is asymptotically the same as that in the classical Born--Infeld model \cite{BI}. 

Using \eq{6.32} or \eq{6.45},  we may compute the free charge explicitly:
\be
q_{\mbox{\tiny free}}(r)=\int_{|\x|\leq r}\rho_{\mbox{\tiny free}}\,\dd \x=\frac{2q}{1+\sqrt{1+2\left(\frac{r_0}r\right)^4}}.
\ee
Thus, at $r=r_0$ and $r=2r_0$, the free charges are
\be
q_{\mbox{\tiny free}}(r_0)=\frac{2q}{1+\sqrt{3}}\approx (0.73205)\,q,\quad q_{\mbox{\tiny free}}(2r_0)=\frac{8q}{4+3\sqrt{2}}\approx (0.97056275)\,q.
\ee
 
Motivated by a specialized $f(R)$-gravity model defined by the normalized action \cite{Car2,Car3,NO}
\be
S=\int\left(R-\frac{\alpha^4}R\right)\sqrt{|g|}\,\dd x,
\ee
in terms of the Ricci scalar $R$ associated with a gravitational spacetime metric $g$, a generalized Born--Infeld
theory with
\be\lb{557}
f(s)=s-\frac\gamma s,
\ee
is considered in \cite{NB,NG} aimed, where $\gamma$ is a parameter, at producing suitable cosmological effects in the
accelerated expansion of a homogeneous and isotropic universe.
An obvious setback of this model is that it does not return to the Maxwell theory in the weak-field limit.
 In view of \eq{2.16} and \eq{557}, we see that 
the nontrivial radial component $E^r$ of the electric field of an electric point
charge satisfies the constitutive equation
\be\lb{558}
\left(1+\frac{4\gamma}{(E^r)^4}\right)E^r=\frac{q}{r^2},\quad q>0.
\ee
If $\gamma>0$, the left-hand side of \eq{558} is bounded from below by $2^{\frac52} 3^{-\frac34}\gamma^{\frac14}$ (say),
so that there is inconsistency in \eq{558}.
If $\gamma=-\beta<0$, then the left-hand side of \eq{558} is an increasing function of $E^r$ which vanishes at $\sqrt{2}\beta^{\frac14}$. Therefore, in this situation, we obtain $E^r>\sqrt{2}\beta^{\frac14}$ for all $r>0$ and
\be\lb{559}
E^r=\frac q{r^2}+\mbox{O}(r^6),\quad r\to0;\quad E^r=\sqrt{2}\beta^{\frac14}+\frac{q^2}{4r^2}+\mbox{O}(r^{-4}),\quad
r\to\infty.
\ee
Since the Hamiltonian energy density \eq{2.13} now assumes the form
\be\lb{560}
{\cal H}=\frac12 (E^r)^2-\frac{6\beta}{(E^r)^2},
\ee
we see by using \eq{559} in \eq{560} that there occurs an energy divergence  at both $r=0$ and $r=\infty$. Besides, it is easily
checked that a magnetic point charge also has a divergent energy.

A rational-function model proposed and investigated by Kruglov \cite{K7,K8,K9,K10}, in our notation, is given by
\be\lb{561}
f(s)=\frac s{1-2\beta s},\quad \beta>0,
\ee
which  introduces an upper-bound cut-off for $s$ in an {\em open} manner, $s<\frac1{2\beta}$,
for the continuity of $f$,  such that the Stone--Weierstrass theorem 
\cite{Stone,Yosida} is not applicable. In other words, unlike that in
the classical Born--Infeld model, \eq{561} cannot be approximated uniformly by a sequence of polynomials over the interval of
concern. An interesting feature of \eq{561} is that it generates regular magnetically charged black holes as shown by Ma \cite{Ma}.
To examine the energy convergence  of an electric point charge, which will be used later to relegate the curvature singularity
of an electrically charged black hole, we insert \eq{561} into \eq{2.16} with \eq{3.3} to get \cite{K8}
\be\lb{562}
\frac{E^r}{\left(1-{\beta} (E^r)^2\right)^2}=D^r=\frac q{r^2}.
\ee
Hence we are led by \eq{562} to the asymptotic estimates \cite{K8}
\be\lb{563}
(E^r)^2 =\frac1\beta +\mbox{O}(r),\quad r\to0; \quad E^r=\frac q{r^2}+\mbox{O}(r^{-4}),\quad r\to\infty.
\ee
Using \eq{561} and \eq{562} in \eq{2.13}, we have
\be\lb{564}
{\cal H}=\frac12(1+\beta [E^r]^2)D^r E^r.
\ee
Combining \eq{563} and \eq{564}, we get 
\be\lb{565}
{\cal H}=\frac q{\sqrt{\beta} r^2}+\mbox{O}(r^{-1}),\quad r\to 0;\quad {\cal H}=\frac{q^2}{2r^4}+\mbox{O}(r^{-8}),\quad r\to\infty,
\ee
resulting in the finiteness of energy of an electric point charge. Note that the sharp estimate of $\cal H$ near $r=0$ in
\eq{565} will be useful for
analyzing the curvature singularity of an electrically charged black hole there.

\section{Fractional-powered nonlinearity}
\setcounter{equation}{0}

Motivated by the classical Born--Infeld model \eq{2.14} and the studies presented earlier, we consider a further extended situation
\be\lb{.1}
f(s)=b^2\left(1-\left[1-\frac s{ p b^2 } \right]^p\right),
\ee
where $0<p<1$. In \cite{K2017b,K2017c}, a more general model than \eq{.1} is proposed and the asymptotic
expression of the electric field of an electric point charge is obtained. 
 In \cite{K5}, this model at $p=\frac34$ is considered to produce dyonic black hole solutions of the
Reissner--Nordstr\"{o}m type. It is seen that \eq{.1} is made to satisfy the condition \eq{2.4} in order to accommodate the Maxwell theory.
Hence, by \eq{2.13}, we get the energy density
\be\lb{.2}
{\cal H}=\left(1-\frac{({\bf E}^2-{\bf B}^2)}{2p b^2}\right)^{p-1}{\bf E}^2+b^2\left(\left[1-\frac{({\bf E}^2-{\bf B}^2)}{ 2 p b^2 } \right]^p-1\right)
\ee
On the other hand, using \eq{2.16} and \eq{2.17} with \eq{.1}, we have
\be\lb{.3}
{\bf D}=\left(1-\frac{{\bf E}^2}{2p b^2}\right)^{p-1}{\bf E},\quad {\bf H}=\left(1+\frac{{\bf B}^2}{2pb^2}\right)^{p-1}{\bf B},
\ee
for the electrostatic and magnetostatic cases, respectively.
Thus, in the  electric point charge situation with \eq{3.3}, we have
\be\lb{.4}
E^r=\sqrt{2p} b+\mbox{O}(r^{\frac2{1-p}})\quad\mbox{as }r\to0;\quad E^r=\frac q{r^2}+\mbox{O}\left(r^{-6}\right)\quad\mbox{as }r\to\infty.
\ee
In view of \eq{.2} and \eq{.4}, we have
\be\lb{.5}
{\cal H}=\mbox{O}\left(r^{-2}\right)\quad \mbox{as }r\to0;\quad {\cal H}=\mbox{O}\left(r^{-4}\right)\quad\mbox{as }
r\to\infty,
\ee
resulting in finiteness of energy. As before, it is seen that the total free electric charge is also the same as the prescribed point charge, $q_{\mbox{\tiny free}}=q$.

In the situation of a point magnetic charge given by \eq{4.34}, we see from \eq{.2} that
\be\lb{.6}
{\cal H}=\mbox{O}\left(r^{-4p}\right)\quad \mbox{as }r\to0;\quad{\cal H}=\mbox{O}\left(r^{-4}\right)\quad\mbox{as }r\to\infty.
\ee
Therefore, by \eq{.6}, we see that the finiteness of the energy of a magnetic point charge is characterized by the condition
\be\lb{.7}
p<\frac34.
\ee
Besides, the free magnetic charge density defined by the second expression in \eq{.3} is
\bea\lb{.8}
\rho_{\mbox{\tiny free}}^m&=&\frac1{4\pi}\nabla\cdot{\bf H}=\frac1{4\pi r^2}\frac{\dd}{\dd r}(r^2 H^r)=
\frac{g^3 (1-p)r^{1-4p}}{2\pi p b^2\left(r^4+\frac{g^2}{2p b^2}\right)^{2-p}}\nn\\
&=&\frac{g(1-p)}{\pi r_0^3}\left(\frac {r_0}r\right)^7\frac1{\left(1+\left[\frac{r_0}r\right]^4\right)^{2-p}},\quad r_0=\frac{g^{\frac12}}{(2p)^{\frac14} b^{\frac12}}.
\eea
So the condition \eq{.7} ensures $r^2\rho^m_{\mbox{\tiny free}}\to0$ as $r\to0$. Furthermore, \eq{.8} gives us the free
magnetic charge contained in the ball $\{|\x|\leq r\}$,
\be
g_{\mbox{\tiny free}}(r)=\int_{|\x|\leq r}\rho_{\mbox{\tiny free}}^m\,\dd \x=\frac g{\left(1+\left[\frac{r_0} r\right]^4\right)^{1-p}},
\ee
which is finite in all range of $p$, $0<p<1$. In particular, the total free magnetic charge is $g_{\mbox{\tiny free}}=g_{\mbox{\tiny free}}(\infty)=g$ in all situations, among which the range 
\be\lb{.10}
\frac34\leq p<1
\ee
spells out  magnetic point charges of infinite energy. In other words, in the parameter range \eq{.10}, a
magnetic point charge is permissible with regard to the free magnetic charge but not permissible with regard to
maintaining a finite energy. Note that, although this calculation is for the magnetic point charge situation, it suggests the same choice
for the electrostatic situation for the `radius' parameter,
\be
r_0=\frac{q^{\frac12}}{(2p)^{\frac14} b^{\frac12}},
\ee
for a point electric charge $q>0$, which contains that of Born--Infeld \cite{BI} when $p=\frac12$.

We next work out the critical case $p=\frac34$ as a concrete example. With this and the  electric point charge assumption \eq{3.3},
the first expression in \eq{.3} reads
\be
\left(\frac{r_0}r\right)^2=\left(1-\left[\frac{r_0^2 E^r}q\right]^2\right)^{-\frac14}\frac{r_0^2 E^r}q,
\ee
whose solution is
\be\lb{.13}
E^r=\frac q{r_0^2}\left(\frac{r_0}r\right)^2\left(\sqrt{\frac14\left(\frac{r_0}r\right)^8+1}-\frac12\left(\frac{r_0}r\right)^4\right)^{\frac12}.
\ee
Thus the potential function  is
\be
\phi(r)=\frac q{r_0}h\left(\frac r{r_0}\right),\quad h(\rho)=\int_\rho^\infty\frac1{\eta^2}\left(\sqrt{\frac1{4\eta^8}+1}-\frac1{2\eta^4}\right)^{\frac12}\,\dd \eta.
\ee
Besides, from \eq{.13}, the free electric charge density is
\be
\rho^e_{\mbox{\tiny free}}=\frac1{4\pi r^2}\frac{\dd}{\dd r}(r^2 E^r)=\frac{\sqrt{2}q}{4\pi r_0^3}\frac{\left(\sqrt{4\left[\frac r{r_0}\right]^8+1}-1\right)^{\frac12}}{\left(\frac r{r_0}\right)^5\sqrt{4\left[\frac r{r_0}\right]^8+1}}.
\ee
Thus, $\rho^e_{\mbox{\tiny free}}\sim \frac1r$ for $r$ near $0$ and $\rho^e_{\mbox{\tiny free}}\sim r^{-7}$ for 
$r$ near infinity. These properties are the same as in the case when $p=\frac12$ \cite{BI}. Moreover, the free electric charge
contained in the ball $\{|\x|\leq r\}$ is
\be\lb{.16}
q_{\mbox{\tiny free}}(r)=\int_{|\x|\leq r} \rho^e_{\mbox{\tiny free}}\,\dd \x=r^2 E^r
=\frac q{\left(\sqrt{\frac14\left(\frac{r_0}r\right)^8+1}+\frac12\left(\frac{r_0}r\right)^4\right)^{\frac12}}.
\ee
This quantity approaches its limiting value $q$ as $r\to\infty$. For example, we list
\be
q_{\mbox{\tiny free}}(r_0)=(0.7862)\, q,\quad q_{\mbox{\tiny free}}(2r_0)=(0.9845)\, q.
\ee

Inserting $p=\frac34$ into \eq{.2},  we rewrite the radially symmetric electrostatic energy density as
\be\lb{.18}
{\cal H}=\left(1-r_0^4\left[\frac{E^r}q\right]^2\right)^{-\frac14} (E^r)^2 +b^2\left(\left(1-r_0^4\left[\frac{E^r}q\right]^2\right)^{\frac34}-1\right).
\ee
Thus, inserting \eq{.16} into \eq{.18}, we see that the energy of a point electric charge is determined by
\bea
\frac E{4\pi}=\int_0^\infty {\cal H} r^2\,\dd r&=&\frac{q^2}{r_0}\,\int_0^\infty\left(
\left(1-h(\rho)\right)^{-\frac14}  h(\rho) +\frac23\left[\left(1-h(\rho)\right)^{\frac34}-1\right]\right)
\rho^2\,\dd\rho,\lb{.19}\\
r_0^4\left(\frac {E^r}q\right)^2&\equiv&h(\rho)=\frac2{\sqrt{1+4\rho^8}+1},\quad \rho=\frac r{r_0}.
\eea
Thus, we may obtain the numerical value of the integral on the right-hand side of \eq{.19} to find
\be
\frac E{4\pi}=(1.277861367)\,\frac{q^2}{r_0},
\ee
which is very close to the corresponding quantity calculated in \cite{BI}, although in our situation a magnetic point charge is not
allowed due to the divergence of the energy of such a charge.

\section{Dyonic point charges}
\setcounter{equation}{0}

In this section, we consider a dyonic point particle, residing at the origin, carrying both electric and magnetic charges, $q,g>0$, whose
electric displacement field $\bf D$ and magnetic  field $\bf B$ are given by \eq{3.3} and \eq{4.34}, respectively, or
\be\lb{7.1}
D^r=\frac q{r^2},\quad B^r=\frac g{r^2},\quad r=|\x|>0,
\ee
in terms of their nontrivial radial components.

As a start, we consider the classical Born--Infeld model where $f(s)$ is given in \eq{.1} with $p=\frac12$. In such a situation, inserting \eq{7.1} into \eq{2.16}, we see that the nontrivial radial component of
the electric field $\bf E$ is found to be \cite{K4}
\be\lb{7.2}
E^r=\frac q{r^2}\sqrt{\frac{r^4+r_m^4}{r^4+r^4_e}},\quad r_e=\sqrt{\frac qb},\quad r_m=\sqrt{\frac gb},
\ee
giving rise to the asymptotics
\be\lb{7.3}
E^r=\frac{g}{r^2}+\mbox{O}(r^2)\quad\mbox{as }r\to0;\quad E^r=\frac q{r^2}+\mbox{O}\left(r^{-6}\right)\quad
\mbox{as }r\to\infty.
\ee
Thus, unlike in the electric point charge situation, the electric field $E^r$ blows up at 
$r=0$ as  a consequence of the presence of the
magnetic charge, although it still obeys the Coulomb law near $r=\infty$. Besides, $E^r$ looks like $B^r$ near $r=0$. As before, we may calculate the
free electric charge density by
\be\lb{7.4}
\rho_{\mbox{\tiny free}}^e=\frac1{4\pi r^2}\frac{\dd}{\dd r}(r^2 E^r)=\frac{q(r_e^4-r_m^4)\,r}{2\pi (r^4+r_e^4)^2}\sqrt{\frac{r^4+r_e^4}{r^4+r_m^4}}.
\ee
So we see that $\rho_{\mbox{\tiny free}}^e\sim r$ for $r\sim0$, which is different from the behavior near an electric point
charge, and $\rho_{\mbox{\tiny free}}^e\sim r^{-7}$, which is the same as that of an electric charge.
Besides, the free electric charge contained in the ball $\{|\x|\leq r\}$ is
given in view of \eq{7.4} and \eq{7.2} by
\be
q_{\mbox{\tiny free}}(r)=\int_{|\x|\leq r} \rho_{\mbox{\tiny free}}^e\,\dd\x=r^2 E^r -(r^2 E^r)_{r=0}=
q\sqrt{\frac{r^4+r_m^4}{r^4+r^4_e}}-g.
\ee
In particular, the total free electric charge is
\be\lb{7.6}
q_{\mbox{\tiny free}}=q_{\mbox{\tiny free}}(\infty)=q-g,
\ee
which mixes the prescribed electric and magnetic charges. On the other hand, inserting $B^r$ in \eq{7.1} into
\eq{2.16}, we get the nontrivial radial component of the magnetic intensity field $\bf H$ to be
\be\lb{7.7}
H^r=\frac g{r^2}\sqrt{\frac{r^4+r_e^4}{r^4+r_m^4}};\quad H^r=\frac q{r^2}+\mbox{O}(r^2)\quad\mbox{as }r\to0;\quad 
H^r=\frac g{r^2}+\mbox{O}\left(r^{-6}\right)\quad \mbox{as }r\to\infty,
\ee
which is completely analogous to \eq{7.2}--\eq{7.3}. In particular, $H^r$ behaves like an electric point charge near $r=0$ and
a magnetic point charge near $r=\infty$. As a consequence, we obtain the free magnetic charge density,
free magnetic charge contained
in the ball $\{|\x|\leq r\}$, and total free magnetic charge,
\bea
\rho_{\mbox{\tiny free}}^m&=&\frac{g(r_m^4-r_e^4)\,r}{2\pi (r^4+r_m^4)^2}\sqrt{\frac{r^4+r_m^4}{r^4+r_e^4}},\\
g_{\mbox{\tiny free}}(r)&=&
g\sqrt{\frac{r^4+r_e^4}{r^4+r^4_m}}-q,\\
g_{\mbox{\tiny free}}&=&g_{\mbox{\tiny free}}(\infty)=g-q=-q_{\mbox{\tiny free}},\lb{7.10}
\eea
respectively, such that $q_{\mbox{\tiny free}}$ and $g_{\mbox{\tiny free}}$ are mutually dependent quantities.

To compute the energy of the dyon, we insert $B^r$ and $E^r$ given in \eq{7.1} and \eq{7.2} into \eq{.2} with $p=\frac12$ to get
\bea\lb{7.11}
{\cal H}&=&\frac{b^2\left(r_e^4 r_m^4+r^4(r_e^4+r_m^4)\right)}{r^4\left(r^4+\sqrt{(r^4+r_e^4)(r^4+r_m^4)}\right)}\nn\\
&=&\frac{\left(\frac{qg}b\right)^2+(q^2+g^2)r^4}{r^4(r^4+\sqrt{(r^4+[\frac qb]^2)(r^4+[\frac gb]^2)})}.
\eea
It is interesting to note that the two special cases of \eq{7.11} we have already covered are 
\bea
{\cal H}|_{g=0}&=&\frac{q^2}{r^2(\sqrt{r^4+r_e^4}+r^2)},\\
{\cal H}|_{q=0}&=&\frac{g^2}{r^2(\sqrt{r^4+r_m^4}+r^2)},
\eea
corresponding to the electric and magnetic point charge cases, respectively, are the {\em only} two situations when \eq{7.11}
gives rise to a finite energy, $\int_{\bfR^3}{\cal H}\,\dd\x<\infty$. In other words, whenever $q\neq0,g\neq0$, the energy density
\eq{7.11} leads to the divergence of the total energy of the dyonic point charge.

Thus, we see that, although both electric and magnetic point charges are permitted in the Born--Infeld theory defined by the
Lagrangian action density \eq{2.3} where $f(s)$ is as given in \eq{.1} with $p=\frac12$,  dyonic point charges are not permitted energetically.

We next consider the $\arcsin$-model defined by \eq{6.18}. Thus \eq{2.16} gives us
\be\lb{7.14}
{\bf D}=\frac{\bf E}{\sqrt{1-\frac14\beta^2({\bf E}^2-{\bf B}^2)^2}}.
\ee
Inserting \eq{7.1} into \eq{7.14} and setting
\be
r_e=\beta^{\frac14}q^{\frac12},\quad r_m=\beta^{\frac14}g^{\frac12},
\ee
we obtain
\be\lb{7.16}
E^r=\frac q{r_e^2}\left(\frac{r_m^4}{r^4}+\frac{2(r_e^4-r_m^4)}{\sqrt{r_e^4(r_e^4-r_m^4)+r^8}+r^4}\right)^{\frac12},
\ee
as obtained earlier in \cite{K4}, provided that $r_e\geq r_m$ or
\be\lb{7.17}
q\geq g.
\ee
This condition also ensures that the $\arcsin$-model restriction $-1<\beta s<1$ is fulfilled. To see this  delicate structure, we
rearrange  \eq{7.14} to obtain
\be\lb{7.17b}
(\beta s)^2+\frac2{\beta{\bf D}^2}(\beta s)-\left(1-\frac{{\bf B}^2}{{\bf D}^2}\right)=0,\quad
\beta s=\frac\beta2\left({\bf E}^2-{\bf B}^2\right),
\ee
which leads to
\be
\beta s=-\frac1{\beta{\bf D}^2}+\sqrt{\frac1{\beta^2{\bf D}^4}+1-\frac{{\bf B}^2}{{\bf D}^2}},
\ee
where we have chosen plus sign in front of the radical root for consistency.
Thus, we see that \eq{7.1} and \eq{7.17} make the quantity under the radical root on the right-hand side of \eq{7.17b} positive
for all $r>0$.
Besides, we also see that, now, the right-hand side of \eq{7.17b} stays non-negative and
\be
\beta s=\frac{1-\frac{{\bf B}^2}{{\bf D}^2}}{ \sqrt{\frac1{\beta^2{\bf D}^4}+1-\frac{{\bf B}^2}{{\bf D}^2}}+
\frac1{\beta{\bf D}^2}}<{1-\frac{{\bf B}^2}{{\bf D}^2}}\leq1,
\ee
as anticipated. Note that we have also deduced the conclusion $s\geq0$ or ${\bf E}^2\geq {\bf B}^2$ as a by-product.

The condition \eq{7.17} indicates that,
in the $\arcsin$-model, a magnetic point charge cannot be switched on independently as a monopole and must be
accompanied by an electric counterpart.

The free electric charge contained in $\{|\x|\leq r\}$ is 
\be\lb{7.18}
q_{\mbox{\tiny free}}(r)=(r^2 E^r)^{r=r}_{r=0}=\frac q{r_e^2}\left({r_m^4}+\frac{2(r_e^4-r_m^4)r^4}{\sqrt{r_e^4(r_e^4-r_m^4)+r^8}+r^4}\right)^{\frac12}-g,
\ee
which is an increasing function of $r>0$. In particular, the total free electric charge is $q_{\mbox{\tiny free}}=q_{\mbox{\tiny free}}
(\infty)=q-g$, which coincides with \eq{7.6}. Furthermore, in view of \eq{7.1} and \eq{2.16} again, we have
\be
H^r=\frac g{r^2}\frac{r_e^2}{\sqrt{r_m^4+\frac{2(r_e^4-r_m^4)r^4}{\sqrt{r^4_e(r^4_e-r_m^4)+r^8}+r^4}}},
\ee
giving rise to free magnetic charge contained in $\{|\x|\leq r\}$,
\be
g_{\mbox{\tiny free}}(r)=(r^2 H^r)_{r=0}^{r=r}=\frac{g r_e^2}{\sqrt{r_m^4+\frac{2(r_e^4-r_m^4)r^4}{\sqrt{r^4_e(r^4_e-r_m^4)+r^8}+r^4}}}-q,
\ee
and the total free magnetic charge $g_{\mbox{\tiny free}}=g_{\mbox{\tiny free}}(\infty)=g-q$ as before which coincides with \eq{7.10} as well.
Besides, in view of \eq{2.13}, \eq{7.1}, and \eq{7.16}, we have
\bea\lb{7.20}
{\cal H}&=&D^r E^r-\frac1\beta\arcsin \frac\beta2((E^r)^2-(B^r)^2)\nn\\
&=&\frac{q^2}{r_e^2 r^2}\left(\frac{r_m^4}{r^4}+\frac{2(r_e^4-r_m^4)}{\sqrt{r_e^4(r_e^4-r_m^4)+r^8}+r^4}\right)^{\frac12}-\frac{q^2}{r^4_e}\arcsin \frac{(r_e^4-r_m^4)}{\sqrt{r_e^4(r_e^4-r_m^4)+r^8}+r^4}.
\eea
Asymptotically, we have
\be
{\cal H}=\frac{qg}{r^4}+\mbox{O}(1)\quad\mbox{as }r\to0;\quad {\cal H}=\frac{q^2+g^2}{2r^4}+\mbox{O}\left(r^{-12}
\right)\quad\mbox{as }r\to\infty,
\ee
leading to  divergence of the energy, whenever $qg\neq0$, at the origin where the dyon resides.

For the logarithmic model \eq{6.30}, the first equation in \eq{2.16} now reads
\be 
{\bf D}=\frac{\bf E}{1-\frac1{2b^2}({\bf E}^2-{\bf B}^2)},
\ee
which may be resolved by using \eq{7.1} to give us the result
\be\lb{7.23}
E^r=\frac q{r^2}\frac{(r_m^4+2r^4)}{\sqrt{r_e^4(r_m^4+2r^4)+r^8}+r^4},\quad r_e=\sqrt{\frac qb},\quad r_m=\sqrt{\frac gb}.
\ee
See \cite{K6} for an earlier result.
Hence we can derive the free electric charge formula as in \eq{7.18} and obtain the total free electric charge to be 
$q_{\mbox{\tiny free}}=q-g$. Similarly, the free magnetic charge is $q_{\mbox{\tiny free}}=g-q$ as before.
Moreover,  in view of \eq{2.13}, \eq{7.1}, and \eq{7.23},  we see that the associated energy density is
\bea
{\cal H}&=& D^r E^r+b^2\ln\left(1-\frac1{2b^2}[(E^r)^2-(B^r)^2]\right)\nn\\
&=&\frac {q^2}{r^4}\frac{(r_m^4+2r^4)}{\sqrt{r_e^4(r_m^4+2r^4)+r^8}+r^4}
+b^2
\ln\left(\frac{r_m^4+2r^4}{\sqrt{r_e^4(r_m^4+2r^4)+r^8}+r^4}\right),
\eea
which enjoys the asymptotic behavior
\be\lb{7.25}
r^2{\cal H}=\frac{qg}{r^2}+\mbox{O}(r^2)\quad \mbox{as }r\to 0;\quad {\cal H}=\frac{q^2+g^2}{2r^4}+\mbox{O}\left(r^{-8}\right)\quad\mbox{as }r\to\infty.
\ee
Hence the divergence of the dyon energy at the origin again follows immediately. It is interesting to note that, asymptotically, electric and magnetic charges play rather symmetric roles as in the classical Born--Infeld theory.

For the fractionally powered model \eq{.1}, the equations stated in \eq{2.16} are
\be\lb{7.26}
{\bf D}=\left(1-\frac{({\bf E}^2-{\bf B}^2)}{2p b^2}\right)^{p-1}{\bf E},
\quad {\bf H}=\left(1-\frac{({\bf E}^2-{\bf B}^2)}{2p b^2}\right)^{p-1}{\bf B}.
\ee
Accordingly, we set
\be\lb{7.27}
r_e=\frac1{(2p)^{\frac14}}\sqrt{\frac qb},\quad r_m=\frac1{(2p)^{\frac14}}\sqrt{\frac gb}.
\ee

We have seen that the model does not allow a finite-energy magnetic charge when $p=\frac34$. Thus, naturally,  a dyonic point
charge will not have a finite energy in this situation either. Nevertheless, it is of interest to study how the energy diverges as we
undergo a process of passing from a dyonic to a magnetic point charge field configuration, which we now pursue in the sequel.

With $p=\frac34$ in \eq{7.26} and \eq{7.27}, and inserting \eq{7.1}, we obtain after some calculation the result
\be\lb{7.28}
E^r=\frac{q}{r^2}\left(\frac{2(r_m^4+r^4)}{\sqrt{r^8_e+4(r_m^4+r^4)r^4}+r_e^4}\right)^{\frac12},
\ee
which enjoys the same asymptotic behavior as that for the case of $p=\frac12$ stated in \eq{7.3}.
See \cite{K5} for an earlier result.
 Likewise, the magnetic intensity field in \eq{7.26} where $p=\frac34$ is given by
\be\lb{7.29}
H^r=\frac g{r^2}\left(\frac{\sqrt{r^8_e+4(r_m^4+r^4)r^4}+r_e^4}{2(r_m^4+r^4)}\right)^{\frac12},
\ee
so that it has the same asymptotic properties as stated in \eq{7.7}. Furthermore, it follows from \eq{7.28} and \eq{7.29} that the
free total electric and magnetic charges satisfy the same relation, \eq{7.10}, as before. Finally, inserting \eq{7.1} and \eq{7.28} into \eq{.2}, we have
\bea\lb{7.30}
{\cal H}&=& D^r E^r+b^2\left(\left[\frac{E^r}{D^r}\right]^3-1\right)\nn\\
&=&\frac{q^2}{r^4}\left(\frac{2(r_m^4+r^4)}{\sqrt{r^8_e+4(r_m^4+r^4)r^4}+r_e^4}\right)^{\frac12}
+b^2 \left(\left(\frac{2(r_m^4+r^4)}{\sqrt{r^8_e+4(r_m^4+r^4)r^4}+r_e^4}\right)^{\frac32}-1\right).
\eea
Thus, we obtain the estimates
\bea\lb{7.31}
{\cal H}&=&\frac{qg}{r^4}+b^2\left(\frac34\left[\frac qg\right]+\frac14\left[\frac gq\right]^3-1\right)+\mbox{O}(r^{-4})\quad \mbox{as }r\to0,\lb{7.31a}\\
 {\cal H}&=&\frac{q^2+g^2}{2r^4}-\frac{(q^2-g^2)^2}{24b^2 r^8}+\mbox{O}\left(r^{-12}
\right)\quad\mbox{as }r\to\infty,\lb{7.31b}
\eea
which resemble \eq{7.25}. Note that \eq{7.31a} indicates that, near $r=0$, although there is  symmetry in the leading-order term with respect to the interchange of the charges $q$ and $g$,  an asymmetry appears in subsequent terms. On the other hand,
\eq{7.31b} states that such a symmetry is restored asymptotically near infinity.

In the situation of a magnetic point charge with $q=0$, the expression \eq{7.30} reduces to
\bea
{\cal H}_{q=0}&=&b^2\left(\frac{(r_m^4+r^4)^{\frac34}}{r^3}-1\right),\lb{7.32}\\
 r^2{\cal H}_{q=0}&=&\frac{2g^2}{3r_m r}+\mbox{O}(r^2)\quad\mbox{as }r\to0,\quad r^2{\cal H}_{q=0}=\frac{g^2}{2r^2}+\mbox{O}\left(r^{-6}\right)\quad\mbox{as }r\to\infty.\lb{7.33}
\eea

Thus, we see from \eq{7.31a}, \eq{7.31b}, and \eq{7.33} that the total energy $\int_{\bfR^3}{\cal H}\,\dd\x$ of the dyonic point charge diverges  following the power law $\frac1r$ but that of  a magnetic point charge a  logarithmic law
$\ln\frac1r$, at the origin. In other words, the dyonic energy is `more divergent' than the monopole energy at the spot where the charge resides. Hence, in view of energy divergence, a dyon seems {\em more charged} than a monopole.
This refined energy structure is delicate.

\section{Electrically charged black hole solutions}
\setcounter{equation}{0}

Use $g_{\mu\nu}$ to denote a gravitational metric tensor of signature $(+---)$,  $R_{\mu\nu}$ its Ricci tensor, and $G$ Newton's
gravitational constant. Then the Einstein equations assume the form
\be\lb{9.1}
R_{\mu\nu}=-8\pi G\left(T_{\mu\nu}-\frac12 g_{\mu\nu}T\right),
\ee
where $T_{\mu\nu}$ is the energy-momentum tensor of the generalized Born--Infeld model governed by the Lagrangian action 
density 
\be\lb{9.2}
{\cal L}=f(\Lm),\quad \Lm=-\frac14 F_{\mu\nu}F^{\mu\nu},\quad F_{\mu\nu}=
g_{\mu\alpha}g_{\nu\beta}F^{\alpha\beta},
\ee
given by
\be\lb{9.3}
T_{\mu\nu}=-f'(\Lm) F_{\mu\alpha}g^{\alpha\beta}F_{\nu\beta}-g_{\mu\nu}f(\Lm),
\ee
and $T=g^{\mu\nu}T_{\mu\nu}$ is the trace of $T_{\mu\nu}$. In this situation, the Born--Infeld equations
associated with \eq{9.2} are
\be\lb{9.4}
\frac1{\sqrt{-g}}\pa_\mu\left(\sqrt{-g}P^{\mu\nu}\right)=0,\quad P^{\mu\nu}=f'(\Lm)F^{\mu\nu}.
\ee

Subsequently, we will be interested in charged black hole solutions to the coupled system of equations \eq{9.1} and \eq{9.4} for
which the spacetime line element in the ordered spherical coordinates $(x^\mu)=(t,r,\theta,\phi)$ reads
\be\lb{9.5}
\dd s^2=g_{\mu\nu}\dd x^\mu\dd x^\nu=A(r)\dd t^2-\frac{\dd r^2}{A(r)}-r^2(\dd\theta^2+\sin^2\theta\,\dd\phi^2).
\ee
With \eq{9.5}, the nontrivial components of the Ricci tensor are
\be\lb{9.6}
R_{00}=-\frac{A}{2r^2}(r^2 A')',\quad
R_{11}=\frac{1}{2r^2A}(r^2 A')',\quad
R_{22}=(rA)'-1,\quad
R_{33}=\sin^2\theta\, R_{22},
\ee
where and in the sequel the prime $'$ denotes differentiation with respect to the radial variable $r$ which should not be confused with that for $f(s)$
with respect to $s=\Lm$ in \eq{2.4}, as well as elsewhere, and should be clear in various contexts.

On the other hand, in the spherically symmetric electrostatic situation, the field tensor $P^{\mu\nu}$ assumes the form
\be
P^{01}=-P^{10}=-D^r,\quad P^{\mu\nu}=0\quad\mbox{for other values of }\mu,\nu,
\ee
where $D^r$ is a scalar function depending only on $x^1=r$ and represents the radial component of
the electric displacement field $\bf D$ considered earlier. Thus, consistency in \eq{9.4} implies
\be\lb{9.8}
F^{01}=-F^{10}=-E^r,\quad F^{\mu\nu}=0\quad\mbox{for other values of }\mu,\nu,
\ee
where $E^r$ is a scalar function depending only on $r$ and representing the radial component of the
electric field $\bf E$ as before. Consequently, from \eq{9.2}, we get
\be\lb{9.9}
F_{01}=-F_{10}=E^r,\quad F_{\mu\nu}=0\quad\mbox{for other values of }\mu,\nu;\quad \Lm=\frac12 (E^r)^2.
\ee
In view of \eq{9.5}, \eq{9.8}, and \eq{9.9}, we see that the nontrivial components of \eq{9.3} are
\be\lb{9.10}
T_{00}=A(f'(\Lm)(E^r)^2-f(\Lm)),\, T_{11}=-\frac1A(f'(\Lm) (E^r)^2-f(\Lm)),\, T_{22}=r^2 f(\Lm),\, T_{33}=\sin^2\theta T_{22},
\ee
giving rise to
\be\lb{9.11}
T=2f'(\Lm)(E^r)^2-4f(\Lm).
\ee
Thus, it follows from \eq{9.6}, \eq{9.10}, and \eq{9.11} that \eq{9.1} becomes
\bea
(r^2 A')'&=&16\pi G r^2 f(\Lm),\lb{9.12}\\
(rA)'&=& 1-8\pi G r^2 (f'(\Lm)(E^r)^2-f(\Lm)).\lb{9.13}
\eea

On the other hand, from $\sqrt{-g}=r^2\sin\theta$ and \eq{9.8}--\eq{9.9}, we see that the generalized Born--Infeld equations
\eq{9.4} is reduced into
\be\lb{9.14}
(r^2 f'(\Lm)E^r)'=0.
\ee
In view of \eq{9.13} and \eq{9.14}, we have
\bea
(r^2 A')'&=&r((rA)'')\nn\\
&=&-8\pi G r\left((r^2 f'(\Lm)E^r)' E^r+r^2 f'(\Lm) E^r(E^r)'-r^2 f'(\Lm)\Lm'-2rf(\Lm)\right)\nn\\
&=&16\pi G r^2 f(\Lm),
\eea
which recovers \eq{9.12}. Hence the reduced form of \eq{9.1} is actually \eq{9.13}.
Besides, note that,  in the present context, the Hamiltonian energy density is ${\cal H}=T^0_0=g^{00}T_{00}$, which assumes the form
\be\lb{9.16}
{\cal H}=f'(\Lm)(E^r)^2 -f(\Lm),
\ee
which happens to coincide with \eq{2.13}.
Hence, with \eq{9.16}, we can integrate \eq{9.13} to obtain
  the solution
\be\lb{9.17}
A(r)=1-\frac{2GM}r+\frac{8\pi G}r\int_r^\infty \rho^2 {\cal H}(\rho)\,\dd\rho,
\ee
where $M>0$ is an integration constant.

To understand the meaning of the integral on the right-hand side of \eq{9.17}, we recall that 
\be\lb{9.18}
E=\int {\cal H}\sqrt{-g}\,\dd r\dd\theta\dd\phi=4\pi\int_0^\infty {\cal H}(r)r^2\,\dd r
\ee
is the total static energy of the electric field which is the same as the flat-space Born--Infeld energy. Thus, if we denote by $E(r)$ the static energy of the electric field distributed
in the region $\{|{\bf x}|\geq r\}$ such that $E=E(0)$, then \eq{9.17} gives us
\be\lb{9.19}
A(r)=1-\frac{2GM}r+\frac{2G E(r)}r.
\ee
Finiteness of energy implies $E(r)\to0$ rapidly as $r\to\infty$, in general, such that $M$ may be identified with the mass of
the gravitational field \cite{Bre}. An event horizon of the solution occurs at a positive root of \eq{9.19} whose existence is ensured by 
the condition
$a(0)<0$, where $a(r)\equiv rA(r)$, namely,
\be\lb{9.20}
M>E.
\ee
 Since the Kretschmann invariant of \eq{9.5} is
\be\lb{9.21}
K=\frac{(r^2 A'')^2+4(rA')^2+4(A-1)^2}{r^4},
\ee
implicating the only curvature singularity at $r=0$, we find that a naked singularity may occur when \eq{9.20} is violated.

It is interesting to see that in \eq{9.19}
and \eq{9.20} the gravitational mass $M$ and electric energy $E$ are placed at `equal' footings. 

In the electric 
point charge situation, when $\bf D$ is given by \eq{3.3}, the electric field $\bf E$ follows the relation \eq{3.4} if $f(s)$ is defined by \eq{3.1}. So we have
\be\lb{9.22}
E^r=\frac q{r^2}+\mbox{O}(r^{-4}),\quad r\to\infty.
\ee
Inserting \eq{9.22} into \eq{9.16}, we get
\be\lb{9.23}
{\cal H}(r)=\frac{q^2}{2r^4}+\mbox{O}(r^{-8}),\quad r\to\infty.
\ee
Thus we can apply \eq{9.23} in 
 the formula \eq{9.17} to arrive at
\be\lb{9.24}
A(r)=1-\frac{2GM}r+\frac{4\pi G q^2}{r^2}+\mbox{O}(r^{-6}),\quad r\to\infty,
\ee
extending the Reissner--Nordstr\"{o}m electrically charged black hole solution. It is interesting to 
emphasize that, in leading orders, the solution is independent of the details of the model \eq{3.1}.

We now consider the quadratic example \eq{4.1} to refine the solution \eq{9.24}. In such a situation, we use the results \eq{4.6}
and \eq{4.7} to obtain
\bea\lb{9.25}
{\cal H}(r)&=& \frac1{2a} \left( \frac{p(r)}6+\frac2{3p(r)}-\frac23\right)+\frac3{4a}\left(\frac{p(r)}6+\frac2{3p(r)}-\frac23\right)^2\nn\\
&=&\frac{q^2}{2r^4}-\frac{aq^4}{4r^8}+\frac{a^2 q^6}{2r^{12}}-\frac{3a^3 q^8}{2r^{16}}+\mbox{O}(r^{-20}),\quad
r\to\infty.
\eea
Combining \eq{9.17} and \eq{9.25}, we have
\be\lb{xx8.26}
A(r)=1-\frac{2GM}r+\frac{4\pi G q^2}{r^2}\left(1-\frac{aq^2}{10r^4}+\frac{a^2 q^4}{9r^8}-\frac{3a^3 q^6}{13 r^{12}}\right)
+\mbox{O}(r^{-18}),\quad r\to\infty.
\ee
Thus, in terms of higher-order terms, the solution is seen to become model dependent. The solution
\eq{xx8.26} was also obtained in
\cite{K2017}.

As another concrete example, we consider the fractional-powered model \eq{.1} with $p=\frac34$. Thus, in view of \eq{.19} and
\eq{9.19}, we have
\bea
A(r)&=&1-\frac{2GM}r+\frac{2G E(r)}{r},\quad E(r)=F\left(\frac r{r_0}\right),\lb{9.27a}\\
F(\rho)&=&\frac{4\pi q^2}{ r_0}\int_\rho^\infty\left(
\left(1-h(\eta)\right)^{-\frac14}  h(\eta) +\frac23\left[\left(1-h(\eta)\right)^{\frac34}-1\right]\right)
\eta^2\,\dd\eta,\lb{9.27}\\
h(\eta)&=&\frac2{\sqrt{1+4\eta^8}+1},\quad r_0=\left(\frac23\right)^{\frac14}\left(\frac qb\right)^{\frac12}.\lb{9.28}
\eea
Consequently, we obtain the following asymptotic form for the metric factor $A(r)$:
\be
A(r)=1-\frac{2 GM}r+\frac{4\pi G q^2}{r^2}\left(1-\frac{q^2}{120 b^2 r^4}+\frac{q^4}{3888 b^4 r^8}\right)+\mbox{O}(r^{-14}),\quad r\to\infty,
\ee
which is contained in a general dyonic solution obtained in \cite{K5}.

Similarly, for the logarithmic Born--Infeld model \eq{6.30}, the metric factor $A(r)$ of an electrically charged black hole of mass
$M$ and charge $q$ is given by \eq{9.27a} where 
\bea
F(\rho)&=&\frac{4\pi q^2}{r_0}\int_\rho^\infty\left(\frac{2p_0(\eta)}{1-p_0(\eta)}+\ln(1-p_0(\eta))\right)\eta^2\,\dd\eta,\\
p_0(\eta)&=&\frac2{(\sqrt{2+\eta^4}+\eta^2)^2},\quad r_0=\sqrt{\frac qb},
\eea
in view of \eq{6.35} or \eq{x5.47}--\eq{x5.48}. Thus, asymptotically, we get
\be
A(r)=1-\frac{2 GM}r+\frac{4\pi G q^2}{r^2}\left(1-\frac{q^2}{20 b^2 r^4}+\frac{q^4}{108 b^4 r^8}\right)+\mbox{O}(r^{-14}),\quad r\to\infty.
\ee

Not surprisingly, all these solutions share the Reissner--Nordstr\"{o}m solution as their common leading part; but surprisingly, 
on the other hand, the
terms beyond their leading part follow the same alternating pattern in signs and enjoy the same order of approximation
in truncation errors.

\section{Curvature singularities of electrically charge black holes}\lb{sec10}
\setcounter{equation}{0}

As an initial comparison, we recall that for the classical Reissner--Nordstr\"{o}m black hole of mass $M$ and electric charge $q$, the metric factor reads
\be\lb{10.1}
A(r)=1-\frac{2GM}r+\frac{4\pi G q^2}{r^2},
\ee
such that the  Kretschmann scalar \eq{9.21} assumes the form 
\be\lb{10.2}
K=\frac{48 G^2}{r^8}\left(M^2 r^2-8\pi Mq^2 r+\frac{56\pi^2 q^4}3\right).
\ee
In other words, the presence of electricity elevates the Schwarzschild curvature singularity of the blow-up type $r^{-6}$,
in absence of electricity, to the type $r^{-8}$, as is clearly implicated in \eq{10.1} by the appearance of an inverse square term
of the radial variable.

Interestingly,  note that in the situation of the Born--Infeld theory we may rewrite \eq{9.19} as
\be\lb{10.3}
A(r)=1-\frac{2GM} r+\frac{2GE}r-\frac{8\pi G}r\int_0^r {\cal H}(\rho)\rho^2\,\dd\rho,
\ee
where the integral on the right-hand side of \eq{10.3} gives rise to the energy of the electric charge in $\{|{\bf x}|\leq r\}$ which is usually
ensured to be bounded by the structure of the theory. Thus $A(r)$ is of the type $r^{-1}$ as that for the Schwarzschild metric.
In other words, convergence of electric energy `relegates' the curvature singularity of the Reissner--Nordstr\"{o}m metric
to the level of the Schwarzschild metric.

As an illustration, we consider the classical Born--Infeld theory, \eq{.1} with $p=\frac12$, so that
\be\lb{10.4}
{\cal H}=\frac{(E^r)^2}{\sqrt{1-\frac1{b^2}(E^r)^2}}-b^2\left(1-\sqrt{1-\frac1{b^2}(E^r)^2}\right), \quad E^r=\frac q{\sqrt{r^4+a^4}},\quad a=\sqrt{\frac qb},\quad D^r=\frac q{r^2},
\ee
resulting in
\be\lb{10.5}
{\cal H}=\frac{q^2}{r^2(\sqrt{r^4+a^4}+r^2)}.
\ee
Inserting \eq{10.5} into \eq{9.18} and \eq{10.3}, we have
\bea
E&=&\frac{4\pi q^2}a\int_0^\infty \frac{\dd x}{\sqrt{x^4+1}+x^2}=\frac{4\pi^{\frac52} q^2}{3\Gamma^2(\frac34) \,a},\\
A(r)&=&1-\frac{2G}r(M-E)-\frac{8\pi G q^2}{ar}\int_0^{\frac ra}\frac{\dd x}{\sqrt{x^4+1}+x^2},\lb{10.7}
\eea
such that $A(r)$ behaves like that for the Schwarzschild metric, $A(r)\sim r^{-1}$, for $r$ near  zero, as asserted.

In order to examine \eq{9.21}, we use \eq{9.12} and \eq{9.13} to obtain the general relations
\bea
r^2 A''&=& 2(A-1)+16\pi G r^2 f'(\Lm) (E^r)^2,\lb{10.8}\\
rA'&=&1-A-8\pi G r^2 {\cal H}.\lb{10.9}
\eea

As a consequence of \eq{10.5} and  \eq{10.7}--\eq{10.9}, we see that \eq{9.21} gives us
\be\lb{10.10}
K=\frac{16 G^2}{r^4}\left(\frac{3(M-E)^2}{r^2}+\frac{8\pi(M-E) qb}r+16\pi^2 q^2b^2-\frac{32\pi^2 q b^3 r^2}3+\mbox{O}(r^4)\right),\quad r\to0.
\ee
This expression indicates that, in general, the Kretschmann curvature of the charged black hole solution behaves like 
that of the Schwarzschild solution, of the same blow-up type $r^{-6}$, at the singularity, and it  recovers the 
formula for the Kretschmann curvature in the zero electric charge limit, $q=0$ and $E=0$, of the Schwarzschild black hole,
\be\lb{10.11}
K=\frac{48 G^2 M^2}{r^6}.
\ee
More interestingly, in the critical situation when the mass $M$ is equal to the electric energy,
\be\lb{10.12}
M=E,
\ee
the expression \eq{10.10} becomes
\be\lb{10.13}
K=\frac{2^8\pi^2 G^2 qb^2}{r^4}\left(q-\frac{2 b r^2}3+\mbox{O}(r^4)\right),\quad r\to0.
\ee
Thus, the order of the blow-up of the curvature drops from $6$ to $4$, so that the singularity is considerably `relegated' or `regularized'. 

In the rest of this section, we continue to consider the critical situation \eq{10.12} and the subsequent singularity relegations
for the charged black hole solutions obtained in the previous sections. Thus, now, \eq{10.3} simplifies into
\be\lb{10.14}
A(r)=1-\frac{8\pi G}r\int_0^r{\cal H}(\rho)\rho^2\dd\rho.
\ee

For the quadratic model \eq{4.1}, the Hamiltonian energy density of an electric point charge satisfies \eq{4.5}. Applying this in
\eq{10.14}, we find that $A(r)-1$ behaves like $r^{-\frac23}$ for $r$ small. Inserting this into \eq{9.21}, we have
\be\lb{10.15}
K(r)=\mbox{O}\left(r^{-\left(5+\frac13\right)}\right), \quad r\to0,
\ee
which improves upon \eq{10.11}. For the exponential model \eq{6.1}, we may rewrite the Hamiltonian energy density bound
\eq{6.8} for a point charge as
\be
{\cal H}(r)=\mbox{O}(r^{-(2+\vep)}),\quad r\to0,
\ee
where $\vep>0$ is arbitrarily small. Thus $A(r)-1$ is like $r^{-\vep}$ for $r$ small. Using this in \eq{9.21}, we get
\be\lb{10.17}
K(r)=\mbox{O}(r^{-(4+\vep)}),\quad r\to 0,
\ee
where we have conveniently replaced $2\vep$ by $\vep$ because $\vep$ is arbitrarily small. This improves upon \eq{10.15}. For
the arcsin-type model \eq{6.18}, the behavior \eq{6.23} of the Hamiltonian density of a point charge and \eq{10.14} lead to 
$A(r)-1=\mbox{O}(1)$ for $r$ small. Hence \eq{9.21} gives us
\be\lb{10.18}
K(r)=\mbox{O}(r^{-4}),\quad r\to0,
\ee
which improves upon \eq{10.17} and coincides with \eq{10.13} for the classical Born--Infeld model. Likewise, for the logarithmic
model \eq{6.30}, we find from \eq{6.35} that ${\cal H}(r)=\mbox{O}(r^{-2})$ for $r$ small. Hence we have the behavior \eq{10.18}
again since $A(r)-1=\mbox{O}(1)$ for $r$ small. Furthermore, for the full spectrum of the fractional-powered model \eq{.1} with $0<p<1$, the estimate \eq{.5} indicates
that the Kretschmann curvature $K$ of a charge black hole metric enjoys the same property \eq{10.18} of
a relegated singularity under the critical mass-energy condition \eq{10.12}. This result contains \eq{10.13} as a special case.
For the rational-function model \eq{561}, we may similarly use the estimates in \eq{565} to establish \eq{10.18} for
an electrically charged black hole.

\section{Dyonic black hole solutions}\lb{sec11}
\setcounter{equation}{0}

Consider a prescribed dyonic point charge at the origin giving rise to a radially symmetric electric and magnetic field distribution
so that \eq{2.2} and \eq{2.7} become
\be\lb{11.1}
(F^{\mu\nu})=\left(\begin{array}{cccc}0&-E^r&0&0\\ E^r&0&0&0\\0&0&0&-B^r\\0&0&B^r&0\end{array}
\right),\,
(P^{\mu\nu})=(f'(\Lm)F^{\mu\nu})=\left(\begin{array}{cccc}0&-D^r&0&0\\ D^r&0&0&0\\0&0&0&-H^r\\0&0&H^r&0\end{array}
\right),
\ee
where in view of \eq{9.2}, \eq{9.5}, and \eq{11.1}, we have
\be
\Lm=\frac12((E^r)^2-r^4\sin^2\theta (B^r)^2).
\ee

From \eq{9.5} and \eq{11.1}, we see that \eq{9.4} at $\nu=0$ gives us 
$
\pa_r\left(r^2 \sin\theta D^r\right)=0,
$
resulting in
\be
D^r=\frac q{r^2},
\ee
which is an electric point charge. Besides, by the same reason, the equation \eq{9.4} at $\nu=2$ reads $\pa_\theta
(r^2\sin\theta H^r)=0$, rendering
\be
H^r=\frac{H(r)}{\sin\theta},
\ee
where $H(r)$ is a function to be determined.  Thus, consistency in \eq{11.1} leads to the relations
\be\lb{11.5}
B^r=\frac{B(r)}{\sin\theta}, \quad \Lm=\frac12\left((E^r)^2-r^4 B^2(r)\right),\quad f'(\Lm) E^r=D^r,\quad f'(\Lm) B=H.
\ee
In particular, we see that $H$ relates to $B$ but is otherwise arbitrary. This structure offers convenience in generating solutions for
our purposes.

Now consider the Einstein equations \eq{9.1}. Note that, among the nontrivial components of the energy-momentum tensor $T_{\mu\nu}$, the quantities $T_{00}$ and $T_{11}$ are as given in \eq{9.10} with $\Lm$ as given in \eq{11.5}, and the other quantities
are
\be\lb{x10.6}
T_{22}=r^6 f'(\Lm) B^2+r^2f(\Lm),\quad T_{33}=\sin^2\theta T_{22}.
\ee
Thus
\be
T=2f'(\Lm)\left((E^r)^2-r^4 B^2\right)-4f(\Lm)=4\left(f'(\Lm)\Lm-f(\Lm)\right),
\ee
which extends \eq{9.11}. In view of these quantities and \eq{9.6}, we see that the system \eq{9.1} becomes
\bea
(r^2 A')'&=&16\pi G r^2 \left(r^4 f'(\Lm) B^2+f(\Lm)\right),\lb{11.8}\\
(rA)'&=& 1-8\pi G r^2 (f'(\Lm)(E^r)^2-f(\Lm)).\lb{11.9}
\eea

To establish the compatibility of \eq{11.8} and \eq{11.9}, we differentiate \eq{11.9} with respect to $r$ to get
\bea
(rA)''&=&-8\pi G \left((r^2 f'(\Lm) E^r)' E^r+r^2 f'(\Lm) E^r (E^r)'\right)\nn\\
&& +8\pi G\left(2rf(\Lm)+r^2 f'(\Lm) \left[E^r (E^r)'-2r^3 B^2-r^4 BB'\right]\right)\nn\\
&=&16\pi G r f(\Lm)-8\pi G r^5 f'(\Lm)(2B^2+rBB'),\lb{11.10}
\eea
where we have used the condition $\pa_r(r^2 f'(\Lm) E^r)=\pa_r (r^2 D^r)=0$.  Consequently, it follows from \eq{11.10} that
\be\lb{11.11}
(r^2 A')'=r(rA)''=16\pi G r^2 f(\Lm)-8\pi G r^6 f'(\Lm)(2B^2+rBB'),
\ee
which is compatible with \eq{11.8} if and only if $B$ satisfies the equation
\be\lb{11.12}
B'=-\frac{4B}r.
\ee
This equation has the solution
\be\lb{11.13}
B(r)=\frac g{r^4},
\ee
where $g$ is an arbitrary constant resembling the magnetic charge, which may be taken to be positive for convenience. In fact,
the expression \eq{11.13} also follows from the Bianchi identity for the electromagnetic field $F_{\mu\nu}$, which in the current context reads
\be
F_{23,1}=0\quad\mbox{or}\quad \pa_r(r^4 \sin^2\theta \,B^r)=\pa_r (r^4 \sin\theta\, B)=0.
\ee

It is interesting and important to note that \eq{11.13} is universally valid  to ensure both the compatibility 
of \eq{11.8} and \eq{11.9} and the fulfillment of the Bianchi identity, and thus, model independent. 

In summary, we see that the gravitation metric factor $A$ for a black hole of a dyonic point charge is given by the general formula
\bea\lb{11.15}
&&A(r)=1-\frac{2GM}r+\frac{8\pi G}r\int_r^\infty {\cal H}(\rho)\rho^2\,\dd\rho\nn\\
&&=1-\frac{2GM}r+\frac{8\pi G}r\int_r^\infty \left(f'\left[\frac12\left((E^\rho)^2 -\frac{g^2}{\rho^4}\right)\right](E^\rho)^2
-f\left[\frac12\left((E^\rho)^2 -\frac{g^2}{\rho^4}\right)\right]\right)\rho^2\,\dd\rho,\quad\quad\quad
\eea
following from an integration of \eq{11.9}, which is of the same form of \eq{9.17} when magnetism is absent.

Formally, the expression \eq{11.15} gives the asymptotic flatness of the metric, $A(r)\to1$ as $r\to\infty$. Besides, under the general
condition 
\be\lb{ah}
M<4\pi \int_0^\infty {\cal H}(r) r^2\,\dd r,
\ee
which is valid automatically if the energy on the right-hand side of \eq{ah} diverges, we have $A(r)\to\infty$ as $r\to0$. Hence, if $A(r)$ has a global minimum at some $r_0>0$ where $A(r_0)\leq0$, there exist event
horizons: if $A(r_0)=0$, we have an extremal black hole metric, and if $A(r_0)<0$, there are inner and outer event horizons at
$r_-$ and $r_+$, respectively, with $r_-<r_0<r_+$.  So, from $A'(r_0)=0$, we have
\be
M=4\pi \left({\cal H}(r_0) r_0^3+\int_{r_0}^\infty {\cal H}(\rho)\rho^2\,\dd\rho\right).
\ee
Assume $A(r_0)<0$ such that the inner and outer horizons are present. Then the behavior of $A(r)$
described  implies $A'(r_+)\geq0$. This property allows us to calculate the Hawking black hole temperature \cite{FK,K4,MW}
\be\lb{Th}
T_{\mbox{\tiny H}}=\frac{A'(r_+)}{4\pi}=\frac1{4\pi r_+}-2G r_+{\cal H}(r_+),
\ee
where we have used the condition $A(r_+)=0$ or
\be
r_+=2GM-8\pi G \int^\infty_{r_+} {\cal H}(\rho)\rho^2\,\dd \rho
\ee
to eliminate the integral. In the limiting situation when the electromagnetic sector is absent, ${\cal H}=0$, $r_+$ becomes the Schwarzschild radius $r_{\mbox{\tiny s}}=2GM$ and \eq{Th} recovers the classical Hawking temperature $T_{\mbox{\tiny H}}
=\frac1{8\pi GM}$ \cite{Hawking}.

As a quick illustration, we may insert the Hamiltonian density for a dyonic charge given in \eq{7.11} into \eq{11.15} to obtain
\be\lb{11.16}
A(r)=1-\frac{2GM}r+\frac{8\pi G}r\int_r^\infty\frac{\left(\frac{qg}b\right)^2+(q^2+g^2)\rho^4}{\rho^2(\rho^4+\sqrt{(\rho^4+[\frac qb]^2)(\rho^4+[\frac gb]^2)})}\,\dd\rho.
\ee
By asymptotic expansion, we get from \eq{11.16} the formula
\be\lb{11.17}
A(r)=1-\frac{2GM}r+\frac{4\pi G}{r^2}\left(q^2+g^2+\frac1{10b^2 r^4}\left(q^2 g^2-\frac14(q^2+g^2)^2\right) +\mbox{O}
\left(\frac1{b^4 r^8}\right)\right),\, r\to\infty.
\ee
Thus, the classical Reissner--Nordstr\"{o}m solution is recovered in the Maxwell theory limit, $b\to\infty$.

As another example, we consider the dyonic black hole for the $\arcsin$-model \cite{K3,K4} with Hamiltonian density \eq{7.20}. Hence, inserting \eq{7.20} into \eq{11.15}, we have
\bea
&&A(r)=1-\frac{2GM}r\nn\\
&&+\frac{8\pi G q^2}{r r_e^2}\int_r^\infty\left(\left(\frac{r_m^4}{\rho^4}+\frac{2(r_e^4-r_m^4)}{\sqrt{r_e^4(r_e^4-r_m^4)+\rho^8}+\rho^4}\right)^{\frac12}-\frac{\rho^2}{r_e^2}\arcsin \frac{(r_e^4-r_m^4)}{\sqrt{r_e^4(r_e^4-r_m^4)+\rho^8}+\rho^4}\right)\,\dd\rho, \nn\\
&&
\eea
which gives us the asymptotic expression 
\be\lb{11.19}
A(r)=1-\frac{2GM}r+\frac{4\pi G}{r^2}\left(q^2+g^2-\frac{\beta^2}{216 r^8}(q^2-g^2)^3+\mbox{O}\left(\frac{\beta^4}{r^{16}}\right)\right),\quad r\to\infty.
\ee
It is interesting to note that, unlike the solution \eq{11.17} which is symmetric in $q$ and $g$, this solution is asymmetric in $q$ and $g$
in higher orders. 

For the fractional-powered model \eq{.1}, we have seen that there is a range of the power $p$, \eq{.10}, where the symmetry
regarding finiteness of energy is broken between electric and magnetic point charges. Thus, it will be interesting to know how this
broken symmetry is observed in the context of black hole metrics. For this purpose, we take the borderline situation, $p=\frac34$,
as an example. Thus, we may insert the result \eq{7.30} into \eq{11.15} to arrive at
\be\lb{11.20}
A(r)=1-\frac{2GM}r+\frac{8\pi G b^2}{r}\int_r^\infty \left(\frac{3r_e^4}{2\rho^2}F^{\frac12}(\rho)+\rho^2\left(F^{\frac32}(\rho)-1\right)\right)\,\dd\rho,
\ee
where
\be
F(\rho)=\frac{2(r_m^4+\rho^4)}{\sqrt{r^8_e+4(r_m^4+\rho^4)\rho^4}+r_e^4}.
\ee
Thus, inserting \eq{7.27} and using asymptotic expansion, we obtain the expression
\be
A(r)=1-\frac{2GM}r+\frac{4\pi G}{r^2}\left(q^2+g^2-\frac{(q^2-g^2)^2}{60 b^2 r^4}+\frac{(q^2+5g^2)(q^2-g^2)^2}{1944
b^4 r^8}+\mbox{O}\left(\frac1{b^6 r^{12}}\right)\right),
\ee
for $r$ large. Hence, we see that the electric-magnetic symmetry is maintained up to the order $r^{-6}$ but broken at the order
$r^{-10}$ for $r\to\infty$. In other words, in this model, the electric-magnetic asymmetry is `deeply hidden' gravitationally.

\section{Curvature singularities of magnetically charged black holes}
\setcounter{equation}{0}

In this section, we aim to relegate the curvature singularities of black hole solutions induced from finite-energy magnetic point charges or monopoles in the same spirit of electric point charges as seen in Section \ref{sec10}. In the present context, though,
the magnetic field is universally given by \eq{11.13}, which gives rise to the Hamiltonian density explicitly when electricity
is switched off, so that 
 \eq{11.15} may be computed directly by the much reduced formula
\be\lb{12.1}
A(r)
=1-\frac{2GM}r-\frac{8\pi G}r\int_r^\infty
f\left( -\frac{g^2}{2\rho^4}\right)\rho^2\,\dd\rho.
\ee
Since now the energy carried by the monopole is
\be\lb{12.2}
E=-4\pi\int_0^\infty f\left( -\frac{g^2}{2r^4}\right)r^2\,\dd r,
\ee
we may rewrite \eq{12.1} as
\be\lb{12.3}
A(r)=1-\frac{2G}r (M-E)+\frac{8\pi G}r\int_0^r f\left(-\frac{g^2}{2\rho^4}\right)\rho^2\dd\rho.
\ee

For the exponential model \eq{6.1} considered in \cite{K}, we have
\bea
A(r)&=&1-\frac{2G}r(M-E)-\frac{4\pi G g^2}r\int_0^r \frac1{\rho^2}\e^{-\frac{\beta g^2}{2\rho^4}}\,\dd\rho,\lb{12.4}\\
 E&=&2\pi \int_0^\infty \frac{g^2}{r^2}\e^{-\frac{\beta g^2}{2r^4}}\,\dd r
=\frac{\pi^2 g^{\frac32}}{(2\beta)^{\frac14}\Gamma\left(\frac34\right)},\lb{12.5}
\eea
for the gravitational metric factor and monopole energy, respectively. Since the integral on the right-hand side of \eq{12.4}
vanishes faster than any power of $r$ as $r\to0$, we see that the Kretschmann curvature follows the sharp estimate
\be\lb{x11.6}
K=\frac{48G^2(M-E)^2}{r^6}+\mbox{O}(r^\sigma),\quad r\to0,
\ee
where $\sigma>0$ may be arbitrarily large. In particular, if $M\neq E$, we attain the same curvature singularity for $K$ as that
for the Schwarzschild black hole as stated in \eq{10.11}.

If $M>E$, then $A(r)\to -\infty$ when $r\to 0$ and $A(r)\to 1$ when $r\to\infty$. Such a picture ensures the existence of an event horizon. 

If
$M<  E$, then $A(r)\to\infty$ as $r\to0$ and $A(r)\to1$ as $r\to\infty$. Hence $A(r)$ attains its global minimum at some $r_0>0$
where $A'(r_0)=0$. Thus, now, an event horizon exists if and only if $A(r_0)\leq0$. 
If $A(r_0)=0$, then we can apply the condition $A'(r_0)=0$ to get
\be\lb{12.6}
r_0^2=4\pi G g^2 \e^{-\frac{\beta g^2}{2r_0^4}},
\ee
which is a necessary condition for the occurrence of an extremal black hole. This condition, when combined with $A(r_0)=0$ or
\be\lb{12.7}
r_0={2G}(M-E)+{4\pi G g^2}\int_0^{r_0} \frac1{\rho^2}\e^{-\frac{\beta g^2}{2\rho^4}}\,\dd\rho,
\ee
implies $A'(r_0)=0$ as well. In other words, the conditions \eq{12.6}--\eq{12.7} are necessary and sufficient for the occurrence of
an extremal black hole. As a consequence, if the solution $r_0$ to \eq{12.6}, if any, violates \eq{12.7} such that either $A(r_0)<0$ or
$A(r_0)>0$, then there are event horizons both below and above $r=r_0$ or there is no event horizon, respectively.
In particular, if we relate $A(r)$ and $A'(r)$ by the equation
\be\lb{x11.9}
A(r)=-rA'(r)+\left(1-\frac{4\pi G g^2}{r^2} \e^{-\frac{\beta g^2}{2r^4}}\right),
\ee
and note that the maximum of the function ${\e^{-\frac1{\eta^{4}}}}/{\eta^2}$ in $\eta>0$ is attained at $\eta=2^{\frac14}$ which is
${\e^{-\frac12}}/{\sqrt{2}}$ so that when
\be\lb{x11.10}
\frac{\sqrt{\frac\beta2}}{4\pi G g}>\frac{\e^{-\frac12}}{\sqrt{2}}\quad\mbox{or}\quad\frac{\sqrt{\beta\e}}{4\pi Gg}>1
\ee
the equation \eq{12.6} has no solution and that the second term on the right-hand side of \eq{x11.9} stays always positive, we see that $A>0$ where $A'=0$. Consequently, there is no event horizon. In other words, the condition \eq{x11.10} spells out a 
situation when the curvature singularity at $r=0$ becomes naked.

In the critical situation $M=E$ where $E$ is as given in \eq{12.5}, we see from \eq{12.4} that the
metric factor $A(r)$ is simplified into
\be
A(r)=1-\frac{4\pi G g^2}r\int_0^r \frac1{\rho^2}\e^{-\frac{\beta g^2}{2\rho^4}}\,\dd\rho,\lb{12.8}
\ee
which satisfies $A(r)\to1$ as $r\to0$ and as $r\to\infty$. As before, let $r_0>0$ be such that $A(r)$ attains its global minimum there. Then $A'(r_0)=0$ such that $r_0$ satisfies the equation
\be
\int_0^{r_0} \frac1{\rho^2}\e^{-\frac{\beta g^2}{2\rho^4}}\,\dd\rho= \frac1{r_0}\e^{-\frac{\beta g^2}{2r_0^4}}.
\ee
 In this situation, the existence of an event
horizon is ensured by the necessary and sufficient condition $A(r_0)\leq0$ or
\be\lb{12.9}
r_0\leq {4\pi G g^2}\int_0^{r_0} \frac1{\rho^2}\e^{-\frac{\beta g^2}{2\rho^4}}\,\dd\rho=\frac{4\pi G g^2}{r_0}\e^{-\frac{\beta g^2}{2r_0^4}},
\ee
so that the equality leads to the extremal case. In particular, when \eq{x11.10} is fulfilled, the condition \eq{12.9} will never occur.
Hence there is no event horizon. However,
in all circumstances under the condition $M=E$, it is seen
from \eq{x11.6} that the Kretschmann curvature is regular at $r=0$. Consequently, we have obtained black holes free of 
curvature singularity in the model.

For the logarithmic model \eq{6.30}, the formula \eq{12.3} leads to
\bea
A(r)&=&1-\frac{2G}r(M-E)-\frac{8\pi G b^2}r\int_0^r \ln\left(1+\frac{g^2}{2b^2\rho^4}\right)\rho^2\,\dd\rho,\lb{12.8}\\
E&=&4\pi b^2 \int_0^\infty \ln\left(1+\frac{g^2}{2b^2 r^4}\right)r^2\,\dd r=\frac{2^{\frac74}\pi^2b^{\frac12}g^{\frac32}}3.\lb{12.9}
\eea
Moreover, the integration on the right-hand side of \eq{12.8} may be carried out to give us the explicit result 
\bea
A(r)&=&1-\frac{2G}r(M-E)-\frac{2^{\frac94}\pi G b^{\frac12}g^{\frac32}}rF\left(\frac{2^{\frac14}b^{\frac12}r}{g^{\frac12}}\right),\\
F(\rho)&=&\int_0^\rho\ln\left(1+\frac1{\eta^4}\right) \eta^2\,\dd \eta
=\frac{ \rho^3}3\ln\left(1+\frac1{\rho^4}\right)+\frac{\sqrt{2}}6\ln\left(\frac{1-\sqrt{2}\rho+\rho^2}{1+\sqrt{2}\rho+\rho^2}\right)\nn\\
&&\quad +\frac{\sqrt{2}}3\left(\arctan(1+\sqrt{2}\rho)-\arctan(1-\sqrt{2}\rho)\right).
\eea
Thus, we have
\be\lb{12.14}
A(r)=1-\frac{2G}r(M-E)-\frac{8\pi G b^2}3\left(\frac43+\ln\left[\frac{g^2}{2b^2}\right]-4\ln r\right)r^2+\mbox{O}(r^6),\quad r\to0.
\ee
For this solution, we may draw similar conclusions 
when $M\neq E$ about the event horizons as those for the exponential model, which we omit here.

When $M=E$, however, we have from \eq{12.14} that $A(r)-1=\mbox{O}(|\ln r|r^2)$ for $r$ small. Using this in \eq{9.21}, we obtain
\be\lb{12.15}
K=\mbox{O}([\ln r]^2),\quad r\to0.
\ee
Thus, although curvature blow-up is still present at $r=0$, it is greatly quenched, in contrast to that in the electric point charge situation as described in \eq{10.18}.

For the $\arctan$-model \eq{6.12}, the formula \eq{12.3} gives us
\bea
A(r)&=&1-\frac{2G}r(M-E)-\frac{8\pi G}{r\beta}\int_0^r \arctan\left(\frac{\beta g^2}{2\rho^4}\right)\rho^2\,\dd\rho,\lb{12.16}\\
E&=&\frac{4\pi}\beta \int_0^\infty \arctan\left(\frac{\beta g^2}{2 r^4}\right)r^2\,\dd r\nn\\
&=&2\pi\left(\frac2\beta\right)^{\frac14}g^{\frac32}\int_0^\infty \arctan \left(\frac1{\eta^4}\right)\, \eta^2\,\dd \eta
\approx (1.36822)\,2\pi\left(\frac2\beta\right)^{\frac14}g^{\frac32}.\lb{12.17}
\eea
By asymptotic expansion of the integrand on the right-hand side of \eq{12.16}, we have
\be
A(r)=1-\frac{2G}r(M-E)-\frac{4\pi^2 G}{3\beta} r^2+\frac{16\pi G}{7\beta^2 g^2} r^6+\mbox{O}(r^{14}),\quad r\to0,
\ee
which improves upon \eq{12.14} because the logarithmic factor in front of the term containing $r^2$ disappears. Consequently,
at the critical mass, $M=E$, we have
\be
A(r)=1-\frac{4\pi^2 G}{3\beta} r^2+\frac{16\pi G}{7\beta^2 g^2} r^6+\mbox{O}(r^{14}),\quad r\to0,
\ee
so that the Kretschmann curvature \eq{9.21} is  regular at $r=0$ and assumes the form
\be
K(r)=\frac{128\pi^4 G^2}{3\beta^2}-\left(\frac{2048\pi^3 G^2}{3\beta^3 g^2}\right)r^4
+\left(\frac{268288\pi^2 G^2}{49\beta^4 g^4}\right)r^8+\mbox{O}(r^{12}),\quad r\to0.
\ee

Note that, unlike the exponential model, the $\arctan$-model does not permit a finite-energy electric point charge.

Thus, it will be
interesting to investigate the fractional-powered model \eq{.1} for $0<p<\frac34$ which allows finite-energy electric {\em and} magnetic point charges. In fact, with \eq{.1} and \eq{12.3},
 we have
\bea
A(r)&=&1-\frac{2G}r(M-E)-\frac{8\pi G b^2}{r}\int_0^r \left(\left[1+\frac{g^2}{2pb^2 \rho^4}\right]^p-1\right)\rho^2\,\dd\rho,\lb{12.21}\\
E&=&{4\pi b^2} \int_0^\infty  \left(\left[1+\frac{g^2}{2pb^2 r^4}\right]^p-1\right)  r^2\,\dd r\nn\\
&=&2^{\frac54}\pi b^{\frac12}p^{-\frac34}g^{\frac32}\int_0^\infty\left(\left[1+\frac1{\eta^4}\right]^p-1\right)\eta^2\dd \eta.\lb{12.22}
\eea
In the range $0<p<\frac34$, the only known exact value of the integral
in \eq{12.22} is at $p=\frac12$, which is the classical Born--Infeld model.

Examining \eq{12.21}, we see that lower values of $p$ give rise to higher vanishing orders for the tail term. 
More precisely, from \eq{12.21}, we have 
\be\lb{11xA}
A(r)=1-\frac{2G}r(M-E)+\mbox{O}\left(r^{-2(2p-1)}\right),\quad r\to0,\quad 0<p<\frac34.
\ee
Such information is
useful for us to understand the curvature singularity under the critical mass-energy condition for the model with respect to $p$. For example,
for $p=\frac12$ and $p=\frac14$, we have the explicit asymptotic expressions
\bea
A(r)&=&1-\frac{2G}r(M-E)-8\pi G\left(gb-\frac{b^2}3 r^2+\frac{b^3}{10g}r^4+\mbox{O}(r^8)\right),\lb{12.23}\\
A(r)&=&1-\frac{2G}r(M-E)-8\pi G\left(\frac{g^{\frac12}b^{\frac32}}{2^{\frac34}}r-\frac{b^2}3 r^2+\frac{b^{\frac72}}{3\cdot
2^{\frac{19}4}g^{\frac32}} r^5+\mbox{O}(r^9)\right),\lb{12.24}
\eea
respectively, for $r$ small. Thus, in view of \eq{9.21} and \eq{12.23}--\eq{12.24}, we get
\be\lb{12.25}
K(r)=\mbox{O}\left(r^{-4}\right),\quad p=\frac12;\quad K(r)=\mbox{O}\left(r^{-2}\right),\quad p=\frac14,
\ee
as $r\to0$, under the critical mass-energy condition, $M=E$. More generally, under this critical condition, 
we may insert \eq{11xA} into \eq{9.21} to get
\be\lb{11xK}
K=\mbox{O}\left(r^{-8p}\right),\quad r\to0,\quad 0<p<\frac34,
\ee
 which contains the special cases stated in \eq{12.25} and confirms our verdict that small values of $p$ 
lead to gain in curvature regularity in the fractional-powered model. Note that the borderline limit $p\to\frac34$ in \eq{11xK}
corresponds to the Schwarzschild curvature singularity, $K\sim r^{-6}$, which is of a massive origin.

The regular magnetically charged black holes of Ma \cite{Ma} arising from the rational-function model of Kruglov \cite{K7,K8}
also belong to the family of solutions subject to the critical mass-energy condition.

It is of interest to note that, on the other hand,
 magnetic monopoles may also elevate singularities by boosting the curvature blow-up rates. To see how this plays out to be, we consider a special case of the model \eq{3.1}:
\be\lb{12.26}
f(s)=s+\beta s^m,\quad s=\Lm,\quad \beta>0,
\ee
where $m\geq3$ is an odd integer. With radial component of the magnetic field of a monopole given by $B^r=\frac{g}{r^2}$, we obtain
from \eq{12.1} and \eq{12.26} the exact result
\be\lb{12.27}
A(r)=1-\frac{2GM}r +\frac{4\pi G g^2}{r^2}\left(1+\frac{\beta g^{2(m-1)}}{(4m-3)2^{m-1} r^{4(m-1)}}\right).
\ee
Thus the associated Kretschmann curvature \eq{9.21} is found to be
\be
K=\frac{4096\pi^2 G^2 \beta^2 g^{4m}}{(4m-3)^2 2^{2(m-1)} r^{8m}}\left(\frac3{32}-\frac{7m}{16}+\frac{17 m^2}{16}-\frac{3m^3}2+m^4\right)+\mbox{O}\left(\frac1{r^{4(m+1)}}\right),\quad r\to0.
\ee
In other words, we see that the curvature blow-up rate at $r=0$ may be made arbitrarily high with a suitable choice of the
power integer $m$.

\section{Bardeen type black hole metrics}
\setcounter{equation}{0}

The metric factor $A(r)$ of the Bardeen black hole in our context reads
\be\lb{13.1}
A(r)=1-\frac{2GM r^2}{(r^2+g^2)^{\frac32}},
\ee
which is seen to be free of curvature singularity, with the Kretschmann invariant given by
\be
K=\frac{12G^2 M^2(8g^8-4g^6 r^2+47 g^4 r^4-12 g^2 r^6+4r^8)}{(g^2+r^2)^7}.
\ee
In \cite{AG1}, a Born--Infeld type field-theoretical interpretation is provided as a possible source accounting for the rise of such a regular black hole solution, based on a magnetic point charge formalism. See also \cite{AG2} for an earlier work based on an
electric point charge model. The former is simpler and more direct in achieving an enhanced regularity because it allows a direct prescription of the charge as we discussed in
the previous two sections. Although the these two approaches are technically different, they share the common feature that they
are both built over the idea of maintaining the critical mass-energy condition, $M=E$, as we emphasized previously. In this section,
we develop this idea further based on our general framework formulated in Section \ref{sec11}.

Motivated by \cite{AG1}, we now consider the model defined by
\be\lb{13.3}
f(s)=-\alpha \left(\frac{\beta\sqrt{-2s}}{1+\beta \sqrt{-2s}}\right)^{p},\quad s=\Lm,
\ee
where $\alpha,\beta,p$ are positive parameters. Note that the minus sign under the radical root excludes electricity but includes
magnetism such that, unlike the classical Born--Infeld model \cite{BI} and
the models studied in this paper elsewhere, this model cannot recover the Maxwell theory as its limiting situation. Since
\be\lb{13.4}
f'(s)=p\alpha\beta^2\frac{(\beta\sqrt{-2s})^{p-2}}{(1+\beta\sqrt{-2s})^{p+1}},
\ee
we see regularity of the model at $s=0$ requires the condition $p\geq2$. In the magnetic monopole situation of our interest, we have
$\Lm=
-\frac{g^2}{2r^4}$, where $g>0$ is the magnetic charge. So we arrive at the associated Hamiltonian energy density
\be
{\cal H}=-f(\Lm)=\frac{\alpha(\beta g)^p}{(r^2+\beta g)^p}.
\ee
Thus, to ensure a finite monopole energy,
\be
E=4\pi \int_0^\infty {\cal H}(r) r^2\,\dd r<\infty,
\ee
we need to observe the condition
$
p>\frac32,
$
in our subsequent discussion. With this preparation, we see that \eq{12.3} leads to the expressions
\bea
A(r)&=&1-\frac{2G}r(M-E)-\frac{8\pi G \alpha(\beta g)^p}r\int_0^r \frac{\rho^2}{(\rho^2+\beta g)^p}\,\dd\rho,\lb{13.7}\\
E&=&4\pi \alpha(\beta g)^p \int_0^\infty \frac{\rho^2}{(\rho^2+\beta g)^p}\,\dd\rho=4\pi\alpha (\beta g)^{\frac32}
\int_0^\infty\frac{\eta^2}{(\eta^2+1)^p}\,\dd \eta,\quad p>\frac32.\lb{13.8}
\eea
Consequently, under the critical mass-energy condition $M=E$, we may conveniently
 rewrite \eq{13.7} and \eq{13.8}, respectively, as
\be\lb{13.9}
A(r)=1-\frac{8\pi G\alpha (\beta g)^{\frac32}}r F_p\left(\frac r{\sqrt{\beta g}}\right),\quad F_p(\rho)\equiv
\int_0^{\rho}\frac{\eta^2}{(\eta^2+1)^p}\,\dd \eta;\quad E=4\pi\alpha(\beta g)^{\frac32} F_p(\infty).
\ee

At this point, it is useful to recall the Chebyshev theorem which states that the necessary and sufficient condition for the
indefinite integral
\be
\int (a+b\eta^m)^p \eta^n\,\dd \eta,
\ee
where $m,n,p$ are rational numbers, to be an elementary function is that at least one of  the three quantities, $p, \frac{n+1}m,p+\frac{n+1}m$, is an integer \cite{MZ,T},
although, in general, it is given by a hypergeometric function. Applying this theorem to our integration on the right-hand side of \eq{13.7} or \eq{13.9}, we see that such a condition takes the form
\be
p\quad \mbox{or}\quad p+\frac12\quad \mbox{is an integer for $p$ given in \eq{13.3}.}
\ee

As an illustration, for $p=2,3,4$, we obtain
\bea
F_2(\rho)&=&\frac12\arctan \rho -\frac \rho{2(\rho^2+1)};\quad E=\pi^2\alpha(\beta g)^{\frac32},\lb{13.12}\\
 F_3(\rho)&=&\frac18\arctan \rho-\frac{\rho(1-\rho^2)}{8(\rho^2+1)^2};\quad E=\frac14\pi^2\alpha(\beta g)^{\frac32},\\
F_4(\rho)&=&\frac3{48}\arctan \rho-\frac{\rho(3-8\rho^2-3\rho^4)}{48(\rho^2+1)^3}; \quad E=\frac1{8}\pi^2\alpha(\beta g)^{\frac32},
\eea
and for  $p=\frac52,\frac72,\frac92$, we get the results
\bea
F_{\frac52}(\rho)&=&\frac{\rho^3}{3(\rho^2+1)^{\frac32}};\quad E=\frac{4\pi\alpha(\beta g)^{\frac32}}3,\lb{13.15}\\
F_{\frac72}(\rho)&=&\frac{\rho^3(2\rho^2+5)}{15(\rho^2+1)^{\frac52}};\quad E=\frac{8\pi\alpha(\beta g)^{\frac32}}{15},\\
F_{\frac92}(\rho)&=&\frac{\rho^3(8\rho^4+28\rho^2+35)}{105(\rho^2+1)^{\frac72}};\quad E=\frac{32\pi\alpha(\beta g)^{\frac32}}{105},
\eea
among which \eq{13.15} was found in \cite{AG1} that recovers the Bardeen black hole \eq{13.1} with setting $\beta=g$ and 
$\alpha=\frac{3M}{4\pi g^3}$.

Most generally, for any $p>\frac32$, the expression for $A(r)$ in  \eq{13.9} gives us the asymptotic expansion
\be
A(r)=1-8\pi G\alpha r^2\left(\frac13-\left[\frac{p}{5\beta g}\right]r^2+\mbox{O}(r^4)\right),\quad r\to0.
\ee
In view of this and \eq{9.21}, we have 
\be
K=\frac{512\pi^2 G^2 \alpha^2}3-\left(\frac{512\pi^2 G^2\alpha^2 p}{\beta g}\right)r^2+\mbox{O}(r^4),\quad r\to0.
\ee
In particular, we see that curvature regularity holds in the full range of $p$, $p>\frac32$, for the model \eq{13.3}.

Note that, from \eq{2.16} and \eq{13.4}, we may express the nontrivial radial component $H^r$ of the magnetic intensity field $\bf H$ as
\be
H^r=f'\left(-\frac{g^2}{2r^4}\right) \frac{g}{r^2}=p\alpha\beta\left(\frac{\beta g}{r^2+\beta g}\right)^{p-1}\frac{r^4}{(r^2+\beta g)^2},
\ee
which is regular and behaves like $r^4$ at $r=0$. Furthermore, the free magnetic charge contained in the region $\{|\x|\leq r\}$ is
\be
g_{\mbox{\tiny free}}(r)=\frac1{4\pi}\int_{|\x|\leq r}\nabla\cdot{\bf H}\,\dd\x=\int_0^r\frac{\dd}{\dd\rho}(\rho^2 H^\rho)\,\dd\rho=p\alpha\beta\left(\frac{\beta g}{r^2+\beta g}\right)^{p-1}\frac{r^6}{(r^2+\beta g)^2}.
\ee
Hence the total free magnetic charge is
\be
g_{\mbox{\tiny free}}=\lim_{r\to\infty}
g_{\mbox{\tiny free}}(r)=\left\{\begin{array}{lll}&\infty,\quad&\frac32<p<2,\\ &2g\alpha\beta^2,\quad &p=2,\\ &0,\quad &p>2,\end{array}\right.
\ee
which spells out a unique situation for the model \eq{13.3} to have a finite non-vanishing total free magnetic charge. In particular, 
the Bardeen black hole \eq{13.1} corresponds to a vanishing total free magnetic charge.

For arbitrary $M>0$, the metric factor $A(r)$ given in \eq{13.7} has the asymptotic expansion
\be
A(r)=1-\frac{2GM}r+\frac{8\pi G\alpha (\beta g)^p}{r^{2(p-1)}}\left(\frac1{2p-3}-\frac{p(\beta g)}{(2p-1)r^2}+
\frac{p(p+1)(\beta g)^2}{2(2p+1)r^4}+\mbox{O}(r^6)\right),
\ee
for $r$ large and $p>\frac32$.

We next consider electrically charged Bardeen type black holes. In this situation, the metric factor $A$ can be obtained from integrating \eq{10.9} in which
\be
{\cal H}=f'\left(\frac12 (E^r)^2\right)(E^r)^2-f\left(\frac12(E^r)^2\right),
\ee
is the Hamiltonian energy density of the electrostatic field generated from an electric point charge given by the
radial electric displacement
field $D^r=\frac q{r^2}$, related through the constitutive relation
\be\lb{13.25}
f'\left(\frac12 (E^r)^2\right) E^r=D^r.
\ee
In general, \eq{13.25} is difficult to solve for $E^r$ in terms of $D^r$, so that it is a more challenging task to construct electrically charged
black holes than magnetically charged ones. In \cite{AG2}, it is shown that such a difficulty may be by-passed through an implicit
function theory approach which we now revisit.

Following \cite{AG2}, we now use the notation
\be\lb{13.26}
F=-\frac14 F_{\mu\nu}F^{\mu\nu},\quad P=-\frac14 P_{\mu\nu}P^{\mu\nu},\quad P^{\mu\nu}=f'(F)F^{\mu\nu},\quad
P_{\mu\nu}=f'(F)F_{\mu\nu}.
\ee
Hence we have
\be\lb{13.27}
\left(f'(F)\right)^2 F=P.
\ee
In the electric point charge situation, we have
\be\lb{13.28}
F=\frac12 (E^r)^2,\quad P=\frac12 (D^r)^2,\quad {\cal H}=2f'(F) F-f(F)\equiv {\cal H}(F),
\ee
so that \eq{13.27} is equivalent to \eq{13.25}. The nice idea in \cite{AG2} is to express $\cal H$ directly in terms of $P$ through
the implicit equation \eq{13.27}, such that, in turn, to express the Lagrangian action density ${\cal L}=f(F)$ in terms 
of $P$ as well. 

In fact, from the last relation in  \eq{13.28}, and then \eq{13.27}, we have
\be
\frac{\dd{\cal H}}{\dd P}\frac{\dd P}{\dd F}=\frac{\dd{\cal H}}{\dd F}=\frac1{f'(F)}\frac{\dd}{\dd F}\left( (f'(F))^2 F\right)=\frac1{f'(F)}\frac{\dd P}{\dd F}.
\ee
Hence $\frac1{f'(F)}=\frac{\dd {\cal H}}{\dd P}$ so that the last relation in \eq{13.28} leads to
\be\lb{13.30}
{\cal L}=f(F)=\frac2{f'(F)} \left(f'(F)\right)^2 F-{\cal H}(F)=2P\,\frac{\dd{\cal H}(P)}{\dd P}-{\cal H}(P),
\ee
as arrived at in \cite{AG2}. Besides, from \eq{13.25}, we also have
\be\lb{13.30b}
E^r=\frac{D^r}{f'(F)}=D^r\,\frac{\dd {\cal H}}{\dd P}.
\ee

To obtain the Bardeen type {\em electrically} charged black holes, we take 
\be\lb{13.31}
{\cal H}(P)=\alpha \left(\frac{\beta\sqrt{2P}}{1+\beta \sqrt{2P}}\right)^{p},
\ee
as we have done for the monopole situation in \eq{13.3} before, which gives us the Lagrangian action density by \eq{13.30} as
\be\lb{13.32}
{\cal L}=\alpha \left(\frac{\beta\sqrt{2P}}{1+\beta \sqrt{2P}}\right)^{p}\left(\frac p{1+\beta\sqrt{2P}}-1\right).
\ee

In the electric point charge situation, $P=\frac{q^2}{2r^4}$ where $q>0$ is a prescribed electric charge. Inserting this into \eq{13.31} and integrating the resulting equation \eq{10.9}, we obtain $A(r)$ as that for the magnetic charge situation, with
$g$ there simply replaced with $q$ here. For example, use \eq{13.12} in \eq{13.9} with $g$ replaced by $q$, we have the solution
\be\lb{x12.34}
A(r)=1-{4\pi G\alpha \beta q}\left(\frac{\arctan\left(\frac r{\sqrt{\beta q}}\right)}{\frac r{\sqrt{\beta q}}}-\frac{1}{\left(\frac r{\sqrt{\beta q}}\right)^2+1}\right)\equiv 1-4\pi G\alpha\beta q\,A_0(\rho),\quad \rho=\frac r{\sqrt{\beta q}}.
\ee
It is clear that $A(r)\to1$ as $r\to0$ and $r\to\infty$ and $A(r)$ attains its global minimum at
\be
r_0=\sqrt{\beta q}\,\rho_0,
\ee
where $\rho_0>0$ is the unique global maximum point of the function $A_0(\rho)$ defined by \eq{x12.34}. Hence 
$A(r_0)=1-8\pi G \alpha\beta q A_0(\rho_0)$ so that $A(r_0)<0, A(r_0)=0, A(r_0)>0$ correspond to the situations
when there occur inner and outer event horizons, an extremal (clustered, single) horizon, no horizon, respectively, which is
a typical picture for the behavior of a Reissner--Nordstr\"{o}m charged black hole metric.

Note that, a drawback in \eq{13.31} or \eq{13.32} is that the model does not permit magnetic monopoles due to the radical root
involved. Thus, with regard to electricity and magnetism, the models \eq{13.3} and \eq{13.32} are mutually rejecting or exclusive.
To overcome this exclusiveness, that is, to accommodate both electricity and magnetism, we may consider those models for which
the said sign restriction due to electricity and magnetism does not arise as a mathematical obstruction. For example, we may consider
the model \eq{6.1} which allow both electric and magnetic point charges of finite energies. Unfortunately, for this model, it is hard to
resolve the constitutive equation \eq{13.25} for $E^r$ in terms of $D^r$ so that an explicit construction of an electrically charged
black hole free of curvature singularity is still at large. Now, motivated by the Hamiltonian formalism in \cite{AG2}, as discussed here,
based on the quantity $P$, we may directly consider the model
\be\lb{13.33}
{\cal H}=P\e^{-\beta P},\quad \beta>0,
\ee
such that for the situation of an electric point charge, with $P=\frac {q^2}{2r^4}$, we see that $A(r)$ is given by the formulas
\eq{12.4}--\eq{12.5} with $g$ there replaced with $q$ here. Using \eq{13.30b}, we get the nontrivial component of the electric field 
to be
\be
E^r=\frac{q}{r^2}\left(1-\frac{\beta q^2}{2r^4}\right)\e^{-\frac{\beta q^2}{2r^4}},
\ee
which is clearly seen to be asymptotically Coulomb in the limit $r\to\infty$.
In particular, under the critical mass-energy, $M=E$, the black hole
solution is free of singularity with the metric factor
\bea
A(r)&=&1-\frac{4\pi G q^2}r\int_0^r\frac1{\rho^2}\e^{-\frac{\beta q^2}{2\rho^4}}\,\dd\rho\nn\\
&=&1-\frac{4\pi G q^2}{r^2}\left(1-\frac{\beta q^2}{10 r^4}+\frac{\beta^2 q^4}{72 r^8}+\mbox{O}\left(r^{-12}\right)\right),\quad r\to\infty.
\eea 
Besides, with \eq{13.30}, the Lagrangian action density of the exponential model \eq{13.33} assumes the form
\be
{\cal L}=P\e^{-\beta P}(1-2\beta P).
\ee

We now show that the model \eq{13.33} allows a finite-energy magnetic point charge. In such a situation, we note that
the nontrivial radial components of the magnetic field and magnetic intensity field, $B^r$ and $H^r$, respectively, follow
the following relations:
\be\lb{13.40}
f'(F) B^r=H^r,\quad F=-\frac12 (B^r)^2,\quad P=-\frac12(H^r)^2.
\ee
Hence we have
\bea\lb{13.41}
\frac{g^2}{r^4}=(B^r)^2&=&\frac{(H^r)^2}{(f'(F))^2}={(H^r)^2}\left(\frac{\dd{\cal H}}{\dd P}\right)^2\nn\\
&=&(H^r)^2\e^{-2\beta P}(1-\beta P)^2=(H^r)^2 \left(1+\frac\beta2[H^r]^2\right)\e^{\beta (H^r)^2},
\eea
in view of \eq{13.33} and \eq{13.40}. As a consequence, we have $(H^r)^2\to\infty$ as $r\to0$, which enables us to deduce
from \eq{13.41} that
\be\lb{13.42}
\e^{\beta (H^r)^2}\leq \frac1{r^4}\quad \mbox{or}\quad \beta(H^r)^2\leq -4\ln r,\quad 0<r<r_1,
\ee
for some $r_1>0$. Using \eq{13.40} and \eq{13.42} in \eq{13.33}, we get
\be
|{\cal H}|=\frac12(H^r)^2\e^{\frac12\beta(H^r)^2}\leq \frac{2|\ln r|}{\beta r^2},\quad 0<r<r_1.
\ee
This estimate establishes the local convergence of the energy of the magnetic field generated from a magnetic point charge at $r=0$.

Furthermore, from \eq{13.41} again, we see that $H^r\to0$ as $r\to\infty$. Applying this property in \eq{13.41}, we deduce
$H^r\sim r^{-2}$ for $r$ large. Thus, from \eq{13.33} and \eq{13.40}, we see that ${\cal H}\sim r^{-4}$, which establishes
the convergence of the magnetic energy near infinity.

Thus, we conclude that the total energy of a magnetic point charge in the model \eq{13.33} is indeed finite as well.

\section{Hayward type black holes}
\setcounter{equation}{0}

Another interesting and much studied charged black hole solution free of curvature singularity is given by the metric factor \cite{BP,Hay}
\be\lb{h1}
A(r)=1-\frac{2m r^2}{r^3+2l^2 m},
\ee
where $m,l>0$ are constants, which shares the same spirit of the Bardeen black hole \cite{Hay} and is sometimes referred to as the Hayward black hole \cite{F,Kumar}.

To see how \eq{h1} may arise as a magnetically charged black hole, we insert it into \eq{12.1} to obtain
\be\lb{h2}
{8\pi G}\int_r^\infty f\left(-\frac{g^2}{2\rho^4}\right)\rho^2\dd\rho=\frac{2m r^3}{r^3+2l^2 m}-{2GM}.
\ee
Thus, by \eq{12.2} and \eq{h2}, we read off the monopole energy
\be\lb{h3}
E=-4\pi \int_0^\infty f\left(-\frac{g^2}{2\rho^4}\right)\rho^2\dd\rho=M.
\ee
That is, we uncover again the critical mass-energy condition $M=E$. Letting $r\to\infty$ in \eq{h2}, we also get the recognition
$m=GM$.
Besides, differentiating \eq{h2}, we have
\be\lb{h4}
f(s)=-\alpha \left(\frac{\beta (-2s)^{\frac34}}{1+\beta(-2s)^{\frac34}}\right)^2,\quad \alpha=\frac3{8\pi G l^2},\quad\beta=
\frac{2l^2 m}{g^{\frac32}},\quad s=-\frac{g^2}{2r^4}.
\ee
Consequently, comparing \eq{h4} with the generalized Bardeen model \eq{13.3}, we arrive at the following generalized Hayward
model
\be\lb{h5}
f(s)=-\alpha \left(\frac{\beta (-2s)^{\frac34}}{1+\beta(-2s)^{\frac34}}\right)^p,\quad s=\Lm,
\ee
where $\alpha,\beta,p>0$ are parameters. Regularity of $f'(s)$ at $s=0$ requires $p\geq\frac43$. It is clear that this family of models exclude electric point charges as in the family of the Bardeen type models given by \eq{13.3}. Furthermore, using \eq{h5}
with $s=-\frac{g^2}{2r^4}$  in
\eq{h3}, we can integrate to find
\be\lb{h6}
E=\frac{4\pi\alpha\beta g^{\frac32}}{3(p-1)}, \quad \mbox{resulting in}\quad \alpha=\frac{3(p-1)M}{4\pi \beta g^{\frac32}}.
\ee

Inserting \eq{h5} with $s=-\frac{g^2}{2r^4}$ into \eq{12.3} with $M=E$ and integrating, we obtain the general expression
\be\lb{h7}
A(r)=1-\frac{2GM}{ r}\frac{\left(\left[\frac r{r_0}\right]^3+1\right)^{p-1}-1}{\left(\left[\frac r{r_0}\right]^3+1\right)^{p-1}},\quad r_0=\beta^{\frac13}g^{\frac12},\quad p\geq\frac43,
\ee
in view of \eq{h6}. Consequently, asymptotically, we have
\bea
A(r)&=&1-\frac{2GM}{r_0}\left((p-1)\left[\frac r{r_0}\right]^2-\frac {p(p-1)}2\left[\frac r{r_0}\right]^5+\frac{p(p^2-1)}6\left[\frac r{r_0}\right]^8
+\mbox{O}(r^{11})\right),\, r\to0,\quad\quad\\
A(r)&=&1-\frac{2GM}{r}\left(1-\left[\frac {r_0}r\right]^{3(p-1)}+(p-1)\left[\frac {r_0}r\right]^{3p}+\mbox{O}(r^{-3(p+1)})\right),\quad r\to\infty,
\eea
so that near infinity the solution looks like a Schwarzschild black hole.

Below are a few concrete examples, at $p=2,3,4$:
\bea
A(r)&=&1-\frac{2GM r^2}{r^3+r_0^3},\quad p=2,\\
A(r)&=&1-\frac{2GM r^2\left(r^3+2r_0^3\right)}{\left(r^3+r_0^3\right)^2},\quad p=3,\\
A(r)&=&1-\frac{2GM r^2\left(r^6+3r_0^3 r^3+3r_0^6\right)}{\left(r^3+r_0^3\right)^3},\quad p=4,
\eea
among which the result at $p=2$ corresponds to the familiar Hayward black hole metric \eq{h1} as  shown in \cite{BP,Hay}.
The regularity of these solutions are self-evident.

To obtain the Hayward type electrically charged black holes, we may use the notation \eq{13.26} to write down the Hamiltonian 
energy density
\be
{\cal H}(P)=\alpha\left(\frac{\beta(2P)^{\frac34}}{1+\beta(2P)^{\frac34}}\right)^p,\quad p\geq\frac43,
\ee
in view of \eq{h5}. Thus, applying \eq{13.30}, we arrive at the Lagrangian action density of the model
\be
{\cal L}=\alpha\left(\frac{\beta(2P)^{\frac34}}{1+\beta(2P)^{\frac34}}\right)^p
\left(\frac{3p}{2\left(1+\beta(2P)^{\frac34}\right)}-1\right),
\ee
which is analogous to \eq{13.32}, such that electrically charged black holes are given by the same metric factor \eq{h7} with
the magnetic charge $g>0$ being simply replaced with an electric charge $q>0$, since now the electrostatic situation
is defined by $P=\frac{q^2}{2r^4}$.

\section{Charged black holes in an exponential model}
\setcounter{equation}{0}
\setcounter{theorem}{0}

A simple but rich-in-content model we would like to consider is given by the exponential function
\be\lb{13.44}
f(s)=\frac1\beta (\e^{\beta s}-1),\quad\beta>0,\quad s=\Lm,
\ee
which has previously been studied in \cite{H1,H2} in connection to asymptotic Reissner--Nordstr\"{o}m black hole solutions.
An obvious advantage of  such a model is that it is free of sign restriction and regular everywhere. Note that, alternatively, taking $p\to\infty$ in fractional-powered model \eq{.1}, we obtain 
\be\lb{13.44b}
\lim_{p\to\infty} b^2\left(1-\left[1-\frac s{pb^2}\right]^p\right)=\frac1\beta \left(1-\e^{-\beta s}\right),\quad \beta=\frac1{b^2},
\ee
which may be viewed as a companion to \eq{13.44}. However, it is clear that the model \eq{13.44b} does not allow a finite-energy magnetic
point charge. In other words,  we now lose the electromagnetic symmetry with \eq{13.44b}. However, with \eq{13.44}, both
electric and magnetic point charges are of finite energies, thus the electromagnetic symmetry is restored, as we shall see below.

In the electric point charge situation, the constitutive equation \eq{13.25} becomes
\be
\e^{\frac\beta2(E^r)^2}E^r=D^r,
\ee
which may be recast into an elegant nonlinear equation of the form
\be
\e^W W=\delta,\quad W=\beta(E^r)^2,\quad\delta=\beta (D^r)^2,
\ee
so that $\beta (E^r)^2=W(\delta)=W(\beta [D^r]^2)$ is determined by the classical Lambert $W$ function \cite{CG}. For our purpose,
we note that $W(x)$ is analytic for $x>-\frac1\e$ and has the Taylor expansion
\be\lb{13.47}
W(x)=\sum_{n=1}^\infty \frac{(-n)^{n-1}}{n!}x^n,
\ee
about $x=0$, and the asymptotic expansion
\be\lb{13.48}
W(x)=\ln x-\ln\ln x+\frac{\ln\ln x}{\ln x}+\cdots,\quad x>3.
\ee
Now, in terms of this $W$-function, the Hamiltonian energy density of the electric point charge with $D^r=\frac q{r^2}$ reads
\be\lb{13.49}
{\cal H}=\frac1\beta\left(\e^{\frac12 W\left(\frac{\beta q^2}{r^4}\right)}\left[W\left(\frac{\beta q^2}{r^4}\right)-1\right]+1\right),
\ee
giving rise to the total electric energy
\bea\lb{13.50}
E=\int {\cal H}\,\dd\x&=&\frac{4\pi}\beta \int_0^\infty \left(\e^{\frac12 W\left(\frac{\beta q^2}{r^4}\right)}\left[W\left(\frac{\beta q^2}{r^4}\right)-1\right]+1\right)r^2\,\dd r\nn\\
&=&\frac{\pi q^{\frac32}}{\beta^{\frac14}}\int_0^\infty \frac{\left(\e^{\frac12 W\left(x\right)}\left[W\left(x\right)-1\right]+1\right)}{x^{\frac74}}\,\dd x.
\eea
Using the asymptotic expansions \eq{13.47} and \eq{13.48} in \eq{13.50}, we see that the finiteness of the electric energy follows.
A numerical evaluation of the integral in \eq{13.50} gives us the approximation $E\approx (6.570480628)\,\frac{\pi q^{\frac32}}{\beta^{\frac14}}$.
Besides, it is also useful to write down the estimates for the electric field $E^r$ here,
\be
(E^r)^2=\frac1\beta\ln\left(\frac{\beta q^2}{r^4}\right)+\mbox{O}\left(\ln\ln\frac1r\right),\quad r\to0;\quad E^r=\frac q{r^2}
+\mbox{O}\left(r^{-4}\right),\quad r\to\infty.
\ee
Consequently, we may calculate the free electric charge contained in $\{|\x|\leq r\}$ and the total charge as before:
\be
q_{\mbox{\tiny free}}(r)=\left(r^2 E^r\right)=\frac{r^2}{\sqrt{\beta}}\sqrt{W\left(\frac{\beta q^2}{r^4}\right)},\quad r>0;\quad
q_{\mbox{\tiny free}}=q_{\mbox{\tiny free}}(\infty)=q.
\ee

From \eq{11.15}, \eq{13.49}, and \eq{13.50}, the metric factor $A(r)$ for an electrically charged black hole is found to be
\bea\lb{13.53}
A(r)&=&1-\frac{2GM}r+\frac{8\pi G}{\beta r}\int^\infty_r    \left(\e^{\frac12 W\left(\frac{\beta q^2}{\rho^4}\right)}\left[W\left(\frac{\beta q^2}{\rho^4}\right)-1\right]+1\right)\rho^2\,\dd\rho\nn\\
&=&1-\frac{2G}r(M-E)-\frac{8\pi G}{\beta r}\int_0^r   \left(\e^{\frac12 W\left(\frac{\beta q^2}{\rho^4}\right)}\left[W\left(\frac{\beta q^2}{\rho^4}\right)-1\right]+1\right)\rho^2\,\dd\rho.
\eea
Thus, from \eq{13.47} and the upper line in \eq{13.53}, we have
\be\lb{13.54}
A(r)=1-\frac{2GM}r+\frac{4\pi G q^2}{r^2}\left(1-\frac{\beta q^2}{20 r^4}+\mbox{O}\left(r^{-8}\right)\right),\quad r\to\infty,
\ee
and, from \eq{13.48} and the lower line in \eq{13.53}, we have
\be\lb{13.55}
A(r)=1-\frac{2G}r(M-E)+\frac{32\pi Gq}{\beta^{\frac12}}\left(\ln r+\mbox{O}(1)\right),\quad r\to0.
\ee
In particular,  under the critical mass-energy condition $M=E$,  we see from \eq{13.55} that there holds
\be
K=\mbox{O}\left(\frac{(\ln r)^2}{r^4}\right),\quad r\to0,
\ee
for the Kretschmann curvature. So an electric black hole singularity still persists but is seen to be relegated.

For a magnetic point charge with $B^r=\frac {g}{r^2}$ ($g>0$) for the radial component of the magnetic field, we apply
\eq{13.44} to \eq{12.1} and \eq{12.3} to get
\bea\lb{13.57}
A(r)&=&1-\frac{2GM}r+\frac{8\pi G}{\beta r}\int_r^\infty \left(1-\e^{-\frac{\beta g^2}{2\rho^4}}\right)\rho^2\dd\rho\nn\\
&=&1-\frac{2G}r (M-E)-\frac{8\pi G}{\beta r}\int_0^r \left(1-\e^{-\frac{\beta g^2}{2\rho^4}}\right)\rho^2\dd\rho,\\
E&=&\frac{4\pi}\beta\int_0^\infty  \left(1-\e^{-\frac{\beta g^2}{2r^4}}\right)r^2\dd r=\frac{2^{\frac54}\pi g^{\frac32}}{\beta^{\frac14}}\int_0^\infty \left(1-\e^{-\frac1{\eta^4}}\right) \eta^2\,\dd \eta.\lb{13.58}
\eea
The integral on the right-hand side of \eq{13.58} may be expressed in terms of the Whittaker $M$ function \cite{Wh} which gives rise to the
approximate value 1.208535. The upper line in \eq{13.57} renders the expression
\be\lb{x13.16}
A(r)=1-\frac{2GM}r+\frac{4\pi G g^2}{r^2}\left(1-\frac{\beta g^2}{20 r^4}+\mbox{O}\left(r^{-8}\right)\right),\quad r\to\infty,
\ee
which assumes the identical form as \eq{13.54} within an exchange of the electric and magnetic charges, $q$ and $g$,
although in the magnetic
situation there is no involvement of the Lambert $W$ function.
The lower line in \eq{13.57}, on the other hand, yields
\bea
A(r)&=&1-\frac{2G}r (M-E)-\frac{8\pi G r^2}{3\beta} +\frac{2^{\frac94}\pi G g^{\frac32}}{\beta^{\frac14} r}F\left(\frac{2^{\frac14}r}{\beta^{\frac14}g^{\frac12}}\right),\lb{13.60}\\
F(\rho)&=&\int_0^{\rho}\e^{-\frac1{\eta^4}}\eta^2\,\dd \eta.\lb{13.61}
\eea
It is clear that the function \eq{13.61} vanishes at $\rho=0$ faster than any power function of $\rho$. Thus, we may rewrite \eq{13.60} as
\be\lb{13.62}
A(r)=1-\frac{2G}r(M-E)-\frac{8\pi G r^2}{3\beta}+\mbox{O}(r^n),\quad r\to0,
\ee
for an arbitrarily large integer $n$. The radial component of the induced magnetic intensity field, $H^r$, follows the constitutive equation 
\be
H^r=\e^{-\frac{\beta g^2}{2r^4}}\, \frac{g}{r^2},
\ee
such that the free total magnetic charge contained in the region $\{|\x|\leq r\}$ and the full space are
\be
g_{\mbox{\tiny free}}(r)=r^2 H^r=g\,\e^{-\frac{\beta g^2}{2r^4}},\quad g_{\mbox{\tiny free}}=g_{\mbox{\tiny free}}(\infty)=g,
\ee
respectively. It is interesting that  the
magnetic intensity field  and the induced free magnetic charge, generated from the prescribed magnetic point charge, vanish rapidly near the point
charge and spread out in space. 

Assuming $M=E$ in \eq{13.62} and inserting  the resulting $A(r)$ into \eq{9.21}, we have
\be
K=\frac{512\pi^2 G^2}{3\beta^2}+\mbox{O}(r^m),\quad r\to0,
\ee
where $m$ is an arbitrarily large integer. Hence the curvature singularity at $r=0$ now disappears.

For a dyonic point charge with $D^r$ and $B^r$ given in \eq{7.1}, we have the constitutive equations
\be\lb{13.66}
\e^{\frac{\beta}2\left([E^r]^2-[B^r]^2\right)}\, E^r = D^r,\quad \e^{\frac{\beta}2\left([E^r]^2-[B^r]^2\right)}\, B^r = H^r.
\ee
Again, we can use the Lambert $W$ function to solve the first equation in \eq{13.66} to obtain
\be\lb{13.67}
\beta (E^r)^2=W\left(\beta(D^r)^2 \e^{\beta(B^r)^2}\right)=W\left(\frac{\beta q^2}{r^4}\,\e^{\frac{\beta g^2}{r^4}}\right).
\ee
Thus, for $r$ small, we apply \eq{13.48} to get
\be\lb{13.68}
\beta (E^r)^2=\frac{\beta g^2}{r^4}+\ln\left(\frac{q^2}{g^2}\right)-\frac{r^4}{\beta g^2}\ln\left(\frac{\beta q^2}{r^4}\right)+\mbox{O}(r^8).
\ee
Consequently, inserting \eq{13.68} into the associated Hamiltonian energy density given as 
\be\lb{13.69}
{\cal H}=\frac1\beta\left( \e^{\frac{\beta}2\left([E^r]^2-[B^r]^2\right)}\left(\beta [E^r]^2-1\right)+1\right),\quad B^r=\frac{g}{r^2},
\ee
we find
\be\lb{13.70}
{\cal H}=\frac{qg}{r^4}+\frac1{2\beta g}\left(2(g-q)+q\ln\left[\frac{q^2 r^4}{\beta g^4}\right]\right)+\mbox{O}(r^4),\quad
r\to0,
\ee
 neglecting irrelevant  truncation error terms. The dominant term in \eq{13.70} is seen to be
\be\lb{13.71}
\frac{q g}{r^4},
\ee
which leads to the divergence of the dyon energy near $r=0$. This term is a clear indication that the energy divergence is a
consequence of a dyonic point charge, or both electric and magnetic charges assigned at a given point,
characterized with $q,g\neq0$. 

Nevertheless, we may obtain a dyonic black hole solution of the model \eq{13.44} as follows. In fact, inserting \eq{13.69} into \eq{11.15}, we have the explicit and exact result
\be
A(r)=1-\frac{2GM}r +\frac{8\pi G}{\beta r}\int_r^\infty \left( \e^{\frac12\left(W\left(\frac{\beta q^2}{\rho^4}\,\e^{\frac{\beta g^2}{\rho^4}}\right)-\frac{\beta g^2}{\rho^4}\right)}
\left(W\left(\frac{\beta q^2}{\rho^4}\,\e^{\frac{\beta g^2}{\rho^4}}\right)-1\right)+1\right)\rho^2\dd\rho.
\ee
Hence, using \eq{13.47}, we get the asymptotic expression
\be
A(r)=1-\frac{2GM}r+\frac{4\pi G}{r^2}\left((q^2+g^2)-\frac\beta{20 r^4}(q^2-g^2)^2+\mbox{O}\left(r^{-8}\right)\right),\quad r\to\infty,
\ee
which contains the Reissner--Nordstr\"{o}m solution in the $\beta\to0$ limit, is symmetric with respect to the electric and magnetic
charges, and covers \eq{13.54} and \eq{x13.16} as special cases. On the other hand, from \eq{13.70}, we see that such an electric and magnetic charge symmetry is broken
near $r=0$.

The free electric charge contained in the region $\{|\x|\leq r\}$ and total free electric charge in the full space are
\be\lb{13.74}
q_{\mbox{\tiny free}}(r)=\int_0^r \frac{\dd}{\dd\rho}\left(\rho^2 E^\rho\right)\,\dd\rho=\frac{r^2}{\sqrt{\beta}}\sqrt{W\left(\frac{\beta q^2}{r^4}\,\e^{\frac{\beta g^2}{r^4}}\right)}-g,\quad q_{\mbox{\tiny free}}=q_{\mbox{\tiny free}}(\infty)=q-g,
\ee
respectively in view of \eq{13.67}, \eq{13.68}, and then \eq{13.47}. Furthermore, we may use \eq{13.47}, \eq{13.67}, and \eq{13.68} to get the asymptotic expansions
\bea
E^r&=&\frac{q}{r^2}\left(1+\frac{\beta(g^2-q^2)}{2r^4}\right)+\mbox{O}(r^{-10}),\quad r\to\infty,\lb{13.75}\\
E^r&=&\frac{g}{r^2}+\frac{r^2}{\beta g}\ln\left(\frac qg\right)+\mbox{O}(r^6),\quad r\to0.\lb{13.76}
\eea
Interestingly, \eq{13.75} implies that $E^r$ follows an electric Coulomb law for $r$ large, and \eq{13.76} indicates that $E^r$ follows
a magnetic Coulomb law for $r$ small. Such a balance explains why the total free electric charge is a combination of both electric and
magnetic charges, as given in \eq{13.74}.
 
Likewise, in view of the second equation in \eq{13.66} and \eq{13.68}, we obtain
\be\lb{Hr35}
H^r=\frac{q}{r^2}-\frac{qr^2}{2\beta g^2}\ln\left(\frac{\beta q^2}{r^4}\right)+\mbox{O}(r^6),\quad r\to0.
\ee
Consequently, in leading orders, the magnetic intensity field of a dyonic point charge behaves like a pure electric point charge at the origin. 
 Moreover, using \eq{13.47} in the second
equation in \eq{13.66} again, we get
\be\lb{Hr36}
H^r=\frac{g}{r^2}\left(1+\frac{\beta(q^2-g^2)}{2r^4}\right)+\mbox{O}(r^{-10}),\quad r\to\infty.
\ee
Thus, the magnetic intensity field $H^r$ behaves like the magnetic field $B^r$ for $r$ large and follows the Coulomb law as well.

As a consequence of  \eq{Hr35} and \eq{Hr36},  we see that the associated total free magnetic charge is
\be
g_{\mbox{\tiny free}}=(r^2 H^r)_{r=0}^{r=\infty}=g-q.
\ee

At this juncture, it will be of interest to discuss various energy conditions \cite{HE} related to the occurrence and disappearance of curvature
singularities we have since seen. In fact, in the current spherically symmetric dyon situation, we have, in view of
\eq{9.10} with \eq{11.5} and \eq{x10.6}, 
\be
T^0_0=T^1_1=f'(\Lm) (E^r)^2 -f(\Lm),\quad T^2_2=T^3_3=-r^4f'(\Lm)B^2-f(\Lm),
\ee
giving rise to the density $\rho=T^0_0$, radial pressure $p_1=p_{r}=-T^1_1$, and tangential pressure $p_2=p_3=p_\perp=
-T_2^2=-T^3_3$ \cite{D1,D2}, such that
 the weak energy condition reads $\rho\geq0,\rho+p_i\geq0$ ($i=1,2,3$), which may be reduced into
\be\lb{aa37}
f'(\Lm)(E^r)^2 -f(\Lm)\geq0,\quad f'(\Lm) \left((E^r)^2+r^4 B^2\right)\geq0,
\ee
the dominant energy condition becomes $\rho\geq|p_i|$ ($i=1,2,3$), which combines \eq{aa37} with the additional requirement
$\rho-p_i\geq0$ ($i=1,2,3$), or
\be
f'(\Lm)\left((E^r)^2-r^4B^2\right)-2f(\Lm)\geq0,\lb{aa38}
\ee
and the strong energy condition imposes $\rho+\sum_{i=1}^3 p_i\geq0$, which recasts itself in the form
\be\lb{aa39}
r^4 f'(\Lm)B^2+f(\Lm)\geq0.
\ee

Thus, we can readily check that, for the exponential model \eq{13.44} in the magnetic point charge situation with
$B$ given by \eq{11.13}, $\Lm=-\frac{g^2}{2r^4}$, and $E^r=0$, both the weak and dominant energy conditions
 hold. For example, in this situation the left-hand side of \eq{aa38} assumes the form
\be
-f'(\Lm)r^4B^2-2f(\Lm)=\frac {2h(\tau)}\beta, \quad h(\tau)=1-\e^{-\tau}(1+{\tau}),\quad \tau=\frac{\beta g^2}{2r^4},
\ee
and it is clear that $h(\tau)>0$ for all $\tau>0$. On the other hand, it is evident that the condition \eq{aa39} is violated. In fact,
in this situation, we may rewrite the left-hand side of \eq{aa39} as
\be\lb{aa41}
r^4 f'(\Lm)B^2+f(\Lm)=\frac{h(\tau)}\beta,\quad h(\tau)=(2\tau+1)\e^{-\tau}-1,\quad \tau=\frac{\beta g^2}{2r^4}.
\ee
The function $h(\tau)$ defined in \eq{aa41} decreases for $\tau>\frac12$ and $h(2)<0$ (say). Thus $h(\tau)<0$ for $\tau>2$. In other words,
the strong energy condition is invalid in the region 
\be\lb{aa42}
r<\beta^{\frac14}\sqrt{\frac g2}\quad\mbox{(say)}.
\ee
In contrast, the
electric point charge situation with $\Lm=\frac12 (E^r)^2$ fulfills the strong energy condition, on the other hand. Such a picture explains the
 disappearance and occurrence of the curvature singularity associated with the magnetic and electric point charges, respectively,
in view of the Hawking--Penrose singularity theorems \cite{Haw,HE,HP,MTW,Nab,Pen1,Pen2,Sen,Wald} for which the fulfillment of the strong energy condition is essential. More generally, the strong energy condition \eq{aa39} holds for a dyonic point charge so long as the electric charge 
dominates over the magnetic charge, $q\geq g$, since now $\Lm\geq0$. To see this, we use \eq{13.67} to get
\be
2\beta\Lm=W\left(\frac{\beta q^2}{r^4}\e^{\frac{\beta g^2}{r^4}}\right)-\frac{\beta g^2}{r^4}\geq0,
\ee
in view of the condition $q\geq g$ and the identity $W(\tau\e^\tau)=\tau$ ($\tau\geq0$) for the Lambert $W$ function. Hence
\eq{aa39} follows.

We have previously seen that the $\arctan$-model permits regular magnetically charged black hole solutions. Following the above
discussion, we see that in this situation the weak and dominant energy conditions are both satisfied but the strong energy condition fails in the region \eq{aa42} (say) for a similar reason. Besides, an electric point charge obviously enjoys the strong energy condition 
\eq{aa39}. 

We may also obtain regular electrically charged black hole solutions following the method in \cite{AG2} and discussed in the previous
section. To this end, we use the notation \eq{13.26} to set up the Hamiltonian energy density
\be\lb{14.36}
{\cal H}=\frac1\beta\left(1-\e^{-\beta P}\right).
\ee
Thus, in view of \eq{13.30}, we obtain the Lagrangian action density for the model \eq{14.36} as follows,
\be
{\cal L}=2P\e^{-\beta P}+\frac1\beta\left(\e^{-\beta P}-1\right).
\ee
In this situation, the electrically charged black hole metric factor is given by \eq{13.57}--\eq{13.58} with the magnetic charge $g$
being simply replaced by an electric charge $q$.

Formerly, the exponential model \eq{13.44} may be viewed as taking the limit $p\to\infty$ in the fractional-powered model
\be\lb{e7}
f(s)=\frac1{b^2}\left(\left[1+\frac{s}{pb^2}\right]^p-1\right),\quad p>0, 
\ee
in analogy to \eq{13.44b} with setting $\beta=\frac1{b^2}$. Here we observe that, although the model \eq{13.44} permits
both electric  and magnetic point charges, the model \eq{e7} excludes a magnetic point charge. In fact, when $p$ is a general real
number, a magnetic point charge with $s=-\frac{g^2}{2r^4}$ is unacceptable due to the sign restriction, and when $p$ is an integer 
so that there is no longer a sign restriction, a finite-energy magnetic charge does not exist in view of the discussion made in 
Section 3. Moreover, if $p$ is a rational number in the form $\frac mn$ where $m,n$ are coprime integers with $n$ odd, then there
is no sign restriction. In this case, a magnetic point charge carries a finite energy when $p<\frac34$. This condition is the same
as that obtained for the model \eq{.1} in Section 6. On the other hand, for an electric point charge so that the radial component of its electric displacement field is given
by $D^r=\frac{q}{r^2}$, then the radial component $E^r$ of its electric field is determined by the equation 
\be\lb{e8}
\left(1+\frac{(E^r)^2}{2pb^2}\right)^{p-1} E^r=\frac{q}{r^2}.
\ee
In order to maintain consistency in \eq{e8} in the limit $r\to0$ such that $E^r\to\infty$ as $r\to0$, we have $p>\frac12$.
Thus the situation $p=\frac12$ is excluded. Now assume $p>\frac12$. Then \eq{e8} yields
\be
E^r\sim r^{-\frac2{2p-1}},\quad r\to0;\quad E^r=\frac{q}{r^2}+\mbox{O} (r^{-4}),\quad r\to\infty.
\ee
Consequently, in view of \eq{2.13}, \eq{e7}, and \eq{e8}, we may obtain the following asymptotic estimates for the Hamiltonian energy density
\be
{\cal H}\sim r^{-\frac{4p}{2p-1}},\quad r\to0;\quad {\cal H}\sim r^{-4},\quad r\to\infty.
\ee
Hence $r^2{\cal H}\sim r^{-\frac2{2p-1}}$ for $r$ small. Therefore we conclude that the electric point charge carries a finite energy
if and only if
\be
p>\frac32.
\ee
This is an interesting condition since it recovers the classical Maxwell theory situation at $p=1$ and spells the Maxwell theory out as
one among a family of the Born--Infeld type models which do not permit an electric point charge.

Summarizing our results regarding finite-energy charged black hole solutions arising in the generalized Born--Infeld theory of electromagnetism, we have
\begin{theorem}\lb{th2}
In the context of the Born--Infeld type nonlinear theory of electromagnetism, the gravitational metric of a finite-energy charged black hole solution enjoys a relegated curvature singularity at the same level of that of the Schwarzschild black hole without charge.
In particular, the well-known higher-order, deteriorated, curvature singularity of the classical Reissner--Nordstr\"{o}m black hole metric is seen to arise as a consequence of the divergence of its electromagnetic energy. In the critical coupling situation when the black hole gravitational mass is equal to its electromagnetic energy in the presence of electricity or magnetism, the
curvature singularity may be further relegated from that of the Schwarzschild black hole or even completely removed so that the gravitational metric is free of singularity.
\end{theorem}

We note that our results indicate that the theory \eq{2.3} is unable to provide a model that allows the existence of a finite-energy {\em dyonically charged} point particle. Theorem \ref{th2}, on the other
hand, explains the importance of the finiteness of the electromagnetic energy of a black hole in relegating the associated curvature singularity. Thus, it will be of interest to explore
and extend \eq{2.3}
further to accommodate dyonically charged black hole solutions aimed at achieving relegated and removed curvature singularity.

\section{Cosmology driven by a Born--Infeld scalar field}
\setcounter{equation}{0}
\setcounter{theorem}{0}

In this section, we study the cosmological expansion of a universe
propelled by  a Born--Infeld type scalar-field governed by the Lagrangian action density of the general form 
 \cite{Bab1,Ba1,Bab2,Ad,An,Ba2,Al,Ru,Ba3,GGY} 
\be\lb{a1}
{\cal L}=f(X)-V(\vp),
\ee
where $X=\frac12\pa_\mu\vp\pa^\mu\vp=\frac12g^{\mu\nu}\pa_\mu\vp \pa_\nu\vp$, $\vp$ being a real-valued scalar field, $g_{\mu\nu}$ the gravitational metric tensor, and $V$ is a potential density
function. The wave equation or the Euler--Lagrange equation associated with \eq{a1} is
\be\lb{a2}
\frac1{\sqrt{-g}}\pa_\mu \left(\sqrt{-g}f'(X)\pa^\mu\vp\right)+V'(\vp)=0,
\ee
with the associated energy-momentum tensor given by
\be\lb{a3}
T_{\mu\nu}=\pa_\mu\vp\pa_\nu\vp f'(X)-g_{\mu\nu}(f(X)-V(\vp)).
\ee

For cosmology, we consider an isotropic and homogeneous universe governed by the Robertson--Walker line element
\be\lb{a4}
\dd s^2=g_{\mu\nu}\dd x^\mu\dd x^\nu=\dd t^2-a^2(t)(\dd x^2 +\dd y^2 +\dd z^2),
\ee
in Cartesian coordinates, where $a(t)>0$ is the scale factor or radius of the universe to be determined. With \eq{a4}, we write down the nontrivial components of the associated Ricci tensor
\be\lb{a5}
R_{00}=\frac{3\ddot{a}}a,\quad R_{11}=R_{22}=R_{33}=-2\dot{a}^2 -a\ddot{a},
\ee
where $\dot{a}=\frac{\dd a}{\dd t}$, etc. Correspondingly, we assume that the scalar field is also only time-dependent so that 
\eq{a2} becomes
\be\lb{a6}
\left(a^3 f'(X)\dot{\vp}\right)\dot{}=-a^3 V'(\vp).
\ee
Besides, now, the nontrivial components of \eq{a3} are
\be\lb{a7}
T_{00}=\dot{\vp}^2 f'(X)-\left(f(X)-V(\vp)\right),\quad T_{11}=T_{22}=T_{33}=a^2 \left(f(X)-V(\vp)\right),\quad X=\frac12\dot{\vp}^2.
\ee
From \eq{a7}, we get the trace of $T_{\mu\nu}$ to be
\be\lb{a8}
T=\dot{\vp}^2 f'(X)-4\left(f(X)-V(\vp)\right).
\ee
Thus, in view of \eq{a5}, \eq{a7}, and \eq{a8}, we see that the Einstein equation \eq{9.1} becomes 
\bea
\frac{\ddot{a}}a&=&-\frac{8\pi G}3\left(\frac12\dot{\vp}^2 f'(X)+f(X)-V(\vp)\right),\lb{a9}\\
\frac{\ddot{a}}a+2\left(\frac{\dot{a}}a\right)^2&=&8\pi G\left(\frac12\dot{\vp}^2 f'(X)-f(X)+V(\vp)\right).\lb{a10}
\eea
Combining \eq{a9} and \eq{a10}, we have 
\be\lb{a11}
\left(\frac{\dot{a}}a\right)^2=\frac{8\pi G}3\left(\dot{\vp}^2 f'(X)-f(X)+V(\vp)\right).
\ee
It can be directly checked that \eq{a6} and \eq{a11} imply \eq{a9} and \eq{a10} as well. Hence, the coupled governing
equations \eq{9.1} and \eq{a2}--\eq{a3} are recast into \eq{a6} and \eq{a11}, which are now the focus of attention.

Comparing \eq{a3} with that of a perfect fluid, we may recognize a pair of quantities resembling the energy density $\rho$
and pressure $P$ through $T^{00}=\rho$ and $T^{ii}=-Pg^{ii}=a^{-2} P$, respectively, thus leading to
\be\lb{a12}
\rho=\dot{\vp}^2 f'(X)-\left(f(X)-V(\vp)\right),\quad P=f(X)-V(\vp),
\ee
such that the wave equation \eq{a6} is equivalent to the energy conservation law
\be\lb{a13}
\dot{\rho}+3(\rho+P)\frac{\dot{a}}a=0,
\ee
or $\nabla_\nu T^{\mu\nu}=0$, where $\nabla_\nu$ is the covariant derivative. As a consequence, \eq{a11} assumes the form
\be\lb{a14}
\left(\frac{\dot{a}}a\right)^2=\frac{8\pi G}3\rho,
\ee
which is the celebrated Friedmann equation.

For simplicity, we consider the special situation where the potential density function $V$ is constant, say $V_0$. Then \eq{a6} becomes
\be\lb{a15}
\left(a^3 f'(X)\dot{\vp}\right)\dot{}=0,
\ee
which renders the integral
\be\lb{a16}
f'\left(\frac12\dot{\vp}^2\right)\dot{\vp}=\frac c{a^3},
\ee
where $c$ is taken to be a nonzero constant to avoid triviality, which may be set to be positive for convenience. Express $\dot{\vp}$ in \eq{a16} by $\dot{\vp}=
h\left(\frac c{a^3}\right)$. Then \eq{a11} becomes
\bea\lb{a17}
\dot{a}^2&=&\frac{8\pi G}3\left(\dot{\vp}^2 f'\left(\frac12\dot{\vp}^2\right)- f\left(\frac12\dot{\vp}^2\right)+V_0\right)a^2\nn\\
&=&\frac{8\pi G}3\left(\frac ca \,h\left(\frac c{a^3}\right)-a^2 f\left(\frac12\, h^2\left(\frac c{a^3}\right)\right)+ V_0 a^2\right),
\eea
which governs the scale factor $a$ and may be integrated at least {\em  in theory} to yield a dynamical description of the cosmological
evolution under consideration. Below we work out a few illustrative examples.

In the simplest situation where $f(X)=X$, we have $\dot{\vp}=\frac c{a^3}$ and the equation \eq{a17} reads
\be\lb{a18}
\dot{a}^2=\frac{8\pi G}3\left(\frac{c^2}{2a^4}+V_0 a^2\right).
\ee
Comparing \eq{a18} with the classical Friedmann equation with a cosmological constant, $\Lambda$, of the form
\be
H^2=\frac{8\pi G}3\,\rho=\frac{8\pi G}3\left(\rho_m+\frac{\Lambda}{8\pi G}\right),\quad H=\frac{\dot{a}}a,
\ee
with the material energy density $\rho_m$, which is related to the material pressure $P_m$, of the cosmological fluid, through the
energy conservation law
\be\lb{a20}
\dot{\rho}_m+3(\rho_m+P_m)H=0,
\ee
and the barotropic equation of state
\be\lb{a21}
P_m=w\rho_m,
\ee
we arrive at the identification
\be\lb{a22}
\rho_m=\frac{c^2}{2a^6}=\frac12\dot{\vp}^2=P_m,\quad V_0=\frac\Lambda{8\pi G}.
\ee
Thus, $V_0$ gives rise to a cosmological constant. Since the material energy density and pressure are related to their effective 
counterparts by the standard relations
\be\lb{a23}
\rho=\rho_m+\frac\Lambda{8\pi G},\quad P=P_m-\frac{\Lambda}{8\pi G},
\ee
we find from \eq{a12}, \eq{a22}, and \eq{a23} that the effective material energy density, pressure, and cosmological constant should be defined by
\be\lb{a24}
\rho_m=\dot{\vp}^2 f'(X)-f(X),\quad P_m=f(X),\quad \Lambda =\Lambda(\vp)=8\pi G\,V(\vp).
\ee
It is interesting to note that, in such a  context, the cosmological constant is a field-dependent quantity. In the special situation where $f(X)=X$, we have $w=1$ in \eq{a21}.

Since we focus on a homogeneous and isotropic universe for which fields are only time dependent, we have $X=\frac12\dot{\vp}^2$ and we may rewrite the material energy density $\rho_m$ in terms of pressure $P_m$ as
\be\lb{a25}
\rho_m=2X f'(X)-f(X)=2f^{-1}(P_m) f'\left(f^{-1}(P_m)\right)-P_m,\quad X=f^{-1}(P_m),
\ee
which is a nonlinear scalar-wave-matter version of the barotropic equation of state, which is expected to be nonlinear as well in general.

As a first nonlinear example, we note that the scalar field version of the model \eq{.1} may be
\be\lb{a26}
f(X)= b^2\left(1-\left[1-\frac{X}{pb^2}\right]^p\right),\quad 0<p<1.
\ee
Hence
\be\lb{a27}
X=f^{-1}(P_m)=pb^2\left(1-\left[1-\frac{P_m}{b^2}\right]^{\frac1p}\right).
\ee
Inserting \eq{a26} and \eq{a27} into \eq{a25}, we obtain
\bea\lb{a28} 
\rho_m&=& 2pb^2\left(\frac{P_m}{b^2}+\left[1-\frac{P_m}{b^2}\right]^{1-\frac1p}-1\right)-P_m\nn\\
&=&(2p-1)P_m+2pb^2\left(\left[1-\frac{P_m}{b^2}\right]^{1-\frac1p}-1\right).
\eea
Corresponding to the classical Born--Infeld theory,  we have $p=\frac12$ so that \eq{a28} becomes
\be
\rho_m=\frac{b^2 P_m}{b^2- P_m}\quad\mbox{or}\quad P_m=\frac{b^2\rho_m}{b^2+ \rho_m}.
\ee
This example is well known \cite{CGY}. 
Two other simple concrete examples are 
\be
\rho_m=\frac{P_m(3b^4-  P^2_m)}{3(b^2- P_m)^2},\quad p=\frac13;\quad \rho_m=\frac{P_m(2b^6-2b^2 P_m^2+ P_m^3)}{2(b^2- P_m)^3},\quad p=\frac14.
\ee

An elegantly illustrative example is contained in the model \eq{3.1} when we take
\be\lb{a31}
f(X)= X^p,
\ee
where $p\geq1$ is an integer.  Now
\be\lb{a32}
X=P_m^{\frac1p}.
\ee
Inserting \eq{a31} and \eq{a32} into \eq{a25}, we see that the following linear equation of state holds:
\be
\rho_m=(2p-1)P_m, \quad p>0.
\ee
In particular, for $p=2$, we have $P_m=\frac13\rho_m$ and we arrive at a radiation-dominated universe.

For the $\arctan$-model \eq{6.12}, the equation of state for the corresponding scalar field driven universe can also be derived conveniently. In this situation, we
have the following pair of relations
\be\lb{a34}
f(X)=\frac1\beta\arctan (\beta X),\quad X=\frac1\beta\tan (\beta P_m),\quad  \beta>0,
\ee
such that \eq{a25} renders the nonlinear equation
\bea
\rho_m&=&\frac2\beta\tan(\beta P_m)\,\frac1{1+\tan^2(\beta P_m)}-P_m\nn\\
&=&P_m-\frac{4\beta^2}3 P_m^3+\frac{4\beta^4}{15} P_m^5+\mbox{O}\,(\beta^6 P_m^7),\quad\mbox{
for  $P_m$ small.}
\eea

For the logarithmic model \eq{6.30}, we have
\be\lb{a36}
f(X)=-b^2\ln\left(1-\frac X{b^2}\right),\quad X=b^2\left(1-\e^{-\frac {P_m}{b^2}}\right).
\ee
Thus, we obtain from \eq{a25} and \eq{a36} the result
\bea
\rho_m&=&2b^2\left(\e^{\frac {P_m}{b^2}}-1\right)-P_m\nn\\
&=&P_m+\frac{P_m^2}{b^2}+\mbox{O}\left(\frac{P_m^3}{b^4}\right),\quad\mbox{for $P_m$ small}.
\eea
We may also express $P_m$ in terms of $\rho_m$ explicitly by the Lambert $W$ function, along with an asymptotic expansion near
for $\rho_m$ small, as follows,
\bea
P_m&=&-b^2 W\left(-2\e^{-\left(\frac{\rho_m}{b^2}+2\right)}\right)-\rho_m-2b^2\nn\\
&=&\rho_m-\frac{\rho_m^2}{b^2}+\frac{5\rho_m^3}{3b^4}+\mbox{O}\left(\rho_m^4\right),\quad \rho_m\to0.
\eea

For the model \eq{557}, although we have seen that it does not allow finite-energy electric and magnetic point charges, we may still
consider its scalar-wave matter content with
\be\lb{1539}
f(X)=X-\frac\gamma X,\quad \gamma>0,
\ee
say. Using \eq{1539} in \eq{a25}, we get the following equation of state,
\be
\rho_m=-P_m+2\sqrt{P_m^2+4\gamma}=\frac{3P_m^2+16\gamma}{2\sqrt{P_m^2+4\gamma}+P_m}\geq 2\sqrt{3\gamma},
\ee
which indicates that the effect of $\gamma$ is that of a cosmological constant.

For the rational-function model \eq{561}, we have
\be\lb{1541}
f(X)=\frac X{1-2\beta X}.
\ee
Hence, using \eq{a25} and \eq{1541}, we obtain the quadratic equation of state
\be
\rho_m=P_m+4\beta P_m^2.
\ee

For the exponential model \eq{13.44}, we have
\be\lb{a39}
f(X)=\frac1\beta\left(\e^{\beta X}-1\right),\quad X=\frac1\beta\ln\left(1+\beta P_m\right).
\ee
Thus, in view of \eq{a25} and \eq{a39}, we get
\bea\lb{a40}
\rho_m&=&\frac2\beta (1+\beta P_m)\ln\left(1+\beta P_m\right)-P_m\nn\\
&=&P_m+\beta P_m^2-\frac13\,\beta^2 P_m^2+\mbox{O}\left(\beta^3 P^4\right),\quad \mbox{for $P_m$ small.}
\eea
Conversely, we may also express $P_m$ in terms of $\rho_m$ by the explicit formula through inverting \eq{a40} via the
Lambert $W$ function again:
\bea\lb{a41}
P_m&=&\frac1\beta\left(\e^{\frac12+W\left(\frac{\beta\rho_m-1}{2\e^{\frac12}}\right)}-1\right)\nn\\
&=& \rho_m-\beta\rho_m^2+\mbox{O}\left(\beta^2\rho_m^3\right),\quad\rho_m\to0.
\eea

Using \eq{a39} in \eq{a16}, we see that the Friedmann equation \eq{a17} governing the scale factor $a$ reads
\bea\lb{aeq}
\dot{a}^2&=&\frac{8\pi G}3(\rho_m+V_0)a^2\nn\\
&=&\frac{8\pi G}3\left(\frac ca\sqrt{\frac1\beta W\left(\frac{\beta c^2}{a^6}\right)}-\frac{a^2}{\beta}\left[\e^{\frac12 W\left(\frac{\beta c^2}{a^6}\right)}-1\right]+V_0 a^2\right).
\eea
 Moreover, from \eq{a16}, we find
\be\lb{aX}
X=\frac12\dot{\vp}^2=\frac1{2\beta}W\left(\frac{\beta c^2}{a^6}\right).
\ee
Inserting \eq{aX} into \eq{a24} or \eq{a25}, we get
\be\lb{aPP}
\rho_m=\frac1\beta \left(1+\left[W\left(\frac{\beta c^2}{a^6}\right)-1\right]\e^{\frac12 W\left(\frac{\beta c^2}{a^6}\right)}\right),\quad P_m=\frac1\beta
\left(\e^{\frac12W\left(\frac{\beta c^2}{a^6}\right)}-1\right),
\ee
which are both positive-valued quantities. Consequently,
although it may be impossible to obtain an explicit integration of the equation \eq{aeq}, we see that the big-bang solution with $a(0)=0$ is monotone and
enjoys the asymptotic estimates
\be\lb{a}
a(t)\sim t^{\frac23},\quad t\to0; \quad a(t)\sim \left\{ \begin{array}{rl}&t^{\frac13},\quad t\to\infty,\quad\mbox{if }V_0=0;\\
                                                             &  \e^{\sqrt{\frac{8\pi G V_0}3}t},\quad t\to\infty,\quad\mbox{if } V_0>0,\end{array}\right.
\ee
in view of the expansions \eq{13.47} and \eq{13.48}. Since the Kretschmann scalar of the line element \eq{a4} is
\be\lb{aK}
K=\frac{3\left(\dot{a}^4-2a\dot{a}^2 \ddot{a}+2a^2\ddot{a}^2\right)}{2a^4},
\ee
we see that \eq{a} and \eq{aK} give us
\be
K(t)\sim t^{-4},\, t\to0;\quad K(t)\sim t^{-4},\, t\to\infty,\, \mbox{if }V_0=0;\quad\lim_{t\to\infty}K(t)=\frac{32\pi^2 G^2 V_0^2}{3},\, \mbox{if }V_0>0.
\ee
In particular, the big-bang moment, $t=0$, is a curvature singularity.

\begin{figure}[h]
\begin{center}
\includegraphics[height=6cm,width=8cm]{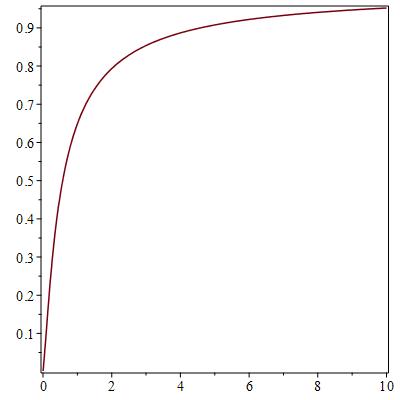}
\caption{A plot of the ratio $\frac{P_m(t)}{\rho_m(t)}$ against the time variable $t$ through $1/W\left(\frac{\beta c^2}{a^6(t)}\right)$ for the exponential model \eq{a39} which indicates that the scalar wave in such a context
monotonically interpolates the dust model at $t=0$ and the stiff matter model at $t=\infty$.}
\label{F2}
\end{center}
\end{figure}

In view of  \eq{aPP},  \eq{a}, and the monotonicity of  the scale factor, we see that both $\rho_m(t)$ and
$P_m(t)$ are decreasing functions of the cosmic time $t$, with the limiting behavior
\be
\lim_{t\to0}\rho_m=\infty,\quad \lim_{t\to0}P_m=\infty,\quad \lim_{t\to\infty}\rho_m=0,\quad\lim_{t\to\infty}P_m=0.\lb{aa}
\ee
The behavior \eq{aa} depicts clearly a big-bang universe which starts from a high-density and high-presssure state and dillutes into
a zero-density and zero-pressure state.
Besides,
\be
\lim_{t\to0}\frac{P_m}{\rho_m}=0,\quad \lim_{t\to\infty}\frac{P_m}{\rho_m}=1.\lb{ab}
\ee
On the other hand,
rewriting $\rho_m$ and $P_m$ given in \eq{aPP} as $\rho_m(W)$ and $P_m(W)$, we have
\be
\frac{\dd}{\dd W}\left(\frac{P_m}{\rho_m}\right)=-\frac{\e^{\frac W2}\left(1+\frac W2-\e^{\frac W2}\right)}{\left(1+(W-1)\e^{\frac W2}\right)^2}<0,\quad W>0.
\ee
Hence the ratio $\frac{P_m(t)}{\rho_m(t)}$ monotonically increases in $t>0$ and interpolates between  its limits 
stated in \eq{ab} at $t=0$ and $t=\infty$, respectively.
This property indicates that the universe starts 
as a dust gas and eventually approaches a stiff matter state, as illustrated in Figure \ref{F2}.

Furthermore, from \eq{a9}, we have
\be
\frac{\ddot{a}}a=-\frac{4\pi G}3\left(\rho_m+3P_m -2V_0\right).
\ee
Hence, if $V_0=0$, then $a(t)$ is globally concave down, and so the expansion of the universe undergoes a decelerated process; if $V_0>0$, $a(t)$ is initially concave down and then concave up, and so the expansion of the universe first experiences a
decelerated process but then an accelerated process after passing through a point of inflection at the moment $t=t_0$ where $t_0>0$ satisfies
\be
\rho_m(t_0)+3P_m(t_0)=2V_0,
\ee
or
\be\lb{t0}
\left(W\left(\frac{\beta c^2}{a^6(t_0)}\right)+2\right)\e^{\frac12W\left(\frac{\beta c^2}{a^6(t_0)}\right)}=2(1+\beta V_0),
\ee
by \eq{aPP}.
It is clear that \eq{t0} has a unique positive solution $t_0$, which may  more explicitly be expressed in term of the Lambert $W$ function as
\be
\frac{\beta c^2}{a^6(t_0)}=W^{-1}\left(2\left[W([1+\beta V_0]\e)-1\right]\right).
\ee

Note that an important checkpoint \cite{GS,K2} for accepting or rejecting a cosmological model is to examine whether the resulting 
adiabatic squared speed of sound, $c^2_{\mbox{\tiny s}}$, would stay in the 
correct range $0\leq c^2_{\mbox{\tiny s}}<1$. Recall that it follows from the classical Newton mechanics that
$
c^2_{\mbox{\tiny s}}$ is the rate of change of the fluid pressure with respect to density. Hence we have
\be\lb{ac}
c^2_{\mbox{\tiny s}}=\frac{\dd P_m}{\dd\rho_m}=\frac{P'_m(X)}{\rho_m'(X)}=\frac{f'(X)}{f'(X)+2Xf''(X)},
\ee
in view of \eq{a24} or \eq{a25}. Thus, inserting \eq{a39} into \eq{ac}, we find
\be
c^2_{\mbox{\tiny s}}=\frac1{1+2\beta X}=\frac1{1+\beta \dot{\vp}^2},
\ee
which indicates the the exponential model \eq{a39} meets the required relevance criterion for the range of $c^2_{\mbox{\tiny s}}$.
More generally, it suffices to demand $f''(X)>0$ for $X>0$. Such a condition is easily examined for many models, including
the polynomial model \eq{3.1}.

For the $\arctan$-model \eq{a34}, however, \eq{ac} assumes the form
\be
c^2_{\mbox{\tiny s}}=\frac{1+\beta^2 X^2}{1-3\beta^2 X^2}.
\ee
Hence there are regions where the criterion fails as also observed in view of electromagnetism \cite{K2}.

On the other hand, for the logarithmic model \eq{a36}, we have
\be
c^2_{\mbox{\tiny s}}=\frac{b^2-X}{b^2+X};
\ee
for the fractional-power model \eq{a26}, we have
\be
c^2_{\mbox{\tiny s}}=\frac{pb^2 -X}{(pb^2-X)+2(1-p)X};
\ee
 for the exponential model \eq{6.1}, we have
\be
f(X)=X\e^{\beta X},\quad c^2_{\mbox{\tiny s}}=\frac{1+\beta X}{1+5\beta X+2\beta^2 X^2};
\ee
and, for the rational-function model \eq{1541}, we have
\be
c^2_{\mbox{\tiny s}}=\frac{1-2\beta X}{1+6\beta X}.
\ee
Thus we may conclude that for all these four later models, $c^2_{\mbox{\tiny s}}$ lies in the anticipated correct range.

In terms of \eq{a25}, the ratio
\be\lb{aax61}
w_m(X)\equiv\frac{P_m(X)}{\rho_m(X)}=\frac{f(X)}{2Xf'(X)-f(X)}
\ee
is also an informative quantity regarding the fluid nature of the scalar-wave matter concerned which is easier to calculate. For example, for the quadratic model \eq{4.1}, we have $f(X)=X+aX^2$ such that
\be
w_m(X)=\frac{1+aX}{1+3a X},
\ee
which indicates that the model interpolates between stiff and radiation-dominated matters, characterized by $w_m=1$ and 
$w_m=\frac13$, respectively; for the logarithmic model \eq{a26}, we have
\be
w_m(X)=-\frac{\ln(1-Y)}{\frac{2Y}{1-Y}+\ln(1-Y)},\quad Y=\frac X{b^2}\in[0,1),
\ee
which bridges between stiff and dust matters, like what described in the exponential model; for the $\arctan$-model \eq{a34}, however, the density function
\be
\rho_m(X)=\frac{2X}{1+(\beta X)^2}-\frac1\beta\arctan(\beta X)
\ee 
cannot stay positive in the full range of $X$. In the range where $\rho_m(X)$ does stay positive, $w_m(X)$ goes from one to infinity as
$X$ approaches the zero $X_0$ of $\rho_m(X)$ from the left, which is about $\frac{1.4}\beta$;
for the rational-function model \eq{561}, $\rho_m(X)$ stays positive and 
\be
w_m(X)=\frac{1-2\beta X}{1+2\beta X},
\ee
which indicates an interesting link between a stiff-matter fluid and that of a cosmological constant;
for the exponential model \eq{6.1},
we have $f(X)=X\e^{\beta X}$ and \eq{aax61} gives us the result
\be
w_m(X)=\frac1{1+2\beta X},
\ee
which is pleasantly simple and shows that the model interpolates stiff and dust matters again; for the fractional-powered model \eq{.1}, it can be shown that $\rho_m(X)$ stays positive and $w_m(X)$ assumes its values between stiff and dust matters as well.

With these examples presented, we are interested in the question whether we are able to identity a Born--Infeld model to realize
an {\em arbitrarily prescribed}  equation of state,
\be\lb{a42}
P_m=w(\rho_m)\quad\mbox{or}\quad \rho_m=v(P_m),
\ee
at least in theory. In fact, by virtue of \eq{a24}, \eq{a25}, and \eq{a42}, we see that the function $f$ in \eq{a1} satisfies
the differential equation
\be\lb{a43}
2X\frac{\dd f}{\dd X}-f(X)=v(f(X))\quad \mbox{or}\quad 2X\frac{\dd f}{\dd X}=v(f)+f,
\ee
which is a first-order separable equation and may readily be solved in general. Note also that this equation is invariant under the
rescaling $X\mapsto \gamma X$ ($\gamma>0$).

Summarizing this discussion, we state

\begin{theorem}\lb{th3}
Adapting the Born--Infeld type action principle to describe scalar-wave matters or k-essences,  theoretically, the equation of state of any cosmological fluid model may be realized by
obtaining
the nonlinearity profile function of a k-essence model by solving a first-order separable ordinary differential equation defined by the equation of state.
\end{theorem}

This theorem may serve as a starting point in the study of a cosmological model in the context of k-essences.

As an example, we consider a generalized Chaplygin fluid model \cite{Av,Carn,CGY,CGY2,Son} with the equation of state
\be\lb{a44}
P_m=P_0+w\rho_m-\frac A{\rho_m^{\frac1k}},\quad k=1,2,\dots,
\ee
where $P_0$, $w>-1$, and $A>0$ are constants and $k=1$ is the classical case. In the linear equation of state limit, $A=0$ and $w\neq0$,
we can use $v(f)=\frac {f-P_0}w$ in \eq{a43} to get 
\be
f(X)=\frac{P_0}{1+w^2}+\alpha X^{\frac12\left(1+\frac1w\right)},
\ee
 where $\alpha>0$ is an integration
constant. Hence $P_0$ serves the role of a cosmological constant.
In the purely nonlinear limit, $w=0$,
from \eq{a44}, we have
\be
\rho_m=\frac{(-A)^k}{(P_m-P_0)^k},
\ee
 or $v(f)=\frac{(-A)^k}{(f-P_0)^k}$ within our formalism. Substituting this into \eq{a43}, we obtain
\be\lb{a46}
2X\frac{\dd f}{\dd X}=\frac{(-A)^k}{(f-P_0)^k}+f.
\ee
An explicit integration of \eq{a46} may be made in the limit when $P_0=0$ to yield
\be
f^{k+1}+(-A)^k=\alpha X^{\frac{k+1}2},
\ee
where $\alpha\neq0$ is an integration constant. Thus we have
\be\lb{a48}
f=f(X)=\left((-1)^{k+1}A^k+\alpha X^{\frac{k+1}2}\right)^{\frac1{k+1}},
\ee
as obtained earlier in  \cite{PL}. In general, an explicit expression for $f(X)$ is not available due to the difficulties associated with
integrating \eq{a43} with \eq{a44}. As an illustration, we take $k=1$ in \eq{a46} to obtain
\be\lb{ax69}
X\frac{\dd (f-P_0)^2}{\dd X}=f^2-P_0 f-A.
\ee
For simplicity, we consider the region away from the equilibrium of the equation \eq{ax69} where $f^2-P_0 f-A>0$. With this condition, we obtain the implicit solution 
\be\lb{ax70}
\frac{\alpha X}{(f^2-P_0f-A)}\left(\frac{\left(f-\frac{P_0}{2}-\sqrt{\left[\frac{P_0}2\right]^2+A}\right)^2}{f^2-P_0f-A}\right)^{\frac{P_0}{\sqrt{P_0^2+4A}}}=1,\quad \alpha>0,
\ee
which is complicated. In the limit $P_0=0$, \eq{ax70} returns to \eq{a48} with $k=1$.
Thus, we see that various generalized
 Chaplygin fluid models may be realized by their corresponding Born--Infeld scalar-wave matter models.

For the quadratic equation of state,
\be\lb{a49}
P_m=A\rho_m^2,\quad A>0,
\ee
which is a special case of a general quadratic model covered in \cite{An1,An2,CGY2,CGQ,Costa},
we have
\be
\rho_m=\sqrt{\frac{P_m}A},
\ee
such that $v(f)=\sqrt{\frac fA}$ in \eq{a43}. That is, $f$ satisfies
\be
2X\frac{\dd f}{\dd X}=\sqrt{\frac fA}+f,
\ee
whose solution is found to be
\be
f(X)=\frac1A\left(\alpha X^{\frac14}-1\right)^2,
\ee
where $\alpha>0$ is an integration constant. As before, it may be checked that the choice of the parameter $\alpha$ does not affect the equation of state, \eq{a49}.

Sometimes the equation of state of a cosmological model is implicitly given in terms of the scale factor $a$. For example, the equation
\be\lb{ab72}
\rho_m=\frac A{a^3}+\frac B{a^{\frac32}},\quad A,B>0,
\ee
directly relating the matter density to the scale factor, defines
a two-fluid model \cite{CGY}, which clearly states  that $\rho_m\to\infty$ as $a\to0$, as desired. On the other hand, we may
rewrite the conservation law \eq{a20} as
\be\lb{ab73}
P_m=-\rho_m-\frac a3\,\frac{\dd \rho_m}{\dd a}.
\ee
Hence, inserting \eq{ab72} into \eq{ab73}, we find
\be\lb{ab74}
P_m=-\frac B{2a^{\frac32}}.
\ee
Combining \eq{ab72} and \eq{ab74}, we arrive at the equation of state for the two-fluid model \eq{ab72}:
\be\lb{ab75}
\rho_m=v(P_m)=-2P_m+\frac{4 A P_m^2}{B^2}.
\ee
Using \eq{ab75} in \eq{a43}, we get the differential equation 
\be
2X\frac{\dd f}{\dd X}=f(\beta f-1),\quad \beta =\frac{4A}{B^2},
\ee
which may be solved to yield
\be\lb{ab77}
f(X)=\frac{X^{\frac12}}{\alpha+\beta X^{\frac12}},
\ee
where $\alpha>0$ is an integration constant. In view of \eq{a25}, it may be checked that \eq{ab77} gives rise to
the same equation of state, \eq{ab75}, for any value of $\alpha$. 

Note that, once the equation of state of the model is identified, we can by-pass solving the wave equation \eq{a15} or \eq{a16}
but instead consider the Friedmann equation \eq{a14} subject to the energy conservation law \eq{a20} directly to determine
the dynamic evolution of the scale parameter $a(t)$. For example, for the Chaplygin model \eq{a44} with $P_0=0, 
w=0, k=1$, we have the big bang solution
satisfying the initial condition $a(0)=0$ 
given implicitly by \cite{CGY}
\bea
a^6(t)&=&\frac{C}{A\left(\coth^4\left[3\beta A^{\frac14}t+b(t)\right]-1\right)},\quad \beta=\sqrt{\frac{8\pi G}3},\lb{a53}\\
b(t)&=&\arctan\left(1+\frac CA a^{-6}(t)\right)^{\frac14}-\frac\pi2,\lb{a54}
\eea
where $C>0$ is an integration constant and the cosmological constant or the potential density function is taken to be zero. Although
this solution looks complicated and is of an implicit form, the function $b(t)$ defined in \eq{a54} stays bounded so that large-time
behavior of $a(t)$ can readily be deduced from \eq{a53} to be
\be\lb{a55}
a(t)\sim \left(\frac C{8A}\right)^{\frac16} \e^{\beta A^{\frac14} t},\quad t\to\infty.
\ee
Since the exponential factor in \eq{a55}, namely $\beta A^{\frac14}$, giving rise to the exponential growth law of the scale factor,
may be regarded as `dark energy', thus we conclude that we may identify an appropriate Born--Infeld model, in the current context
which is realized by a Chaplygin fluid model, to achieve the needed exponential growth pattern phenomenologically.

\section{Conclusions and comments}

In this work several new developments are made along the ideas of the Born--Infeld theory of electromagnetism including
electromagnetic asymmetry in view of energy, charges, and constitutive equations, relegation of black hole singularities rendered by the presence of finite-energy
electric and magnetic point charges, and formulation and interpretation of prescribed equations of state in cosmology in terms of
nonlinear scalar-wave matters.
Specifically, these results are summarized as follows.

\begin{enumerate}

\item[(i)] In the general polynomial model \eq{3.1}, a finite-energy electric point charge is always allowed but a magnetic point charge
can only carry infinite energy, indicating an asymmetry between electricity and magnetism. In the concrete quadratic situation given by \eq{4.1}, such an asymmetry is further demonstrated by observing that the free electric charge generated by the electric field
of an electric point charge coincides with the prescribed charge given by the electric displacement field as in
the classical Born--Infeld theory but the free magnetic charge generated by the magnetic intensity
field of a magnetic point charge is  infinite as well.

\item[(ii)] In the exponential model \eq{6.1} and the logarithmic model \eq{6.30}, there is an electromagnetic symmetry as in
the classical Born--Infeld theory, such that both electric and magnetic point charges of finite energies are allowed and that
the free charges coincide with the prescribed ones. On the other hand, in the $\arctan$-model \eq{6.12}, an electric point charge
is not allowed due to an inconsistent constitutive equation, but a finite-energy magnetic point charge is allowed whose free
magnetic charge coincides with the prescribed one. Moreover, in the $\arcsin$-model \eq{6.18}, the situation is just the opposite:
An electric point charge is allowed but a magnetic charge not due to an inconsistent constitutive equation again. These examples
further illustrate the symmetry and asymmetry of electromagnetism embodied in nonlinear electrodynamics in general.

\item[(iii)] In the fractional-powered model \eq{.1}, there is a much refined division concerning the electromagnetic symmetry and
asymmetry given in terms of the power parameter $p$: In the full range of $0<p<1$, a finite-energy electric point charge with
its free electric charge agreeing with its prescribed charge is allowed but a finite-energy magnetic charge is only allowed in part of the range, given by
$0<p<\frac34$. Nevertheless, in the full range of $0<p<1$, the free magnetic charge of the magnetic intensity field
generated by a magnetic point charge coincides with the prescribed charge regardless whether its energy is finite or infinite.
In other words, the electromagnetic asymmetry of the model is now exhibited by an energy rather than charge content.

\item[(iv)] Several models studied here, including the classical Born--Infeld model, the $\arcsin$-model, the logarithmic model,
and the fractional-powered model, indicate that a dyonic point charge blends its 
prescribed electric and magnetic charges by its total free electric and magnetic charges in an asymmetric or skew-symmetric manner,
which is unobservable at the energy level. Another common feature shared by these models is that, both near the spot where the
dyonic charge resides and near infinity, there is a perfect symmetry between the electric and magnetic charges energetically but
asymmetry field-theoretically. In other words, asymptotically, the Hamiltonian energy density in each model is symmetric with respect to an
interchange of the electric and magnetic charges of a dyonic point particle, which appears like a dyon with freely {\em switchable}  electric and
magnetic charges locally near the dyon and asymptotically near infinity. Besides, some detailed properties of the dyon
at $p=\frac34$ in the
fractional-powered model show that the energy of a dyon diverges faster than that of a magnetic monopole, which reveals further
some
delicacy carried by a dyonic point charge.

\item[(v)] It is shown that finite energy achieved in the Born--Infeld theory formalism for an electric point charge relegates
the well-known curvature singularity of the Reissner--Nordstr\"{o}m black hole of the type $K\sim r^{-8}$ to that of the
Schwarzschild black hole of the type $K\sim r^{-6}$,
in terms of the Kretschmann curvature $K$ and radial variable $r$, near $r=0$. Moreover, under a critical mass-energy condition, that is, when the mass $M$
and energy $E$ of an electrically charged black hole are equal, $M=E$ in our context, the singularity may be relegated further,
depending on the specific models considered: For the quadratic model \eq{4.1}, there holds
$K\sim r^{-\left(5+\frac13\right)}$; for the exponential model \eq{6.1},
$K\sim r^{-(4+\vep)}$ where $\vep>0$ is arbitrarily small; for the $\arcsin$-model \eq{6.18}, logarithmic model \eq{6.30}, 
 the entire range of the fractional-powered model \eq{.1}, including the classical Born--Infeld model at $p=\frac12$, 
and the rational-function model \eq{561}, one obtains $K\sim r^{-4}$, which is indeed seen to be
much relegated or regularized from the Schwarzschild singularity.

\item[(vi)] A general formalism of dyonically charged black hole solutions arising in the Born--Infeld type electromagnetism is 
presented which enables a ready access to acquiring such solutions in various concrete situations. In particular, as seen in the
situation of electrically charged black hole solutions,  in general, finite-energy magnetically charged black holes also give rise to the same
curvature singularity as that of a purely massive Schwarzschild black hole, as finite-energy electrically charged black holes. Furthermore, under the same critical mass-energy condition, curvature singularity is significantly relegated more explicitly and 
easily than what may be achieved in electric situations, as a result of how magnetic sector presents itself in the formalism.
Specifically, for the fractional-powered model \eq{.1}, there holds $K\sim r^{-8p}$ for
$0<p<\frac34$, with the upper borderline limt $p\to\frac34$ the Schwarzschild type curvature singularity, $K\sim r^{-6}$,
for a purely massive black hole; for the logarithmic model \eq{6.30}, there holds $K\sim (\ln r)^2$; for the exponential model
\eq{6.1} and the $\arctan$-model \eq{6.12}, the magnetically charged black holes are free of curvature singularity, or become regular. On the other hand, we have seen that magnetically charged black holes may also generate curvature singularities with arbitrarily high
blow-up rates at $r=0$.

\item[(vii)] The Bardeen black hole, a much studied regular black hole solution arising from coupling a Born--Infeld type 
electromagnetic theory with Einstein equations, is recognized as a special case of a rather broad family of regular black hole solutions,
permitting integration in terms of elementary functions in light of the Chebyshev theorem.

\item[(viii)] The Hayward black hole, another well studied regular charged black hole solution, is found to belong to a broad family
of regular black hole solutions arising from the Born--Infeld type electromagnetism. Interestingly, this family of solutions may all
be represented explicitly in terms of rational functions through an elementary integration. 

\item[(ix)] Although the Bardeen and Hayward type charged black holes are regular, they exclude either electricity or magnetism
in the Lagrangian action densities in the magnetically or electrically charged theories. A similar phenomenon has been seen earlier in
the $\arctan$-model \eq{6.12}. The exponential model \eq{13.44}, on the other hand, is shown to give rise to regular 
electrically or magnetically charged black hole solutions, but at the same time, allow finite-energy magnetically or electrically
charged black holes with relegated curvature singularities. Such charged black hole solutions may be constructed
explicitly through the Lambert $W$ function. Specifically, under the critical mass-energy condition, the Kretschmann curvature
$K$ of a magnetically charged black hole enjoys the property $K-K_0\sim r^m$ for $r$ small, where $K_0$ is a constant and
$m$ is an arbitrarily large integer, and $K\sim r^{-4}(\ln r)^2$ for an electrically charged black hole. Not surprisingly, the
constructed regular magnetic black holes violate the strong energy condition, although they fulfill both weak and dominant
energy conditions. Dyonically charged black hole solutions in the exponential model \eq{13.44} are also obtained explicitly which
exhibit the interesting property that their electric field is electric near infinity but magnetic near $r=0$ where the dyonic charge resides
and the magnetic intensity field is magnetic near infinity but electric near $r=0$, enjoying similar $r$-depenent properties and possessing
finite free electric and magnetic total charges, although the energy of a dyonic point charge diverges near $r=0$.
Furthermore, there is an electromagetic symmetry near spatial infinity but an asymmetry near $r=0$, demonstrated by both
electric and magnetic intensity fields and the energy density. In other words, we observe a hidden asymmetry of electromagnetism with a dyonic point charge.

\item[(x)] From the viewpoint of a general formalism, it is shown that the Born--Infeld nonlinear field theory provides a rich source of k-essence models
for cosmological applications including obtaining a desired accelerated expansion dynamics for the
isotropic and homogeneous universe and 
field-theoretical interpretations of various equations of state of hypothetical cosmic fluids.
For the exponential model \eq{13.44}, in particular, it is established that the equation of state relating the pressure and density of the 
cosmological fluid is given by the Lambert $W$ function so that the universe starts as a dust gas and eventually approaches
a stiff-matter state.
\end{enumerate}

It will also be interesting to consider dyonically charged singularity-relegated and regular black hole
solutions and implications and applications of the lines of this work to higher-dimensional and higher-derivative
theories.

\medskip

{\bf Data availability statement}: The data that supports the findings of this study are
available within the article.

 \end{document}